\documentclass[twocolumn,twoside,10pt,aps,pra,superscriptaddress,amsmath,amssymb,longbibliography]{revtex4-1}
\usepackage[switch]{lineno}
\usepackage[english]{babel}

\usepackage{mathrsfs}
\usepackage{graphicx}
\usepackage{amsmath}
\usepackage{amssymb}
\usepackage{amsfonts}
\usepackage{color}
\usepackage{leftidx}
\usepackage{verbatim}

\def\bra#1{\left\langle{#1}\right|}
\def\ket#1{\left|{#1}\right\rangle}
\def\braket#1#2{\left\langle{{#1}}\mathrel{\left|{\vphantom{{#1}{#2}}}\right.\kern-\nulldelimiterspace}{{#2}}\right\rangle}

\begin{document}

\title{Symmetry-protected many-body Ramsey spectroscopy: precision scaling and robustness}
%
%
\author{Sijie Chen}
\affiliation{Institute of Quantum Precision Measurement, State Key Laboratory of Radio Frequency Heterogeneous Integration, College of Physics and Optoelectronic Engineering, Shenzhen University, Shenzhen 518060, China}
\affiliation{Laboratory of Quantum Engineering and Quantum Metrology, School of Physics and Astronomy, Sun Yat-Sen University (Zhuhai Campus), Zhuhai 519082, China}

\author{Jiahao Huang}
\affiliation{Laboratory of Quantum Engineering and Quantum Metrology, School of Physics and Astronomy, Sun Yat-Sen University (Zhuhai Campus), Zhuhai 519082, China}
\affiliation{Institute of Quantum Precision Measurement, State Key Laboratory of Radio Frequency Heterogeneous Integration, College of Physics and Optoelectronic Engineering, Shenzhen University, Shenzhen 518060, China}
\author{Min Zhuang}
\altaffiliation{Email: mmzhuang@szu.edu.cn}
\affiliation{Institute of Quantum Precision Measurement, State Key Laboratory of Radio Frequency Heterogeneous Integration, College of Physics and Optoelectronic Engineering, Shenzhen University, Shenzhen 518060, China}
\author{Chaohong Lee}
\altaffiliation{Email: chleecn@szu.edu.cn, chleecn@gmail.com}
\affiliation{Institute of Quantum Precision Measurement, State Key Laboratory of Radio Frequency Heterogeneous Integration, College of Physics and Optoelectronic Engineering, Shenzhen University, Shenzhen 518060, China}
\affiliation{Quantum Science Center of Guangdong-Hong Kong-Macao Greater Bay Area (Guangdong), Shenzhen 518045, China}
%
\begin{abstract}
Quantum entanglement is a powerful quantum resource for enhancing measurement precision beyond classical limit. 
Here we propose an entanglement-enhanced symmetry-protected destructive many-body Ramsey interferometry for precise parameter estimation.
Through matching the symmetry of input states and Hamiltonian, the spectral shift of Ramsey spectroscopy caused by interparticle interaction, noise, decoherence and experimental imperfection can be mitigated for both time-independent and time-dependent signals, as explored in the companion Letter  [S. Chen, et al., Ramsey Spectroscopy via Symmetry-Protected Destructive Many-Body Interferometry (submitted)].
In this work, we show that symmetric entangled input states can further improve the measurement precision of symmetry-protected Ramsey spectroscopy without affecting the measurement accuracy.
Through inputting spin cat states and applying suitable interaction-based readout operations, the measurement precisions of time-independent and time-dependent signals can both approach the Heisenberg limit.
In particular, we systematically analyze how precision scaling and robustness depend on the input states, parameters, experimental imperfection and decoherence.
This work establishes a practical pathway to Heisenberg-limited quantum metrology with many-body Ramsey interferometry, offering resilience against realistic noise and imperfections.

\end{abstract}

\date{\today}
\maketitle
\section{INTRODUCTION\label{Sec1}}
Quantum metrology employs quantum effects to overcome classical measurement limits, achieving unprecedented precision. 
Ramsey interferometry serves as a foundational technique in this field, enabling precise measurements of transition frequencies, magnetic fields, and accelerations and offering promising applications in atomic clocks~\cite{ADLRMP2015,ADCRMP2009,KHRMP2010,LPRMP2018,MGKRMP2018,ADCRMP2009}, quantum magnetometers~\cite{YSGRMP2000,CLDRMP2017,JFBRMP2020,MWMRMP2020} and quantum gravimeters~\cite{SBRMP2025,CCPRX2025}.
Quantum entanglement in many-body systems is known to provide substantial enhancements in measurement precision~\cite{FCRMP2014,LPRMP2018,JHAPR2024}.
For $N$ uncorrelated particles, the measurement precision is bounded by the standard quantum limit (SQL), scaling as $\propto 1/\sqrt{N}$ due to the central limit theorem. 
Quantum entanglement enables surpassing this limit, with non-Gaussian entangled states — such as spin cat states and Greenberger–Horne–Zeilinger (GHZ) states —playing a pivotal role in quantum metrology~\cite{FCRMP2014,LPRMP2018,JHAPR2024}. 
These states achieve the Heisenberg limit (HL), offering precision scaling as 
$\propto 1/\sqrt{N}$, which represents the ultimate precision bound for parameter quantum-enhanced estimation~\cite{FCRMP2014,LPRMP2018,JHAPR2024}.

In the companion Letter~\cite{SCPR2025}, to mitigate the spectral shift caused by interparticle interaction, noise, decoherence and experimental imperfection, we propose a Ramsey spectroscopy based upon symmetry-protected destructive many-body interferometry (SPDMBI). 
This work provides an in-depth analysis on the measurement precision enhancement and robustness of SPDMBI-based Ramsey spectroscopy with quantum entanglement.
Our protocol can be directly applicable to precise estimation of both time-independent and time-dependent signals~\cite{SCPR2025}. 
However, enhancing measurement precision while preserving accuracy through quantum entanglement remains an outstanding challenge. 
Building on entanglement-enhanced quantum interferometry, we address two fundamental questions: 
(i) Can symmetry-protected destructive Ramsey interferometry be enhanced via entanglement? and
(ii) Can such a protocol simultaneously improve precision and maintain accuracy?

In this article, we propose an entanglement-enhanced symmetry-protected destructive Ramsey interferometry protocol that achieves Heisenberg-limited precision while suppressing interaction-induced frequency shifts.
Especially for spin cat states, we demonstrate that the precision degradation caused by detuning errors with conventional readout can be effectively mitigated by a composite nonlinear readout. 
This process combines two interaction-based operations with two $\pi/2$-pulses and an intermediate $\pi$-pulse, preserving the HL scaling even under realistic experimental imperfections.
For ac field sensing, to prevent the accumulated phase ambiguity that limits the dynamic range of ac field magnitude in conventional many-body quantum lock-in measurement, we propose a novel scheme for ac field sensing via adding an extra dc field.
All the protocols are designed within the framework of SPDMBI, highlighting the broad potential applications of SPDMBI in quantum sensing. 

This article is organized as follows.
In Sec.~\ref{Sec2}, we introduce our general protocol for symmetry-protected destructive many-body Ramsey interferometry (SPDMBI-Ramsey) for both non-entangled and entangled input states.
In Sec.~\ref{Sec3}, we apply our SPDMBI-Ramsey to the measurement of time-independent signal and show how to improve the precision from SQL to HL using spin cat states.
In Sec.~\ref{Sec4}, we apply our SPDMBI-Ramsey to the measurement of time-dependent signal. To measure the ac field magnitude unambiguously, we show how to perform the ac field sensing by applying an extra dc field, in which the ac field magnitude is determined according to the applied dc field magnitude. 
In particular, we demonstrate how to achieve the Heisenberg-limited detection via spin cat states and discuss its robustness against interparticle interaction.
Finally, we give a brief summary and discussion in Sec.~\ref{Sec5}.
\section{GENERAL PROTOCOL\label{Sec2}}
In this section, we provide a brief introduction to SPDMBI.
We consider an ensemble of two-mode bosonic systems with $N$ particles, the two modes can be selected as any desired magnetic levels labeled as spins $\ket{\uparrow}$ and $\ket{\downarrow}$, respectively.
By using the Schwinger representation, the system can be characterized well by the collective spin operators $\hat{J}_x =\frac{1}{2}
\left(\hat{a}^{\dagger}\hat{b}+\hat{a}\hat{b}^{\dagger}\right)$, $\hat{J}_y=\frac{1}{2i}\left(\hat{a}^{\dagger}\hat{b}-\hat{a}\hat{b}^{\dagger}\right)$ and $\hat{J}_z=\frac{1}{2}\left(\hat{a}^{\dagger}\hat{a}-\hat{b}^{\dagger}\hat{b}\right)$, where $\hat{a}$ and $\hat{b}$ denote the annihilation operators for spins $\ket{\uparrow}$ and $\ket{\downarrow}$, respectively.
A system state can be represented in terms of the Dike basis $\ket{J,m}$, with $J=N/2$ and $m=\{-J,-J+1,...,J-1,J\}$.
Given the probe as an ensemble of two-mode bosonic particles, there is a nonlinear interaction $\hat{H}_\textrm{a}=\chi\hat{J}_z^2$~\cite{JHarXiv2022,TParXiv2025} with $\chi$ characterizing the strength, and the coupling between the probe with the target signal $S(s,t)$ is described by
the Hamiltonian
\begin{equation}\label{Hs}
    \hat{H}_\textrm{s}=S(s,t) \hat{J}_z,
\end{equation}
where $S(s,t)$ is a function of $s$ and time $t$ and $s$ is the parameter to be measured.
To infer the information of $s$, we apply an auxiliary control with additional fields $\hat{H}_\textrm{c}=\vec{R}(r,t)\cdot\hat{\vec{J}}$ with an adjustable parameter $r$, and the whole Hamiltonian becomes
\begin{equation}\label{Hs}
    \hat{H}_{w}=\hat{H}_a+\hat{H}_\textrm{s}+\hat{H}_\textrm{c}.
\end{equation}
For both Rabi and Ramsey spectroscopy, after the interaction picture transformation, the Hamiltonian~\eqref{Hs} becomes
\begin{eqnarray}\label{HwI1}
    \hat{H}_\textrm{I}(\delta,t)&=&f_1^\textrm{even}\hat{J}_x+f_2^\textrm{odd}\hat{J}_y+f_3^\textrm{odd}\hat{J}_z\nonumber\\
    & &+e_1^\textrm{even}\hat{J}_y\hat{J}_z+e_2^\textrm{odd}\hat{J}_z\hat{J}_x+e_3^\textrm{odd}\hat{J}_x\hat{J}_y\nonumber\\
    & &+g_1^\textrm{even}\hat{J}_x^2+g_2^\textrm{even}\hat{J}_y^2+g_3^\textrm{even}\hat{J}_z^2,
\end{eqnarray}
where $x_i^\textrm{even}(-\delta,t)=x_i^\textrm{even}(\delta,t)$ and $x_i^\textrm{odd}(-\delta,t)=-x_i^\textrm{odd}(\delta,t)$ $(x=f,g,e)$ are symmetric and antisymmetric functions versus $\delta=s-r$, respectively.
Therefore, Hamiltonian~\eqref{HwI1} has the symmetry
\begin{equation}\label{H_sym1} 
    \hat{U}_\textrm{ex}^{\dagger}\hat{H}_I(\delta,t)\hat{U}_\textrm{ex}=\hat{H}_I(-\delta,t)
\end{equation}
with $\hat{U}_\textrm{ex}=e^{-i\pi\hat{J}_x}$ describing the exchange transformation $\hat{a}\leftrightarrow\hat{b}$.
In spectroscopy, one usually estimate $\delta$ to infer $s$ for given $r$.
For an initial state $\ket{\Psi(0)}=\sum_{m=-J}^{J}C_m(0)\ket{J,m}$ whose density matrix $\hat{\rho}(0)=\ket{\Psi(0)}\bra{\Psi(0)}$, under Markovian noises, the time-evolution can be described by a Lindblad master equation~\cite{HPB2007,DMAIPA2020} (in units of $\hbar=1$)
\begin{equation}\label{Lindblad}
    \frac{\partial\hat{\rho}(\delta,t)}{\partial t}=-i\left[\hat{H}_I(\delta,t),\hat{\rho}(\delta,t)\right]+\mathcal{L}\hat{\rho}(\delta,t),
\end{equation}
where $\mathcal{L}\hat{\rho}=\sum_{k}\gamma_k\left(\hat{\mathcal{L}}_k\hat{\rho}\hat{\mathcal{L}}_k^\dagger-\frac{1}{2}\left\{\hat{\mathcal{L}}_k^\dagger\hat{\mathcal{L}}_k,\hat{\rho}\right\}\right)$ describes the decoherence. 
When the initial state is symmetric, its coefficients satisfy
\begin{equation}\label{Coef}
    C_m(0)=\pm C_{-m}(0).
\end{equation}
Due to $\hat{U}_\textrm{ex}^\dagger\ket{J,m}=i^N\ket{J,-m}$, we have $\hat{U}_\textrm{ex}^{\dagger} \hat{\rho}(0)\hat{U}_\textrm{ex}=\hat{\rho}(0)$ (see {Appendix A} for more details).
Meanwhile, if the Lindblad operator also satisfies 
\begin{equation}\label{LS}
\hat{U}_\textrm{ex}^{\dagger}\mathcal{L}\hat{\rho}\hat{U}_\textrm{ex}=\mathcal{L}\hat{U}_\textrm{ex}^{\dagger}\hat{\rho}\hat{U}_\textrm{ex}
\end{equation}
(such as collective dephasing~\cite{HPB2007} and balanced atom losses~\cite{UDPRL2009,JLNC2019,JHSR2016}), we have
\begin{eqnarray}\label{Ld1}
&&\frac{\partial\hat{U}_\textrm{ex}^\dagger\hat{\rho}(\delta,t)\hat{U}_\textrm{ex}}{\partial t}=\hat{U}_\textrm{ex}^\dagger\partial_t\hat{\rho}(\delta,t)\hat{U}_\textrm{ex}\\\nonumber
&&=-i\hat{U}_\textrm{ex}^\dagger\left[\hat{H}_I(\delta,t),\hat{\rho}(\delta,t)\right]\hat{U}_\textrm{ex}+\hat{U}_\textrm{ex}^{\dagger}\mathcal{L}\hat{\rho}(\delta,t)\hat{U}_\textrm{ex}\\\nonumber
&&=-i\left[\hat{H}_I(-\delta,t),\hat{U}_\textrm{ex}^\dagger\hat{\rho}(\delta,t)\hat{U}_\textrm{ex}\right]+\mathcal{L}\hat{U}_\textrm{ex}^{\dagger}\hat{\rho}(\delta,t)\hat{U}_\textrm{ex}
\end{eqnarray}
and
\begin{eqnarray}\label{L-d1}
\frac{\hat{\rho}(-\delta,t)}{\partial t}=-i\left[\hat{H}_I(-\delta,t),\hat{\rho}(-\delta,t)\right]+\mathcal{L}\hat{\rho}(-\delta,t).
\end{eqnarray}
Comparing Eqs.~\eqref{Ld1} and \eqref{L-d1} and considering the symmetry of $\ket{\Psi(0)}$ and Hamiltonian $\hat{H}_I(\delta,t)$, $\hat{\rho}(\delta,t)$ can be uniquely determined. Therefore, we can obtain $\hat{U}_\textrm{ex}^\dagger\hat{\rho}(\delta,t)\hat{U}_\textrm{ex}=\hat{\rho}(-\delta,t)$ which satisfy $\hat{U}_\textrm{ex}^\dagger\hat{\rho}(\delta,0)\hat{U}_\textrm{ex}=\hat{\rho}(-\delta,0)$.
Moreover, we have the expectation of half-population difference
\begin{eqnarray}\label{Jz}
    \langle\hat{J}_z(t)\rangle_{-\delta}&=&\textrm{Tr}[\hat{\rho}(-\delta,t)\hat{J}_z]=\textrm{Tr}[\hat{U}_\textrm{ex}^\dagger\hat{\rho}(\delta,t)\hat{U}_\textrm{ex}\hat{J}_z]\nonumber\\\nonumber
    &=&\textrm{Tr}[\hat{U}_\textrm{ex}\hat{J}_z\hat{U}_\textrm{ex}^\dagger\hat{\rho}(\delta,t)]=-\textrm{Tr}[\hat{J}_z\hat{\rho}(\delta,t)]\\
    &=&-\langle\hat{J}_z(t)\rangle_{\delta}
\end{eqnarray}
which is antisymmetric about detuning $\delta$ at any time $t$.
In particular, at the resonance point $\delta=0$, the half-population difference always vanishes, i.e. 
\begin{equation}
    \langle \hat J_z(t)\rangle_{\delta=0}=0.
\end{equation}
Therefore, we call the above destructive interferometry as symmetry-protected destructive many-body interferometry (SPDMBI).

\begin{figure}[!htp]
	\includegraphics[width=\columnwidth]{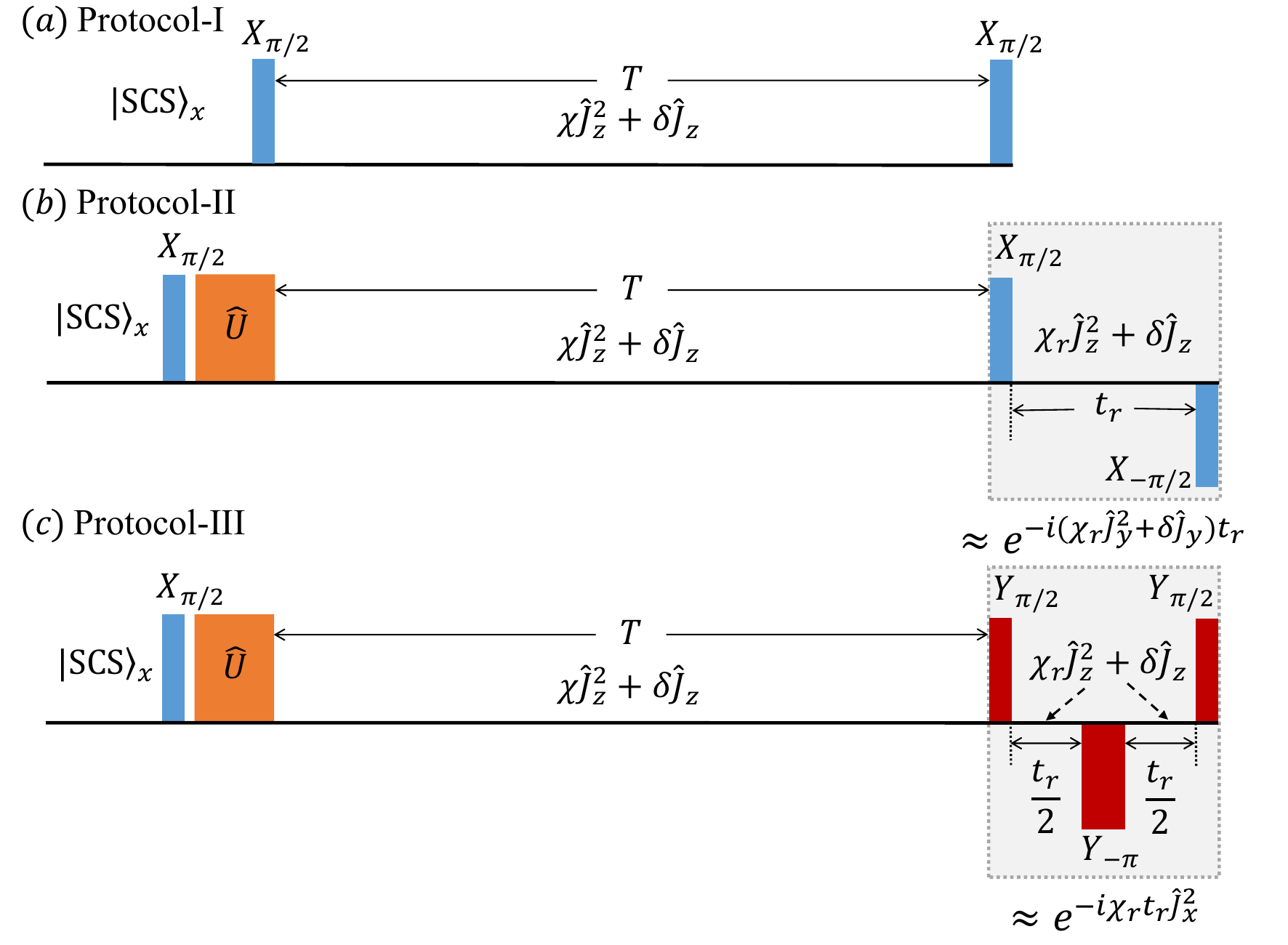}
	\caption{\label{Fig1}(color online).
The SPDMBI-Ramsey protocols for measuring time-independent signal. 
All protocols consist of three stages: initialization, interrogation, and readout.
(a) Protocol-I: For individual particles, a spin coherent state $\ket{\textrm{SCS}}_x$ is input and after the interrogation, a $\pi/2$ pulse along $x$ axis $\hat{U}_\textrm{re}=e^{-i\frac{\pi}{2}\hat{J}_x}$ is applied for readout.
(b) Protocol-II: Start from $\ket{\textrm{SCS}}_x$, an entangled states that satisfies Eq.~\eqref{Coef} (e.g. spin cat states $\ket{\textrm{CAT}(\theta)}$) is prepared by $\hat U$. After an interrogation of $T$, a nonlinear readout operator $\hat{U}_\textrm{re}=e^{-i\frac{\pi}{2}\hat{J}_y^2-i\delta t_r\hat{J}_y}$ is applied by combining two $\pi/2$ pulses along $\pm x$ axis with an interrogation of $t_r=\pi/(2\chi_r)$.
(c) Protocol-III: Start from $\ket{\textrm{SCS}}_x$, an entangled states that satisfies Eq.~\eqref{Coef} is prepared by $\hat U$. After an interrogation of $T$, a nonlinear readout operator $\hat{U}_\textrm{re}=e^{-i\frac{\pi}{2}\hat{J}_x^2}$ is applied by combining two $\pi/2$ pulses along $y$ axis and a $\pi$ pulse along $-y$ axis with two interrogation of $t_r/2=\pi/(4\chi_r)$.
%
%
%
All protocols satisfy SPDMBI under the condition of ideal pulses.
}
\end{figure}

Our SPDMBI can be readily applied to the widely used Ramsey spectroscopy. 
If all the three stages (initialization, interrogation, and readout) satisfy the above symmetry requirements, antisymmetric Ramsey spectra can be obtained, as shown in {Fig.~\ref{Fig1}} for time-independent signal and {Fig.~\ref{Fig7}} for time-dependent signal.
In the initialization stage, a symmetric state $\ket{\Psi(0)}=\sum_mC_m(0)\ket{J,m}$ with $C_m(0)=\pm C_{-m}(0)$ is prepared.
Then, the initial state goes through an interrogation stage with time $T$ and becomes $\ket{\Psi(T)}$.
Finally in the readout stage, another unitary operation $\hat{U}_\textrm{re}$ is performed for recombination.
%
%
Owing to the symmetry protection, our protocol is robust against inaccurate time control and several other noises such as Rabi frequency fluctuation, correlated dephasing and balanced atom losses even when taking the interparticle interaction into account~\cite{SCPR2025}.

Apart from accuracy, the precision of the spectroscopic measurement is also important.
Quantum entanglement is an effective quantum resource for improving measurement precision.
The measurement precision can be further improved by inputting symmetric entangled states for Ramsey spectroscopy.
The symmetric entangled states can be the typical metrologically useful entangled states~\cite{CMarXiv2025} such as spin squeezed state~\cite{MKPRA1993,PYNPJ2025}, twin-Fock state~\cite{MRPRA2007,XLScience2017}, spin cat state~\cite{JHSR2016,SSPRB2021,JHPRA2022,FDPRA2025}, and etc~\cite{CMarXiv2025}.
Among all, the spin cat state, a typical kind of non-Gaussian entangled state, is a promising candidate for entanglement-enhanced metrology. 
By employing spin cat state as input, the measurement precision can be improved to the Heisenberg limit~\cite{LPRMP2018,JHAPR2024}.
In the next sections, we mainly discuss how to realize an entanglement-enhanced symmetry-protected destructive Ramsey interferometry using spin cat states to measure the time-independent and time-dependent signal without affecting the measurement accuracy. 
As an example, we will focus on the dc and ac magnetic field sensing below.  

\section{Time-independent signal measurement via SPDMBI-Ramsey\label{Sec3}}

In this section, we illustrate how to estimate the time-independent signal within our scheme and show how entanglement can improve the measurement precision.
We consider an ensemble of $N$ spin-$1/2$ interacting particles in a static magnetic field $B_z\hat{z}$ to be measured and a control magnetic field $B_\perp[\cos(\omega_r t)\hat{x}+\sin(\omega_r t)\hat{y}]$ with $B_\perp$ and $\omega_r$ respectively representing the magnitude and oscillation frequency of the control field.
Therefore, the interparticle interaction $\hat{H}_a=\chi\hat{J}_z^2$, the signal Hamiltonian $\hat{H}_s=\omega_s\hat{J}_z$ and the control Hamiltonian $\hat{H}_c=\Omega(t)[\cos(\omega_r t)\hat{J}_x+\sin(\omega_r t)\hat{J}_y]$ with the frequency $\omega_s=\gamma_gB_{z}$, the Rabi frequency $\Omega(t)=\gamma_g B_\perp(t)$ and $\gamma_g$ denoting the gyromagnetic ratio.
Transforming $\hat{H}_{w}=\hat{H}_a+\hat{H}_s+\hat{H}_c$ into the interaction picture with respect to $\hat{H}_0=\omega_r\hat{J}_z$ and compensating the linear shift, the Hamiltonian under the interaction scene is 
\begin{equation}\label{Hram}
    \hat{H}_\textrm{Ram}=\chi\hat{J}_z^2+\delta\hat{J}_z+\Omega(t)\hat{J}_x
\end{equation}
with the detuning $\delta=\omega_s-\omega_r$ and the interaction strength $\chi$.


%
%
According to the parameter quantum estimation theory, the precision of the parameter is constrained by the quantum Cram\'{e}r-Rao bound (QCRB)~\cite{CWH1976,FCRMP2014,MSAPX2016,LPRMP2018,JHAPR2024},
\begin{eqnarray}\label{CRB}
\Delta B_z \geq \Delta B_z^\textrm{QCRB}\equiv \frac{1}{\sqrt{\nu F_{Q}^{B_z}}},
\end{eqnarray}
which is characterized by the quantum Fisher information (QFI):
\begin{eqnarray}\label{QFI10}
F_\textrm{Q}^{B_z}&=&4\!\left(\langle \partial_{B_z}\Psi(T)|\partial_{B_z}\Psi(T)\rangle\!-\!|\langle \partial_{B_z}\Psi(T)|\Psi(T)\rangle|^2\right).\nonumber\\
\end{eqnarray}
with $\ket{\Psi(T)}=e^{-i(\chi\hat{J}_z^2+\delta\hat{J}_z)T}\ket{\Psi(0)}$.
$\ket{\Psi(0)}$ is the initial state and $T$ is the evolution time, $\nu$ corresponds to the number of trials, $|\partial_{B_z}\Psi(T)\rangle$ denotes the partial derivative of $\ket{\Psi(T)}$ with respect to the parameter $B_z$.
In our discussion, we choose $\nu=1$ for convenience.
Therefore, according to Eq.~\eqref{QFI10} the expression of the QFI can be simplified to
\begin{eqnarray}\label{QFI11}
    F_\textrm{Q}^{B_z}
    &=& 4T^2\gamma_g^{2} \Delta^2 \hat J_z(0)
\end{eqnarray}
with $\Delta \hat J_z(0)=\sqrt{\bra{\Psi(0)}\hat{J}_{z}^{2}\ket{\Psi(0)} - \left(\bra{\Psi(0)}\hat{J}_{z}\ket{\Psi(0)}\right)^2}$.
Thus, the properties of initial states determine the ultimate measurement precisions.

As shown in the following, $\Delta^2 \hat J_z(0) \propto N$ for individual particles, while for entangled particles, $\Delta^2 \hat J_z(0) \propto N^2$, indicating entanglement has the ability to enhance the measurement precision from SQL to HL.
To further characterize the measurement precision of $B_{z}$, we use the error propagation formula~\cite{CWH1976,NB2021}.
The measurement precision of $B_z$ is
\begin{equation}\label{dBz}
    \Delta B_z=\frac{(\Delta{{\hat J}_{z}})_f}{|\partial{\langle\hat{J}_{z}\rangle_f}/ \partial{B_z}|}=\frac{(\Delta{{\hat J}_{z}})_f}{\gamma_g|\partial{\langle\hat{J}_{z}\rangle_f}/ \partial{\delta}|},
\end{equation}
where
\begin{equation}\label{dJzf}
    (\Delta{{\hat J}_{z}})_f=\sqrt{\langle\hat{J}_z^2\rangle_f-\langle\hat{J}_{z}\rangle_f^2}
\end{equation}
and $\langle\hat{J}_z\rangle_f=\bra{\Psi_f}\hat{J}_z\ket{\Psi}_f$.

\begin{figure}[!htp]
	\includegraphics[width=\columnwidth]{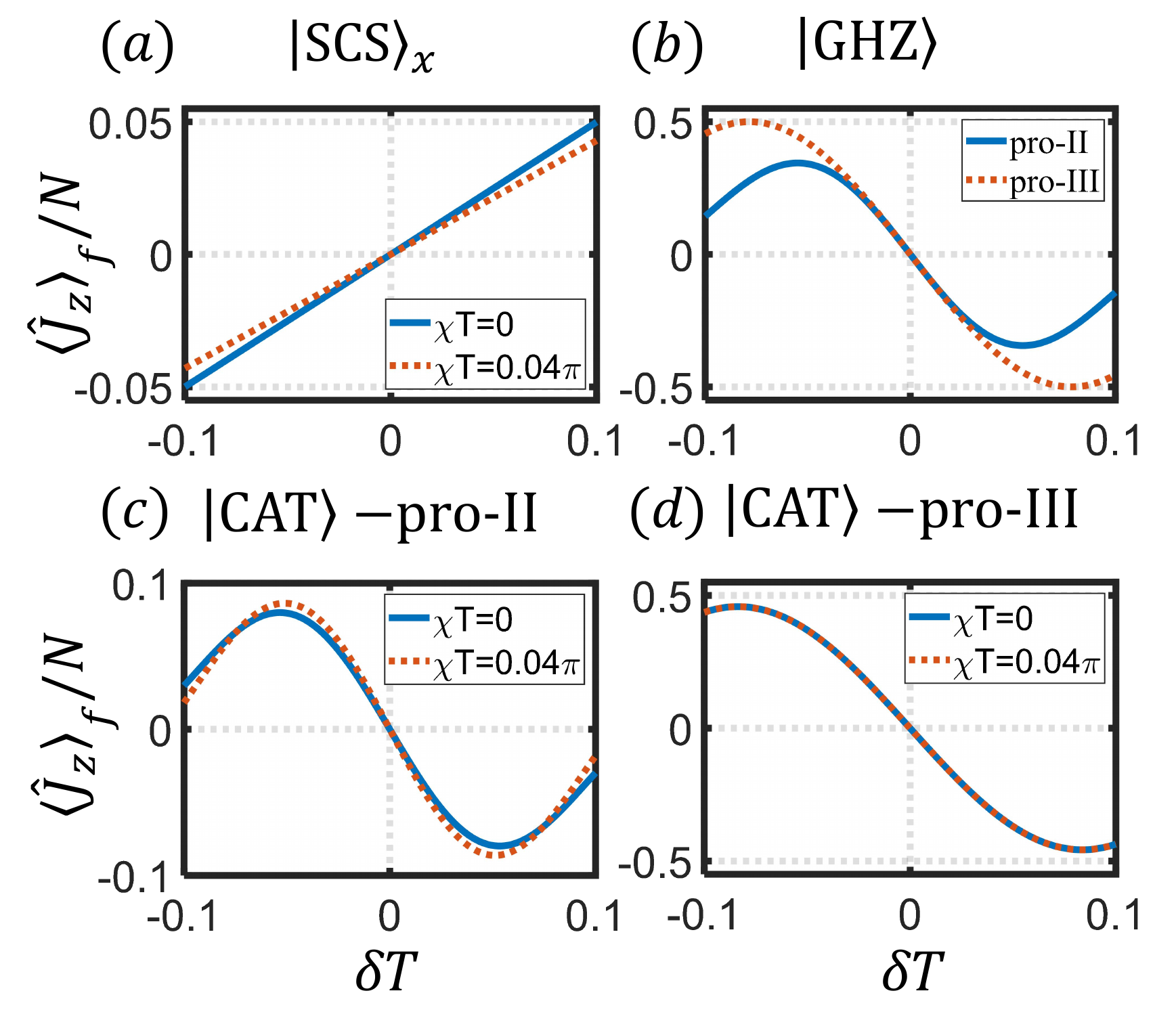}
	\caption{\label{Fig2}(color online).
The SPDMBI-Ramsey spectroscopy for time-independent signal measurement with ideal hard pulse (i.e. $\Omega\to +\infty$ and $\epsilon=0$).
The relative half-population difference $\langle\hat{J}_z\rangle_f/N$ versus $\delta T$ for (a) SCS $\ket{\textrm{SCS}}_x$, (b) GHZ state $\ket{\textrm{CAT}(\theta=0)}$ using protocol-II (blue solid line) and protocol-III (red dotted line), (c) $\ket{\textrm{CAT}}=\ket{\textrm{CAT}(\theta=\pi/8)}$ using protocol-II and (d) $\ket{\textrm{CAT}}$ using protocol-III.
Here, $B_z=1$, $\gamma_g=1$, $T=1$, $\chi_r=0.04\pi$, $t_r=\pi/(2\chi_r)$ and particle number $N=20$.
}
\end{figure}

\begin{figure*}[!htp]
\includegraphics[width=1.5\columnwidth]{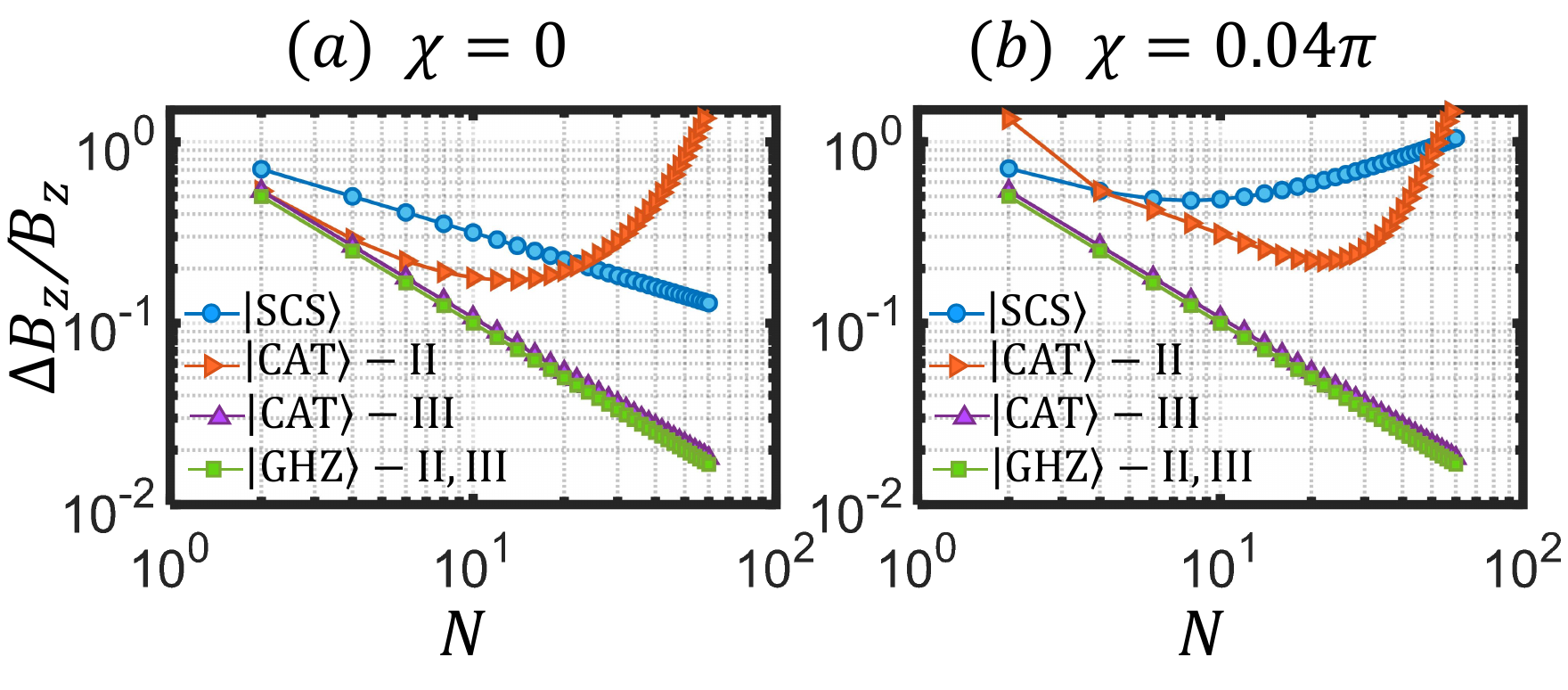}
\caption{\label{Fig3}(color online).
The measurement precision for time-independent signal via SPDMBI.
Precision scaling of $\Delta B_z/B_z$ versus total particle number $N$ for the $\ket{\textrm{SCS}}_x$ (blue circles), $\ket{\textrm{CAT}(\theta=\pi/8)}$ with protocol-II (red right-facing triangles) and protocol-III (purple upward triangle), and $\ket{\textrm{GHZ}}$ (green squares) with  (a) $\chi=0$ and  (b) $\chi=0.04\pi$.
Here, $\gamma_g=1$ and $T=1$.
}
\end{figure*}

\subsection{Individual particles\label{sec31}}
We first consider individual particles without any entanglement.
Suppose that all the particles are prepared in a spin coherent state which can be generally written as
\begin{eqnarray}\label{SCS_dk}
    \ket{\theta,\varphi}_\textrm{SCS}&=&\left[\cos\!\left(\!\frac{\theta}{2}\!\right)e^{-i\varphi/2}\ket{\uparrow}+\sin\!\left(\!\frac{\theta}{2}\!\right)e^{i\varphi/2}\ket{\downarrow}\right]^{\otimes N}\nonumber\\
    &=&\sum_{m=-J}^{J}C_m^{\theta,\varphi}\ket{J,m},
\end{eqnarray}
where $C_{m}^{\theta,\varphi}=\sqrt{\frac{(2J)!}{(J+m)!(J-m)!}}\cos^{J+m}(\frac{\theta}{2})\sin^{J-m}(\frac{\theta}{2})e^{-im\varphi}$.
To implement our SPDMBI, we choose $\ket{\textrm{SCS}}_x\equiv\ket{\pi/2,0}_\textrm{SCS}=e^{-i\frac{\pi}{2}\hat J_y}\ket{J,J}$ as the input state, which can be easily generated by applying a $\pi/2$ pulse along the $y$ axis on the state of all particles in spin-up $\ket{\uparrow}^{\otimes N}=\ket{J,J}$.
We also discuss the influence of the imperfection of preparation $\pi/2$ pulse on our protocol in {Appendix F}.
To finish the Ramsey interferometry, one can use another $\pi/2$ pulse $\hat{U}_\textrm{re}=e^{-i\frac{\pi}{2}{\hat{J}_{x}}}$ for readout, see {Fig.~\ref{Fig1}~(a)}.
Hence the output state before the measurement of the half-population difference can be
written as
\begin{eqnarray}\label{outstate}
    \ket{\Psi}_f&=&\hat{U}_\textrm{re} e^{-i(\chi\hat{J}_z^2+\delta\hat{J}_z)T}\ket{\textrm{SCS}}_x\\\nonumber
    &=&\hat{U}_\textrm{re}\sum_{m=-J}^{J}C_m e^{-i(\delta m+\chi m^2)T}\ket{J,m},
\end{eqnarray}
where $C_m=\sqrt{\frac{(2J)!}{(J+m)!(J-m)!}}\left(\frac{1}{2}\right)^J$.
According to Eq.~\eqref{QFI11}, the QFI for $B_z$ with $\ket{\textrm{SCS}}_x$ is
\begin{eqnarray}\label{FQI_SCS1}
    F_\textrm{Q}^{B_z}=N (T\gamma_g)^{2},
\end{eqnarray}
which means that the ultimate precision bound for $B_z$ with individual particles can only attain the SQL, i.e., $\Delta B_z^\textrm{QCRB}\propto {1}/{\sqrt{N}}$.
Moreover, we also evaluate its precision according to Eq.~\eqref{dBz}.
After some algebra, the expectations of half-population difference and the square of half-population difference on the final state can be explicitly written as
\begin{eqnarray}\label{JzSCS}
    \langle \hat J_{z} \rangle_f = \frac{N}{2}\sin\left(\delta T\right)\left[\cos(\chi T)\right]^{N-1},
\end{eqnarray}
and 
\begin{eqnarray}\label{Jz2SCS}
    &\langle \hat J_{z}^{2} \rangle_{f} =\frac{N(N+1)}{8}-\frac{N(N-1)}{8}\left[\cos(2\chi T)\right]^{N-2}\cos\left(2\delta T\right),\nonumber &\\&&
\end{eqnarray}
see {Appendix B} for the derivation.
As Eq.~\eqref{JzSCS} is exactly antisymmetric with respect to $\delta=0$, one can determine $B_z$ from the antisymmetry of measurement signal $\langle \hat J_{z} \rangle_f$, see {Fig.~\ref{Fig2}~(a)}.
According to Eqs.~\eqref{dBz}, \eqref{JzSCS} and \eqref{Jz2SCS}, one can analytically obtain the measurement precision
\begin{equation}\label{dBz_n}
    \Delta B_z=\frac{\mathcal{F}_\textrm{SCS}}{\gamma_gT\sqrt{N}},
\end{equation}
where 
\begin{widetext}
\begin{equation}
    \mathcal{F}_\textrm{SCS}=\frac{\sqrt{\frac{1}{2}\left[N+1-(N-1)\cos^{N-2}(2\chi T)\right]-\sin^2(\delta T)\left[N\cos^{2N-2}(\chi T)-(N-1)\cos^{N-2}(2\chi T)\right]}}{\left|\cos^{N-1}(2\chi T)\cos(\delta T)\right|}.
\end{equation}
\end{widetext}
From Eq.~\eqref{dBz_n}, the measurement precision $\Delta B_z$ for individual particles exhibits the SQL scaling for $\chi=0$.
As shown in {Fig.~\ref{Fig3}}, we find that $\Delta B_z$ can reach $\Delta B_z^\textrm{QCRB}=\frac{1}{\gamma_gT\sqrt{N}}$ with $\chi=0$ in our protocol.
Furthermore, we numerically calculate the corresponding precision scaling versus particle number $N$ with different $\chi$.
When $\chi\neq0$, the measurement precision scaling will be worse than the SQL.
In {Fig.~\ref{Fig2}~(a)}, the spectral contrast decreases due to interparticle interaction, thus the measurement precision decreases for $\ket{\textrm{SCS}}_x$ compared with the case of $\chi=0$, see {Fig.~\ref{Fig3}~(b) (blue circles)} .

\subsection{Entangled particles\label{sec32}}
Here, we try to use spin cat states to perform an entanglement-enhanced measurement for $B_z$ within this framework.
\begin{figure*}[!htp]
	\includegraphics[width=2\columnwidth]{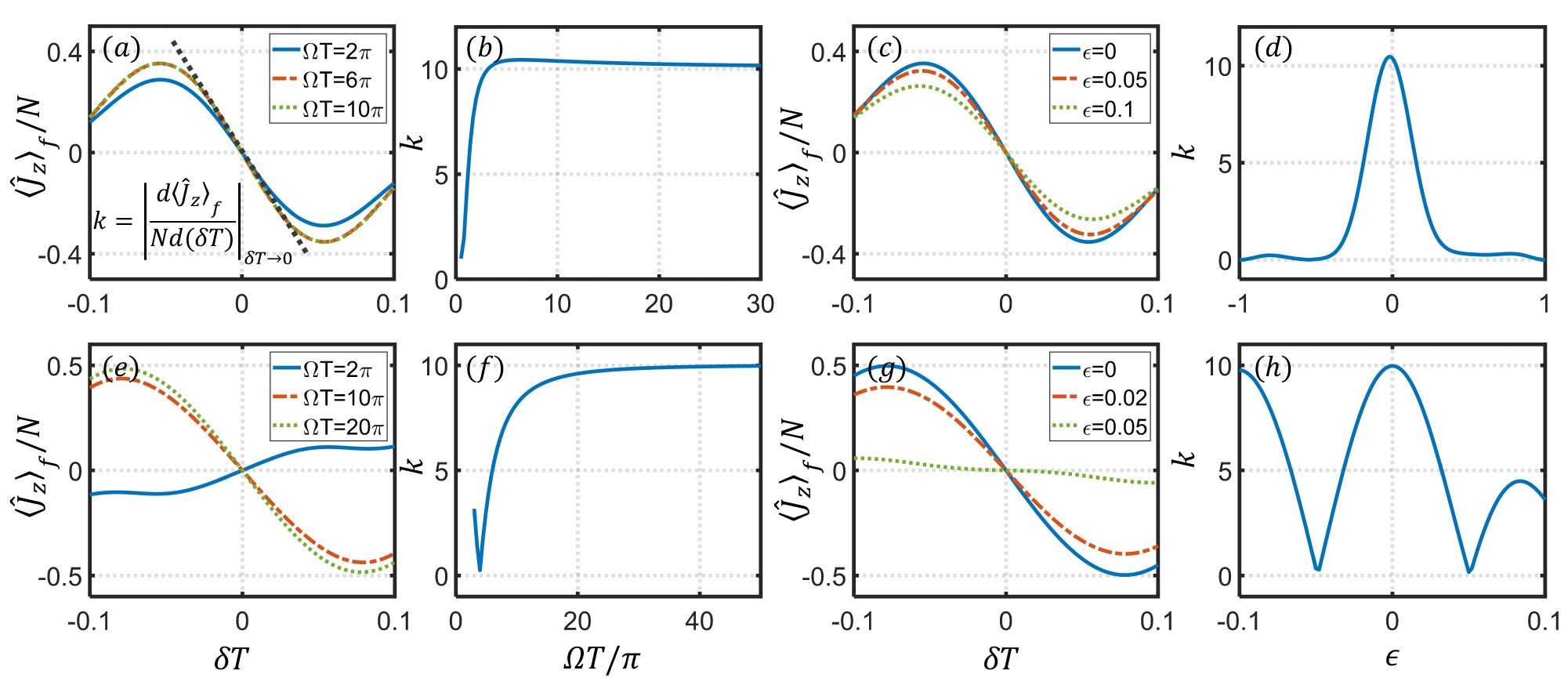}
	\caption{\label{Fig4}(color online).
The robustness of SPDMBI-Ramsey spectroscopy for time-independent signal with GHZ state $\ket{\textrm{GHZ}}$ using protocol-II [(a)-(d)] and protocol-III [(e)-(h)].
(a)(e): The measurement signal $\langle\hat{J}_z\rangle_f/N$ versus the detuning $\delta T$ with different Rabi frequency $\Omega T=2\pi$ (blue solid line), $\Omega T=6\pi$ (red dash-dotted line) and $\Omega T=10\pi$ (green dotted line) of the pulses.
(b)(f): The slope $k$ at the zero point $\delta T=0$ versus the Rabi frequency $\Omega T$.
Here, $T=1$, $\chi=\chi_r=0.04\pi$, $N=20$ and the duration for a $\pi/2$ ($\pi$) pulse is $\pi/(2\Omega)$ ($\pi/\Omega$).
(c)(g): The measurement signal $\langle\hat{J}_z\rangle_f/N$ versus the detuning $\delta T$ with different rotation errors $\epsilon=0$ (blue solid line), $\epsilon=0.05$ (red dash-dotted line) and $\epsilon=0.1$ (green dotted line).
(d)(h): The slope at the zero point $\delta T=0$ versus the rotation errors $\epsilon$.
Here, $T=1$, $\chi=\chi_r=0.04\pi$, $N=20$, Rabi frequency $\Omega=50\pi$ and the duration of a $\pi/2$ ($\pi$) pulse is $(1+\epsilon)\pi/(2\Omega)$ ($(1+\epsilon)\pi/\Omega$).
	}
\end{figure*}
Spin cat state is a typical kind of macroscopic superposition of spin coherent states (MSSCS)~\cite{JHSR2016,JHPRA2018}.
For total particle number $N \geq 6$, when $\theta \leq \pi/8$, the corresponding MSSCS can be regarded as a spin cat state~\cite{JHPRA2018}, which can be written as
%
\begin{eqnarray}\label{cat}
    \ket{\textrm{CAT}(\theta)}
    &=&\frac{1}{\sqrt{2}}\sum_{m=-J}^{J}\left(C_m^{\theta,0}+C_{-m}^{\theta,0}\right)\ket{J,m}.
\end{eqnarray}
%
Importantly, the spin cat state satisfy the requirement of our SPDMBI with $C_m(0)=(C_m^{\theta,0}+C_{-m}^{\theta,0})/\sqrt{2}=C_{-m}(0)$.
Especially when $\theta=0$, the spin cat state $\ket{\textrm{CAT}(0)}=\frac{\ket{J,J}+\ket{J,-J}}{\sqrt{2}}$ corresponds to the well-known GHZ state. We use  $\ket{\textrm{GHZ}}\equiv \ket{\textrm{CAT}(0)}$ hereafter.
In general, the spin cat states and the GHZ state can be prepared by applying suitable operators on the spin coherent state~\cite{CMarXiv2025}.
Our main goal is to illustrate the effect of entanglement for our scheme, thus we ignore the process of state preparation and use the spin cat states as input state directly. 
According to Eqs.~\eqref{QFI11} and \eqref{cat}, the QFI for $B_z$ with a spin cat state can be written as
\begin{eqnarray}\label{FQI_CAT}
    F_{Q}^{B_z}= N^2\left[T\gamma_g\overline{M}(\theta)\right]^{2},
\end{eqnarray}
where the coefficient $\overline{M}(\theta)=\cos\theta\leq1$ does not depend on the particle number $N$~\cite{JHPRA2018} and $\overline{M}(0)=1$ for GHZ state.
However, if one simply uses protocol I and replaces the input $\ket{\textrm{SCS}}_x$ with the spin cat state $\ket{\textrm{CAT}(\theta)}$, the measurement precision cannot support the HL scaling~\cite{LPRMP2018,JHAPR2024,JHSR2016,SSMPRA2018}.

In this regard, we design two protocols for spin cat states, denoted as protocol-II and protocol-III, as shown in {Fig.~\ref{Fig1}~(b) and (c)}.
We find that both protocol-II and protocol-III are suitable for GHZ state.
While for other spin cat states, using protocol-II leads to a reduction of contrast, and protocol-III is more appropriate to saturate the Heisenberg-limited precision bounds, as shown in {Fig.~\ref{Fig3}}.  

For protocol-II, considering the symmetry of the readout Hamiltonian, we choose two $\pi/2$ pulses along $\pm x$ direction with the interval $t_r$.
In order to realize the nonlinear readout, the interparticle interaction between the two $\pi/2$ pulses is $\chi_r$.
For generality, here $\chi_r$ can be different from $\chi$. 
%
For $\pi/2$ pulses, we assume that the Rabi frequency is $\Omega$, and the duration is $\pi/(2\Omega)$.
However, in experiments, there may be a rotation angle error due to inaccurate time control $(1+\epsilon)\pi/(2\Omega)$ with the relative error $\epsilon$.
Therefore, the Hamiltonian for protocol-II can be written as
\begin{equation}\label{Hprob}
    \hat{H}_\textrm{Ram}^{\text{II}}=\chi_b(t)\hat{J}_z^2+\delta\hat{J}_z+\Omega_b(t)\hat{J}_x
\end{equation}
with
\begin{eqnarray}\label{chi}
    \chi_b(t)=\left\{\begin{array}{rl}
    \chi, &0\leq t<T\\
    \chi_r, &T\leq t\leq T+t_r+(1+\epsilon)\pi/\Omega
\end{array}\right.
\end{eqnarray}
and
\begin{small}
\begin{eqnarray}\label{Omega_prob}
    \Omega_b(t)=\left\{\begin{array}{rl}
    0, &0\leq t<T\\
    \Omega, &T\leq t<T+\frac{(1+\epsilon)\pi}{2\Omega}\\
    0, &T+\frac{(1+\epsilon)\pi}{2\Omega}\leq t<T+\frac{(1+\epsilon)\pi}{2\Omega}+t_r\\
    -\Omega, &T+\frac{(1+\epsilon)\pi}{2\Omega}+t_r\leq t\leq T+t_r+\frac{(1+\epsilon)\pi}{\Omega}
\end{array}\right.
\end{eqnarray}
\end{small}
In the limit of $\Omega\to +\infty$, the readout evolution operator reads
\begin{equation}\label{Ureb}
    \hat{U}_\textrm{re}=e^{i\pi/2\hat{J}_x}e^{-i(\chi_r\hat{J}_z^2+\delta\hat{J}_z)t_r}e^{-i\pi/2\hat{J}_x}=e^{-i(\chi_r\hat{J}_y^2+\delta\hat{J}_y)t_r}.
\end{equation}
Especially when $t_r=\pi/(2\chi_r)$, the readout evolution operator 
\begin{eqnarray}
    \hat{U}_\textrm{re}&=&e^{-i\pi/2\hat{J}_y^2-i\pi\delta/(2\chi_r)\hat{J}_y}\nonumber\\
    &=&e^{-i\pi\delta/(2\chi_r)\hat{J}_y}e^{-i\pi/2\hat{J}_y^2}
\end{eqnarray}
becomes a twist along with a rotation along $y$ axis, still satisfying Eq.~\eqref{HwI1}.
The corresponding final state becomes
\begin{eqnarray}\label{CATo1b}
\ket{\Psi}_f&=&e^{-i\delta t_r\hat{J}_y}e^{-i\frac{\pi}{2}{\hat{J}_{y}^2}}e^{-i(\chi\hat{J}_z^2+\delta\hat{J}_z)T}\ket{\textrm{CAT}(\theta)}\\\nonumber
&=&e^{-i\delta t_r\hat{J}_y}\ket{\Psi}_{f_0}.
\end{eqnarray}
When $N$ is an even number, $\ket{\Psi}_{f_0}$ is analytical and can be written as
\begin{eqnarray}\label{CATo2b}
    \ket{\Psi}_{f0}=\sum_{m=-J}^{J}e^{-i\chi T m^2}A_m^{\theta,\delta T}\ket{J,m}.
\end{eqnarray}
The coefficient $A_m^{\theta ,\delta T }$ is
\begin{eqnarray}\label{A1b}
A_m^{\theta ,\delta T }&& = \frac{1}{2}\left[ {{e^{ - i\pi /4}}C_m^{\theta ,\delta T }+ {e^{i\pi /4}}{{\left( { - 1} \right)}^J}C_{ - m}^{\theta ,\delta T}} \right] \nonumber\\ 
&&+\frac{1}{2} \left[ {{e^{ - i\pi /4}}C_{ - m}^{\theta , - \delta T} + {e^{i\pi /4}}{{\left( { - 1} \right)}^J}C_m^{\theta , - \delta T}} \right]
\end{eqnarray}
with $C_{m}^{\theta,\pm \delta T}=C_m^{\theta,0}e^{\mp i m \delta T}$.
Hence the corresponding expectation of half-population difference and the square of half-population difference on the final state are
\begin{widetext}
\begin{eqnarray}\label{Jz_CATb}
&&\langle\hat{J}_z\rangle_f(\chi)=\cos(\delta t_r)\left( { - 1} \right)^{J+1}\sum_{m = 1}^J (-1)^m m {\left( {C_m^{\theta ,0} + C_{ - m}^{\theta ,0}} \right)^2}\sin \left( { 2m\delta T } \right)\\\nonumber
&&~~~~+\sin(\delta t_r)\sum_{m'=1/2}^{J-1/2}(-1)^{J-1/2-m'}\sin(2\chi Tm')\lambda_{m'-1/2}^+(C_{m'-1/2}+C_{-m'+1/2})(C_{m'+1/2}+C_{-m'-1/2})\cos(2m'\delta T)
\end{eqnarray}
and
\begin{eqnarray}\label{Jz2_CATb}
\langle\hat{J}_{z}^2\rangle_f(\chi)&=& \cos^2(\delta t_r)\sum_{m = 1}^J m^2 {\left( {C_m^{\theta ,0} + C_{ - m}^{\theta ,0}} \right)^2}\\\nonumber
& &-\frac{\sin(2\delta t_r)}{4}\sum_{m'=1/2}^{J-1/2}\sin(2\chi Tm')(2m')\lambda_{m'-1/2}^+(C_{m'-1/2}+C_{-m'+1/2})(C_{m'+1/2}+C_{-m'-1/2})\sin(\delta T)\\\nonumber
& &+\frac{\sin^2(\delta t_r)}{4}\left[\sum_{m=-J+1}^J[J(J+1)-m(m-1)]|C_m+C_{-m}|^2\right.\\\nonumber
& &~~~~~~~~~~~~~~~~~\left.+\sum_{m=-J}^{J-2}\cos[4\chi T(m+1)]\lambda_{m+1}^+\lambda_m^+(C_m+C_{-m})(C_{m+2}+C_{-(m+2)})\cos(2\delta T)\right],
\end{eqnarray}
\end{widetext}
respectively (see {Appendix C} for the derivation).
For the GHZ state $\ket{\textrm{GHZ}}$, we have $\langle\hat{J}_z\rangle_f$ and $\langle\hat{J}_z^2\rangle_f$ are both independent on $\chi$ and equal to the corresponding ones of $\chi=0$.
This means that our interaction-based readout can cancel the influence of the nonlinear interaction for the GHZ state.
%
Furthermore, one can obtain the measurement precision $\Delta B_{z}$ according to Eq.~\eqref{dBz}. 
We numerically calculate the precision scaling of $\Delta B_z$ versus total particle number $N$, see Fig.~\ref{Fig3}.
The application of the interaction-based readout guaranteed the robustness against nonlinear interaction for the GHZ state, thus the measurement precision $\Delta B_{z}$ can reach $1/N$-scaling in the vicinity of $\delta= 0$.
However, for spin cat state $\ket{\textrm{CAT}}\equiv\ket{\textrm{CAT}(\theta=\pi/8)}$, the zero point $\delta=0$ is not the optimal detection point~\cite{JHSR2016}, which leads to its precision may be even lower than the SCS when the particle number $N$ is sufficiently large, see Fig.~\ref{Fig3}~(b).
Thus protocol-II is only suitable for the GHZ state and other spin cat states are not. 
%
\begin{figure*}[!htp]
\includegraphics[width=2\columnwidth]{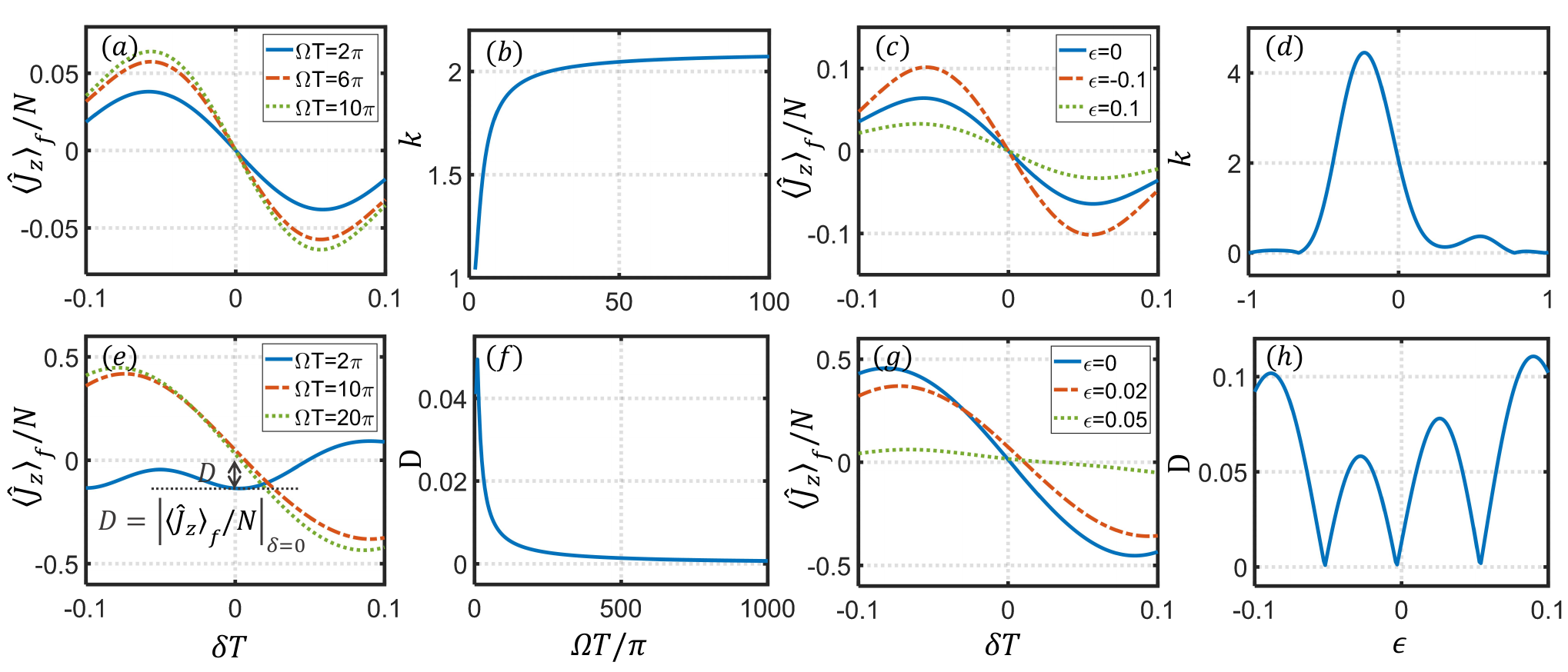}
\caption{\label{Fig5}(color online).
The robustness of SPDMBI-Ramsey spectroscopy for time-independent signal with spin cat state $\ket{\textrm{CAT}(\theta=\pi/8)}$ using protocol-II [(a)-(d)] and protocol-III [(e)-(h)].
(a)(e): The measurement signal $\langle\hat{J}_z\rangle_f/N$ versus $\delta T$ with different Rabi frequency $\Omega T=2\pi$ (blue solid line), $\Omega T=6\pi$ (red dash-dotted line) and $\Omega T=10\pi$ (green dotted line) of the pulses. 
(b)(f): The shift $D=|\langle\hat{J}_z\rangle_f/N|_{\delta=0}$ at the zero point $\delta T=0$ versus the Rabi frequency $\Omega T$.
Here, $T=1$, $\chi=\chi_r=0.04\pi$, $N=20$ and $\epsilon=0$. 
(c)(g): The measurement signal $\langle\hat{J}_z\rangle_f/N$ versus the detuning $\delta T$ with different rotation errors $\epsilon=0$ (blue solid line), $\epsilon=-0.1$ (red dash-dotted line) and $\epsilon=0.1$ (green dotted line).
(d)(h): The slope at the zero point $\delta T=0$ versus the rotation errors $\epsilon$.
Here, $T=1$, $\chi=\chi_r=0.04\pi$, $N=20$, Rabi frequency $\Omega=50\pi$ and the duration of a $\pi/2$ ($\pi$) pulse length is $(1+\epsilon)\pi/(2\Omega)$ ($(1+\epsilon)\pi/\Omega$).
}
\end{figure*}

\begin{figure}[!htp]
	\includegraphics[width=1\columnwidth]{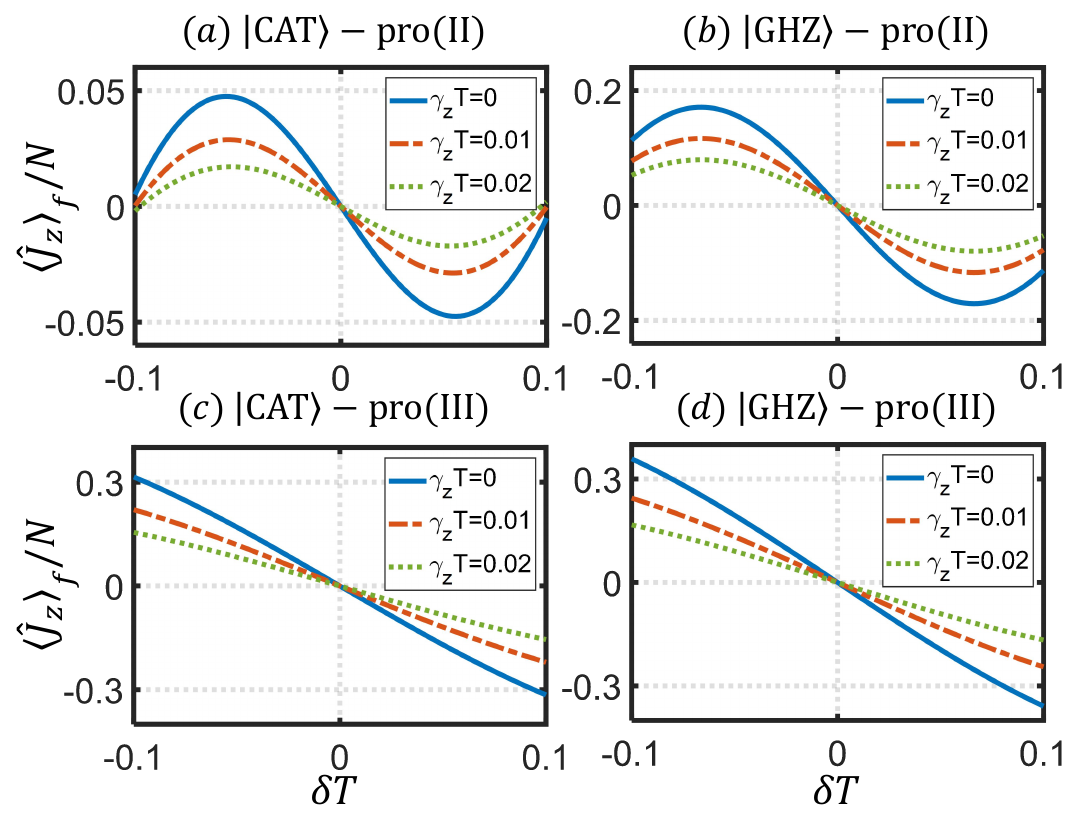}
	\caption{\label{Fig6}(color online).
The robustness of SPDMBI-Ramsey spectroscopy against dephasing $\hat{\mathcal{L}}=\hat{J}_z$ for time-independent signal applying protocol-II (top row) and protocol-III (bottom row).
The measurement signal $\langle\hat{J}_z\rangle_f/N$ versus the detuning $\delta T$ with different dephasing rate $\gamma_z T=0$ (blue solid line), $\gamma_z T=0.01$ (red dash-dotted line) and $\gamma_z T=0.02$ (green dotted line) for (a)(c) spin cat state $\ket{\textrm{CAT}}$ and (b)(d) GHZ state $\ket{\textrm{GHZ}}$.
Here, $T=1$, $\chi=\chi_r=0.04\pi$, $N=8$.
}
\end{figure}

Moreover, we also show the influence of Rabi frequency $\Omega$ of the pulses , the rotation error $\epsilon$ caused by the inaccurate time control and the dephasing during the interrogation.
For both $\ket{\textrm{GHZ}}$ and $\ket{\textrm{CAT}}$, the spectrum is still antisymmetric about $\delta=0$ for finite $\Omega$.
The response of the signal at zero point $k=\frac{1}{N}|\frac{d \langle\hat{J}_z\rangle_f}{d \delta T}|_{\delta\to 0}$ increases as the pulse Rabi frequency and tends to be stable when $\Omega T=5\pi$, see Fig.~\ref{Fig4}~(a) and Fig.~\ref{Fig5}~(a). 
That is, when $\Omega$ is large enough, the result would saturate to the ideal one.
%
%
Similarly, the rotation error $\epsilon$ also does not influence the antisymmetry of the spectra, as shown in {Fig.~\ref{Fig4}~(c) and Fig.~\ref{Fig5}~(c)}, but significantly decreases the response at the zero point, as shown in {Fig.~\ref{Fig4}~(d) and Fig.~\ref{Fig5}~(d)}.
%
To calculate the influence of collective dephasing, we consider the master equation of Eq.~\eqref{Lindblad} with Lindblad operator $\hat{\mathcal{L}}=\hat{J}_z$ and dephasing rate $\gamma_z$. 
We find that the dephasing only leads to the decrease of the signal contrast without affecting the measurement accuracy, which means our protocol is also robust against the collective dephasing, as shown in {Fig.~\ref{Fig6}}~(a) and (b).

In order to improve the measurement precision of spin cat states $\ket{\textrm{CAT}(\theta)}$ to reach the Heisenberg scaling of $1/N$ at $\delta=0$ and suppress the influence of $\delta$ in the readout stage, we propose protocol-III using different nonlinear readout.
This nonlinear readout is made up of two $\pi/2$ pulses and a $\pi$ pulse along the $-y$ direction, as shown in Fig.~\ref{Fig1}~(c).
The corresponding Hamiltonian can be written as
\begin{equation}\label{Hprob}
\hat{H}_\textrm{Ram}^\text{III}=\chi_c(t)\hat{J}_z^2+\delta\hat{J}_z+\Omega_c(t)\hat{J}_y
\end{equation}
where
\begin{eqnarray}\label{chi}
\chi_c(t)=\left\{\begin{array}{rl}
\chi, &0\leq t<T\\
\chi_r, &T\leq t\leq T+t_r+\frac{2(1+\epsilon)\pi}{\Omega}
\end{array}\right.
\end{eqnarray}
and
\begin{eqnarray}\label{Omega_prob}
\Omega_c(t)=\left\{\begin{array}{rl}
0, &0\leq t<T\\
\Omega, &T\leq t<T+\frac{(1+\epsilon)\pi}{2\Omega}\\
0, &T+\frac{(1+\epsilon)\pi}{2\Omega}\leq t<T+\frac{(1+\epsilon)\pi}{2\Omega}+\frac{t_r}{2}\\
-\Omega, &T+\frac{(1+\epsilon)\pi}{2\Omega}+\frac{t_r}{2}\leq t< T+\frac{t_r}{2}+\frac{3(1+\epsilon)\pi}{2\Omega}\\
0, &T+\frac{t_r}{2}+\frac{3(1+\epsilon)\pi}{2\Omega}\leq t<T+t_r+\frac{3(1+\epsilon)\pi}{2\Omega}\\
\Omega, &T+t_r+\frac{3(1+\epsilon)\pi}{2\Omega}\leq t\leq T+t_r+\frac{2(1+\epsilon)\pi}{\Omega}
\end{array}\right.\nonumber\\
\end{eqnarray}
%
For generality, $\chi_r$ can be different from $\chi$. 
%
For ideal hard pulses with $\Omega\to +\infty$ and $\epsilon=0$, the readout evolution operator is
\begin{eqnarray}\label{Ur_nl}
    \hat{U}_\textrm{re}&=&e^{i\frac{\pi}{2}\hat{J}_y}e^{-i(\chi_r\hat{J}_z^2+\delta\hat{J}_z)t_r/2}e^{-i\pi\hat{J}_y}e^{-i(\chi_r\hat{J}_z^2+\delta\hat{J}_z)t_r/2}e^{i\frac{\pi}{2}\hat{J}_y}\nonumber\\
&=&e^{-i\chi_r t_r\hat{J}_x^2},
\end{eqnarray}
%
%
%
When $t_r=\pi/(2\chi_r)$, the readout evolution operator becomes
\begin{equation}
    \hat{U}_\textrm{re}=e^{-i\frac{\pi}{2}\hat{J}_x^2},
\end{equation}
which is a twist along $x$ axis satisfying Eq.~\eqref{HwI1}. 
The corresponding output state is
\begin{eqnarray}\label{CATo1}
\ket{\Psi}_f&=&e^{-i\frac{\pi}{2}{\hat{J}_{x}^2}}e^{-i(\chi\hat{J}_z^2+\delta\hat{J}_z)T}\ket{\textrm{CAT}(\theta)}\\\nonumber
&=&e^{-i\frac{\pi}{2}{\hat{J}_{x}^2}}e^{-i\chi T\hat{J}_z^2}e^{i\frac{\pi}{2}{\hat{J}_{y}^2}}e^{-i\frac{\pi}{2}{\hat{J}_{y}^2}}e^{-iT\delta\hat{J}_z}\ket{\textrm{CAT}(\theta)}\\\nonumber
&=&e^{-i\chi T\hat{J}_z^2}e^{-i\frac{\pi}{2}{\hat{J}_{x}^2}}e^{-iT\delta\hat{J}_z}\ket{\textrm{CAT}(\theta)},
\end{eqnarray}
see {Appendix C} for the derivation.
When $N$ is an even number, the final state $\ket{\Psi}_f$ is analytical and can be written as
\begin{eqnarray}\label{CATo2}
\ket{\Psi}_f=\sum_{m=-J}^{J}e^{-i\chi T m^2}B_m^{\theta,\delta T}\ket{J,m}
\end{eqnarray}
with coefficient 
\begin{eqnarray}\label{A1}
B_m^{\theta ,\delta T }=\left( {C_m^{\theta ,0} + C_{ - m}^{\theta ,0}} \right)\cos \left[ {m\delta T  + (N+1)\pi /4 } \right].
\end{eqnarray}
%
%
%
Hence the corresponding expectation of half-population difference and the square of half-population difference on the final state are
\begin{eqnarray}\label{Jz_CAT}
\langle\hat{J}_z\rangle_f(\chi)= \left( { - 1} \right)^{J+1}\sum\limits_{m = 1}^J m {\left( {C_m^{\theta ,0} + C_{ - m}^{\theta ,0}} \right)^2}\sin \left( { 2m \delta T } \right)\nonumber\\
\end{eqnarray}
and
\begin{eqnarray}\label{Jz2_CAT}
\langle\hat{J}_z^2\rangle_f(\chi)= \sum\limits_{m = 1}^J m^2 {\left( {C_m^{\theta ,0} + C_{ - m}^{\theta ,0}} \right)^2}\nonumber\\
\end{eqnarray}
respectively, see {Appendix C} for the derivation.
Eq.~\eqref{Jz_CAT} is also exactly antisymmetric with respect to $\delta=0$, and one can estimate $B_z$ from the symmetry property of the measurement signal $\langle\hat{J}_z\rangle_f$, see the example for a spin cat state $\ket{\textrm{CAT}(\theta)}$ with $\theta=\pi/8$ in {Fig.~\ref{Fig2}~(d)}.
%
%
Interestingly, Eq.~\eqref{Jz_CAT} and Eq.~\eqref{Jz2_CAT} are independent on $\chi$, showing its robustness against interparticle interaction.
Further, we can analytically obtain the measurement precision
\begin{equation}\label{dBac_CAT}
\Delta B_z=\frac{\tilde{\mathcal{F}}(\theta)}{\gamma_gT},
\end{equation}
where
\begin{equation}\label{tildeD}
\tilde{\mathcal{F}}(\theta)=\frac{\sqrt{\sum_{m=1}^{J}m^2\left(C_m^{\theta,0}+C_{-m}^{\theta,0}\right)^2-\langle\hat{J}_z\rangle_f^2}}{\left|\sum_{m=1}^{J}2m^2\left(C_m^{\theta,0}+C_{-m}^{\theta,0}\right)^2\cos(2m\delta T)\right|}.
\end{equation}
%
%
For GHZ state, $\tilde{\mathcal{F}}(0)={1}/{N}$, the measurement precision $\Delta B_z=\frac{1}{\gamma_g T N}$ can reach $\Delta B_\textrm{ac}^\textrm{QCRB}$ in our protocol.
Further, we can calculate the corresponding precision scaling versus particle number $N$ for $\ket{\textrm{CAT}}$ and $\ket{\textrm{GHZ}}$, see {Fig.~\ref{Fig3}}.
It is indicated that the measurement precision $\Delta B_z$ for spin cat states can exhibit the HL scaling and is robust against interparticle interaction.

Moreover, similar to protocol-II, we also numerically analyze the robustness of protocol-III against the Rabi frequency $\Omega$ of the pulse, the rotation error $\epsilon$ and the dephasing. 
For GHZ state, both $\Omega$ and $\epsilon$ do not affect the symmetry of the spectra but have influence on the response of the signal $k$ at zero point, as shown in {Fig.~\ref{Fig4}~(e)-(h)}.
When $\Omega$ is not large enough, the pulses along $y$ axis breaks the symmetry and the antisymmetric spectrum disappears for spin cat state $\ket{\textrm{CAT}(\theta)}$ with $\theta>0$, as shown in {Fig.~\ref{Fig5}~(e)}.
For convenience, we evaluate the accuracy of the spectra by $D=|\langle\hat{J}_z\rangle_f/N|_{\delta=0}$ here.
Meanwhile, the rotation error $\epsilon$ can also breaks the antisymmetry of spectra, as shown in {Fig.~\ref{Fig5}~(g) and (h)}.

For GHZ state,  we have $\hat{U}_{\pi_y}^\dagger\ket{\textrm{GHZ}}\bra{\textrm{GHZ}}\hat{U}_{\pi_y}=\ket{\textrm{GHZ}}\bra{\textrm{GHZ}}$ with $\hat{U}_{\pi_y}=e^{-i\pi\hat{J}_y}$.
However, for spin cat state,  $\hat{U}_{\pi_y}^\dagger\ket{\textrm{CAT}(\theta=\pi/8)}\bra{\textrm{CAT}(\theta=\pi/8)}\hat{U}_{\pi_y}\neq\ket{\textrm{CAT}(\theta=\pi/8)}\bra{\textrm{CAT}(\theta=\pi/8)}$ with $\hat{U}_{\pi_y}=e^{-i\pi\hat{J}_y}$.
Thus, the results for GHZ state and spin cat state are different (see Appendix A for more details).  
Fortunately, when $\Omega$ becomes large with small $\epsilon$, the antisymmetric spectrum can be achieved for all spin cat states. 
%
In addition, with ideal pulses, the dephasing only leads to the decrease of the signal contrast without affecting the measurement accuracy, which means our protocol is also robust against the collective dephasing, see {Fig.~\ref{Fig6}}~(c) and (d).
\section{Time-dependent signal measurement via SPDMBI-Ramsey}\label{Sec4}
The precise measurement of time-dependent signals is also significant in quantum metrology and sensing.
For example, for an ac magnetic field, one can obtain its frequency via quantum lock-in techniques, and then evaluate its amplitude via proper fitting process~\cite{SKNature2011,MZPRX2021,SSScience2017,JMBScience2017}.
In particular, the many-body quantum lock-in amplifier with spin cat states also satisfies our SPDMBI~\cite{SCPR2025,MZPRX2021}, exhibiting antisymmetric spectra versus the lock-in point with enhanced precision approaching the HL scaling, see Appendix D for more details.
However, for these quantum lock-in techniques, due to the ambiguity resulted from the rotational symmetry of the phase, the amplitude of ac magnetic field cannot be too large~\cite{SKNature2011,SSScience2017,JMBScience2017,EDHPRApplied2022}.
Hence, it remains a challenging endeavor to broaden the measurement range of the amplitude of the ac magnetic field.
In the section, within the framework of our SPDMBI-Ramsey, we propose a novel scheme to estimate the ac signal amplitude in a wide dynamic range using a sequence of periodic pulses and an additional dc magnetic field.

In our protocol, the coupling between the probe and the ac magnetic field is $\hat{H}_\textrm{s}=\gamma_g B_\textrm{ac} \sin(\omega t)\hat{J}_z$.
Here, $\omega$ corresponds to the oscillation frequency and $B_\textrm{ac}$ denotes the amplitude of the ac signal.
In general, the frequency $\omega$ can be extracted via quantum lock-in detection, ranging from single-particle systems~\cite{SKNature2011} to many-body systems~\cite{MZPRX2021}.
In our consideration, we assume that the frequency $\omega$ is known by quantum lock-in detection (see {Appendix D}) and our goal is to measure the parameter $B_\textrm{ac}$.
To measure parameter $B_\textrm{ac}$, we consider the control Hamiltonian $\hat{H}_c=\Omega(t)\hat{J}_x+\gamma_gB_\textrm{dc}\hat{J}_z$.
Taking the nonlinear interparticle interaction $\hat{H}_a=\chi\hat{J}_z^2$ into account, the total Hamiltonian reads 
\begin{eqnarray}
    \hat{H}_{w}&=&\hat{H}_a+\hat{H}_s+\hat{H}_c \nonumber\\
    &=& \chi\hat{J}_z^2+ \Omega(t)\hat{J}_x+\gamma_g [B_\textrm{dc}+ B_\textrm{ac} \sin(\omega t)]\hat{J}_z.
\end{eqnarray}
%
%

\begin{figure}[!htp]
	\includegraphics[width=\columnwidth]{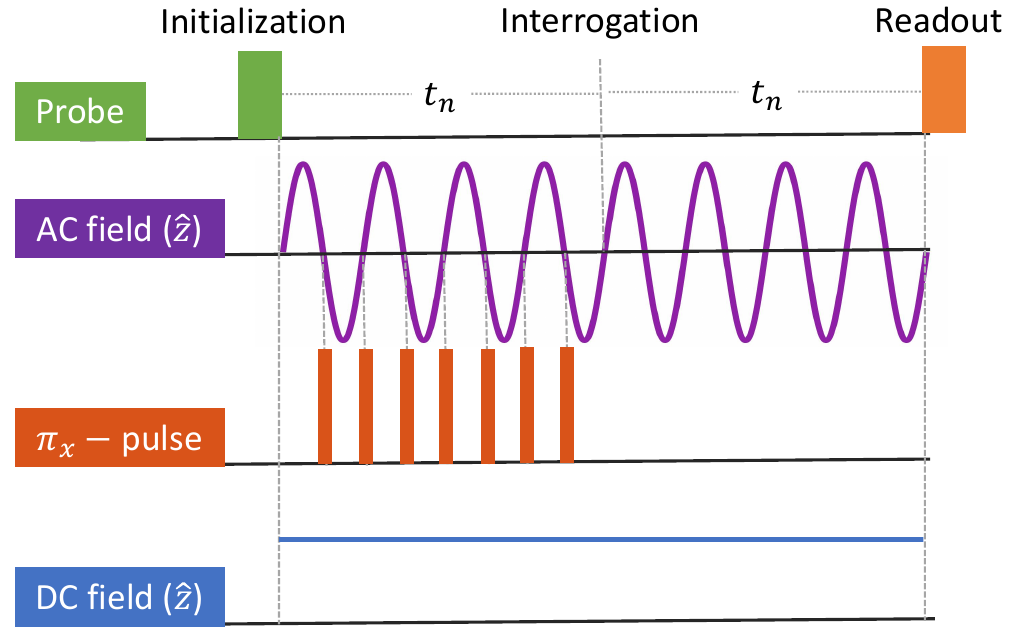}
	\caption{\label{Fig7}(color online).
Protocol for amplitude measurement of ac magnetic field by modulating dc magnetic field.
The scheme consists of three stages: initialization, interrogation, and readout.
First, the system is initialized in a desired input state (green square). 
Then, the interrogation stage is divided into two signal-accumulation processes.
In the first accumulation process, the ac magnetic field (purple line) information is encoded in the accumulated phase $\phi_\textrm{ac,n}$ to be measured via $\pi$ pulse sequences along the $x$ axis (the red thin square).
In the second accumulation process, the ac magnetic field (blue line) information is encoded in the accumulated phase $\phi_\textrm{dc,n}$ as a modulated term.
Finally, the ac magnetic field amplitude $B_{ac}$ can be extracted from a measurement of the half-population difference via the readout.}
\end{figure}

As shown in Fig.~\ref{Fig7}, our protocol can be divided into three stages: initialization, interrogation, and readout.
In the initialization stage, a symmetric state $\ket{\Psi(0)}=\sum_mC_m(0)\ket{J,m}$ with $C_m(0)=\pm C_{-m}$ is prepared.
Then, the input state undergoes an interrogation stage, which is divided into two processes with different $\Omega(t)$.
In the first process with $\Omega(t)=\pi\sum_{j=1}^n\delta(t-j\tau_m)$, it corresponds to a multi-$\pi$-pulse sequence (denoted by thin red lines) being applied at every node of the ac magnetic field for a duration $t_n=2n\pi/\omega$ with $n$ an integer.
In this case, only the information of $B_\textrm{ac}$ is encoded in the accumulated phase $\phi_\textrm{ac,n}\propto B_\textrm{ac}$, and the effective Hamiltonian (in the interaction picture with respect to $\hat{H}_0=\Omega(t)\hat{J}_x$) 
\begin{equation}
    \hat{H}_\textrm{eff}^\textrm{ac}=\frac{2\gamma_g B_\textrm{ac}}{\pi}\hat{J}_z+\chi \hat{J}_z^2
\end{equation}
can be used to describe this process, see {Appendix E} for the derivation.
In the second process with $\Omega(t)=0$, the system freely evolves for the same duration $t_n$, only the information about $B_\textrm{dc} $ is encoded in the accumulated phase $\phi_\textrm{dc,n}\propto B_\textrm{dc}$, thus one can  use another effective Hamiltonian (in the interaction picture with respect to $\hat{H}_0=\Omega(t)\hat{J}_x$) 
\begin{equation}
    \hat{H}_\textrm{eff}^\textrm{dc}=-\gamma_g B_\textrm{dc}\hat{J}_z+\chi \hat{J}_z^2
\end{equation}
to describe this process, see {Appendix E} for the derivation.
Thus the system state before readout is
\begin{eqnarray}\label{psi_out}
    \ket{\Psi(t_n)}&=&e^{-i(2n-1)\pi\hat{J}_x}e^{-i\hat{H}_\textrm{eff}^\textrm{dc}t_n}e^{-i\hat{H}_\textrm{eff}^\textrm{ac}t_n}\ket{\Psi}_\textrm{in}\nonumber\\
&=&e^{-2i\chi t_n\hat{J}_z^2}e^{-in\phi\hat{J}_z}e^{i\pi\hat{J}_x}\ket{\Psi}_\textrm{in},
\end{eqnarray}
with the phase 
\begin{equation}
    \phi=\frac{2\pi\gamma_g}{\omega}  \left(B_\textrm{dc}-\frac{2B_\textrm{ac}}{\pi}\right),
\end{equation}
see {Appendix E} for more details.
According to Eq.~\eqref{psi_out}, one can find the total accumulated phase $n\phi=0$ when $B_\textrm{dc}=\frac{2}{\pi}B_\textrm{ac}$.
Thus one can obtain the parameter $B_\textrm{ac}$ by modulating the strength of dc magnetic field $B_\textrm{dc}$, which is similar to the principle of phase lock-in detection.
%
%
Finally, a unitary operation $\hat{U}_\textrm{re}$ is applied for recombination, and the half-population difference is measured.
The expectation of the half-population difference measurement on the final state is $J_{z,n}=_{\textrm{f,n}}\langle \Psi|\hat{J}_{z}|\Psi\rangle_{\textrm{f,n}}$ with $\ket{\Psi}_\textrm{f,n}=\hat{U}_\textrm{re}\ket{\Psi(t_n)}$.
One can estimate the ac signal amplitude $B_\textrm{ac}$ according to the population measurement which is determined by the accumulated phase $n\phi$.
Since the interferometry process satisfies Eq.~\eqref{HwI1}, one can obtain an antisymmetric Ramsey spectroscopy with symmetric input state.
In the following, we will analyze the measurement precisions of parameter $B_\textrm{ac}$ via our protocol and show how to increase the measurement precision using entangled states.
%

\subsection{Individual particles}\label{Sec41}
We start by considering individual particles without any entanglement.
For input SCS $\ket{\textrm{SCS}}_x$, one can choose $\hat{U}_\textrm{re}=e^{-i\frac{\pi}{2}{\hat{J}_{x}}}$ and the corresponding final state before the measurement is $\ket{\Psi}_\textrm{f,n}=e^{-i\frac{\pi}{2}{\hat{J}_{x}}}e^{-in\phi\hat J_z}e^{-i\frac{\pi}{2}\hat J_y}\ket{J,J}$.
After some algebra, the expectation of half-population difference can be explicitly written as
\begin{eqnarray}\label{Jz_SCS}
J_{z,n}= \frac{N}{2}\sin\left[\frac{4n\gamma_g}{\omega}\left(\frac{\pi}{2} B_\textrm{dc}- B_\textrm{ac}\right)\right] \left[\cos(2\chi t_n)\right]^{N-1}.
\end{eqnarray}

\begin{figure*}[!htp]
	\includegraphics[width=1.8\columnwidth]{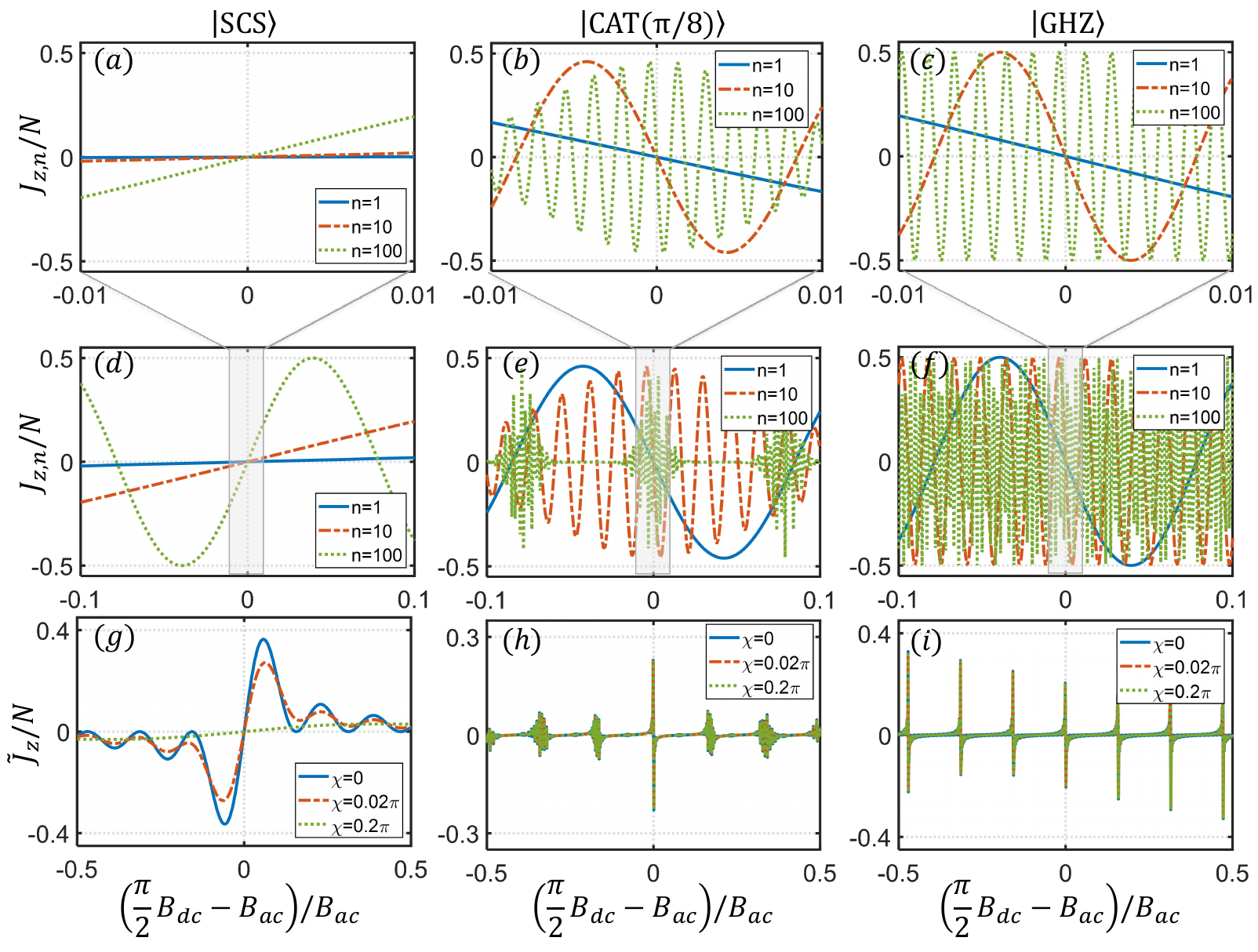}
	\caption{\label{Fig8}(color online).
The output signals of our ac field sensing protocol.
(a)-(f): The measurement signal $J_{z,n}/N$ versus the modulation $\left(\pi B_\textrm{dc}/2-B_\textrm{ac}\right)/B_{\textrm{ac}}$ for (a)(d) SCS, (b)(e) spin cat state, and (c)(f) GHZ state under different evolution cycle number $n$.
The measurement signal $J_{z,n}/N$ is antisymmetric with respect to the phase lock-in point $\phi=0$ with $\phi=\left(\pi B_\textrm{dc}/2-B_\textrm{ac}\right)/B_{\textrm{ac}}=0$.
(g)-(i): The time-averaged signal $\tilde{J}_{z}/N$ versus the modulation $\left(\pi B_\textrm{dc}/2-B_\textrm{ac}\right)/B_{\textrm{ac}}$ for (g) SCS, (h) spin cat state , and (i) GHZ state under different atom-atom interaction $\chi=0,0.02\pi,0.2\pi$ with $n_m=100$.
The time-averaged signal $\tilde{J}_{z}$ is also antisymmetric with respect to $\phi=0$.
Here, $\omega=200\pi$, $\gamma_g=20\pi$, $B_\textrm{ac}=1$, $n_m=100$ and $N=100$.
}
\end{figure*}
As Eq.~\eqref{Jz_SCS} is exactly antisymmetric with respect to the lock-in point $\phi=0$, one can determine $B_\textrm{ac}$ from the antisymmetry of measurement signal $J_{z,n}$, as shown in {Fig.~\ref{Fig8}~(a) and (d)}.
In general, for a large $n$, the pattern of the measurement signal $J_{z,n}$ becomes a high-frequency oscillation and it is not easy to determine the lock-in point from the pattern symmetry in experiments, as shown in {Fig.~\ref{Fig8}~(d)}.
To solve this problem, we can determine the lock-in point via the time-averaged signal 
\begin{eqnarray}\label{tildeJz}
\tilde{J}_{z} =\frac{1}{n_m}\sum_{n=1}^{n_m} J_{z,n}.
\end{eqnarray}
%
For $\chi=0$, one can analytically obtain
\begin{eqnarray}\label{tildeJz_SCS}
\tilde{J}_{z}=\frac{N}{2}\frac{\cos\left(\frac{\phi}{2}\right)-\cos\left(\frac{2n_m+1}{2}\phi\right)}{2\sin(\phi/2)},
\end{eqnarray}
which is also antisymmetric with respect to the lock-in point $\phi=0$, see Fig.~\ref{Fig8}~(g).
Moreover, according to Eqs.~\eqref{Jz_SCS} and \eqref{tildeJz}, we have $\tilde{J}_{z} (\chi)=\frac{N}{2}\sum_{n=1}^{n_m}\sin(n\phi)\left[\cos(2\chi t_n)\right]^{N-1}$ even when $\chi\neq0$.  
Thus the interparticle interaction does not induce any shift on the signal of $\tilde{J}_{z} (\chi)$ but only decreases the contrast of the spectra, see {Fig.~\ref{Fig8}~(g)}.

\begin{figure*}[!htp]
\includegraphics[width=1.55\columnwidth]{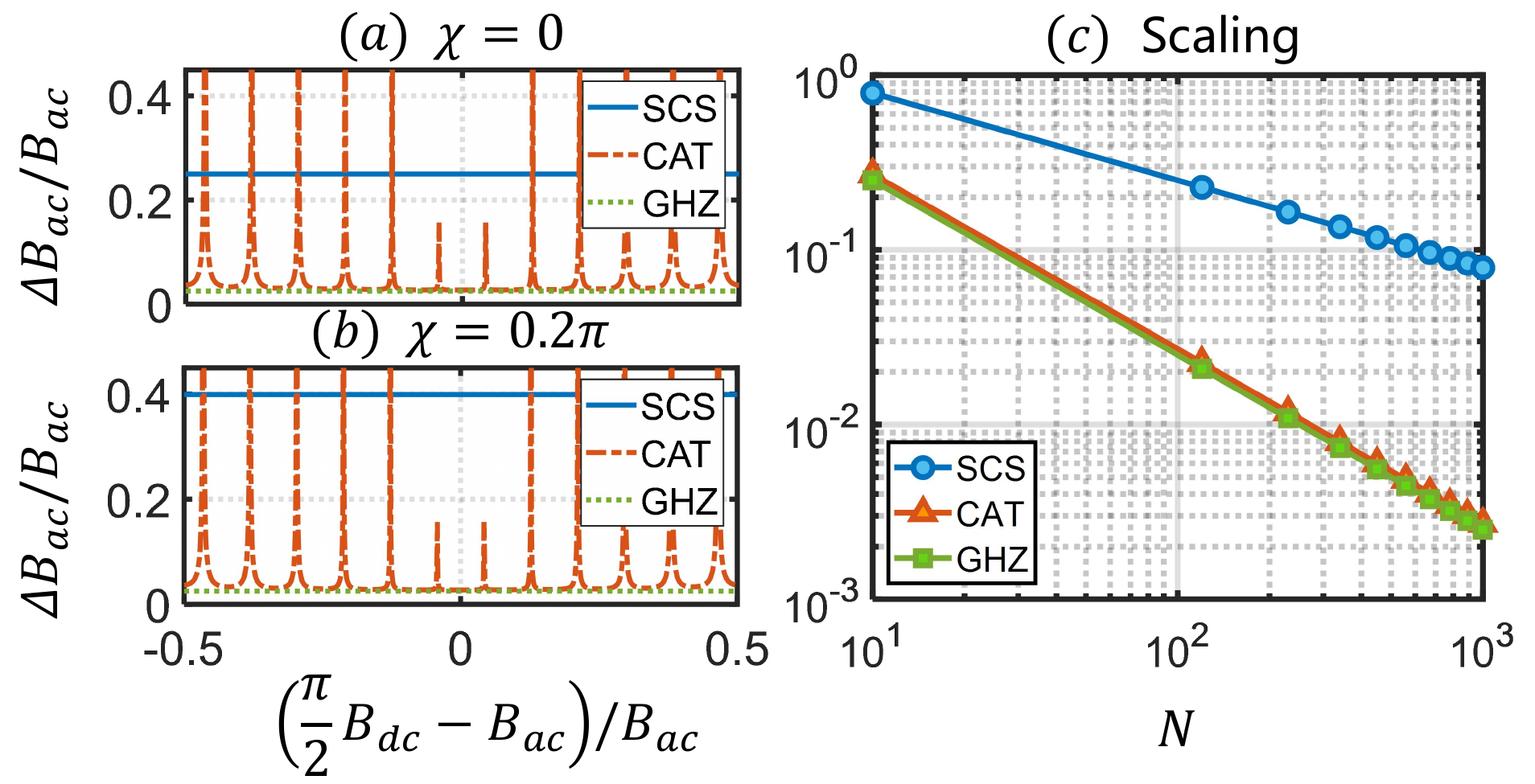}
\caption{\label{Fig9}(color online).
The measurement precision of ac field amplitude with different input states.
Variation of measurement precision versus the modulation $\left(\pi B_\textrm{dc}/2-B_\textrm{ac}\right)/B_{\textrm{ac}}$ for $\ket{\textrm{SCS}}_x$ (solid blue line), $\ket{\textrm{CAT}(\theta=\pi/8)}$ (dashed red line) and $\ket{\textrm{GHZ}}$ (dotted green line) under (a) $\chi=0$ and (b) $\chi=0.2\pi$ with $N=100$.
Precision scaling of $\Delta B_\textrm{ac}$ versus total particle number $N$ for $\ket{\textrm{SCS}}_x$ (blue circles), $\ket{\textrm{CAT}(\theta=\pi/8)}$ (red triangles) and $\ket{\textrm{GHZ}}$ (green squares) with $n=1$.
Here, $\omega/\gamma_g=10$, $B_\textrm{dc}/\textrm{B}_{ac}=2/\pi$.}
\end{figure*}

Then we study the measurement precision.
The QFI for $B_\textrm{ac}$ is $F_{Q}^{B_\textrm{ac}}=N (\frac{4n\gamma_g}{\omega})^{2}$.
Thus the ultimate precision bounds with individual particles for parameter $B_\textrm{ac}$ just can attain the SQL, i.e., $\Delta B_\textrm{ac}^\textrm{QCRB}\propto {1}/{\sqrt{N}}$.
Further, one can analytically obtain $\Delta B_\textrm{ac}$ with $\chi=0$ which reads 
\begin{equation}\label{dBac_n}
\Delta B_\textrm{ac}=\frac{\omega}{4n\gamma_g\sqrt{N}}.
\end{equation}
The measurement precision can saturate the ultimate precision bound in our protocol with $\Delta B_\textrm{ac}=\Delta B_\textrm{ac}^\textrm{QCRB}$.
According to Eq.~\eqref{dBac_n}, we find that the measurement precision does not dependent on the parameter $B_\textrm{ac}$.
Similarly to the case of time-independent signal, however the measurement precision decreases in the presence of interparticle interaction, see the blue line in Fig.~\ref{Fig9}~(b).

\subsection{Entangled particles}
To implement entanglement-enhanced measurement, we consider spin cat states as input in our framework.
Here, we choose an interaction-based operator $\hat{U}_\textrm{re}=e^{-i\frac{\pi}{2}{\hat{J}_{x}^2}}$ (also satisfies Eq.~\eqref{HwI1}) for recombination, and the corresponding final state is 
\begin{eqnarray}\label{Psif_CAT0}
|\Psi\rangle_\textrm{f,n} = e^{-i\frac{\pi}{2}{\hat{J}_{x}^2}}e^{-2i\chi t_n\hat{J}_z^2-in\phi\hat{J}_z}\ket{\textrm{CAT}(\theta)}.
\end{eqnarray}
The final state $|\Psi\rangle_\textrm{f,n}$ has an analytical form when $N$ is an even number, which reads 
\begin{eqnarray}\label{Psif_CAT1}
|{\Psi}\rangle_\text{f,n}=\sum_{m=-J}^{J}e^{-2i\chi t_n m^2}a_m^{\theta,n\phi}\ket{J,m},
\end{eqnarray}
where the coefficient  
\begin{eqnarray}\label{cofficienta}
    a_m^{\theta ,n\phi}=\left( {C_m^{\theta ,0} + C_{ - m}^{\theta ,0}} \right)\cos \left[ {{m\phi  + (N+1)\pi /4}} \right].
\end{eqnarray}
%
The corresponding expectation of half-population difference is
\begin{eqnarray}\label{Jz_CAT1}
J_{z,n}= \left( { - 1} \right)^{J+1}\sum\limits_{m = 1}^J m {\left( {C_m^{\theta ,0} + C_{ - m}^{\theta ,0}} \right)^2}\sin \left( { 2mn\phi } \right).
\end{eqnarray}
According to Eq.~\eqref{Jz_CAT1}, it is obvious that the signal $J_{z,n}$ is independent of $\chi$, and the interparticle interaction neither induces any shift nor decreases the contrast of the measurement signal.
Importantly, due to entanglement, the oscillation frequency of the measurement signal $J_{z,n}$ is dependent on $2m$.
Moreover, Eq.~\eqref{Jz_CAT1} is also exactly antisymmetric with respect to the lock-in point $\phi=0$ for all spin cat states.
Correspondingly, one can determine $B_\textrm{ac}$ from the symmetry property of the measurement signal $J_{z,n} $.
Similarly, the measurement signals $J_{z,n} $ for different evolution time $t_n$ have the same value at the lock-in point, as shown in {Fig.~\ref{Fig8}~(b) and (e)}.
Further, the corresponding time-averaged signal reads
\begin{small}
\begin{eqnarray}\label{tildeJz_CAT1}
    \tilde{J}_{z}&=&\frac{\left( { - 1} \right)^{J+1}}{2}\sum_{m=-J}^{J}m{\left( {C_m^{\theta ,0} + C_{ - m}^{\theta ,0}} \right)^2}\frac{\cos(m\phi)}{\sin(m\phi)}\nonumber\\
    &-&\frac{\left( { - 1} \right)^{J+1}}{2}\sum_{m=-J}^{J}m{\left({C_m^{\theta ,0}+C_{ - m}^{\theta ,0}} \right)^2}\frac{\cos\left[(2n_m\!+\!1)m\phi\right]}{\sin(m\phi)},\nonumber\\
\end{eqnarray}
\end{small}
which is also exactly antisymmetric with respect to the lock-in point, see {Fig.~\ref{Fig8}~(h) and (i)}.

Then we evaluate the measurement precision.
The QFI for $B_\textrm{ac}$ is 
\begin{equation}
    F_{Q}^{B_\textrm{ac}}=N^2\left[\frac{4n\gamma_g}{\omega}\overline{M}(\theta)\right]^{2}=N^2\left[\frac{4n\gamma_g}{\omega}\cos \theta\right]^{2}.
\end{equation}
By using entangled particles prepared in spin cat states, the ultimate precision bounds can be improved to the HL, i.e. $\Delta B_\textrm{ac}^\textrm{QCRB} \propto 1/N$.
In addition, according to Eqs.~\eqref{Jz2_CAT} and \eqref{Jz_CAT1}, the measurement precision can be analytically obtained as
\begin{eqnarray}\label{Delta_CAT1}
    \Delta B_\textrm{ac}=\tilde{\mathcal{F}}(\theta)\frac{\omega}{4n\gamma_g},
\end{eqnarray}
with
\begin{eqnarray}\label{cofficientD}
 \tilde{\mathcal{F}}(\theta)=\frac{\sqrt{\sum_{m=1}^{J}m^2\left(C_m^{\theta,0}+C_{-m}^{\theta,0}\right)^2-(J_{z,n})^2}}{\left|\sum_{m=1}^{J}2m^2\left(C_m^{\theta,0}+C_{-m}^{\theta,0}\right)^2\cos(2mn\phi)\right|}.
 \end{eqnarray}
In particular, we find that $\Delta B_\textrm{ac}=\Delta B_\textrm{ac}^\textrm{QCRB}$ for GHZ state.
%
Furthermore, we study the influence of interparticle interaction on the measurement precision and we find that the interparticle interaction does not affect the measurement precision for spin cat states, see {Fig.~\ref{Fig9}~(a) and (b)}.
We also numerically calculate the corresponding scaling versus particle number $N$ for $\ket{\Psi(\theta=\pi/8)}_{\textrm{CAT}}$ and $\ket{\textrm{GHZ}}$ at the lock-in point, as shown in {Fig.~\ref{Fig9}~(c)}.
%
It is indicated that the measurement precision $\Delta B_\textrm{ac}$ for spin cat states can exhibit the HL scaling.
\section{SUMMARY AND DISCUSSION\label{Sec5}}
We present a general entanglement-enhanced, symmetry-protected Ramsey interferometry protocol that achieves Heisenberg-limited precision while suppressing various spectroscopic shifts.
The protocol is experimentally feasible, requiring neither additional interaction control nor strict timing~\cite{SCPR2025}, and is applicable to diverse quantum systems including Bose-condensed atoms~\cite{EDNature2001,BNPRA2011,GPNJP2018,TVNSR2021}, trapped ions~\cite{MJBPRA2009,PAISR2016,FWMS2021,LDPRAppl2021,KAGScience2021} and nitrogen-vacancy centers~\cite{JHSEL2012,DFPRB2015,CLSR2017,JFBRMP2020,ZQNPJ2022}.
We demonstrate the protocol's effectiveness through measurements of both dc and ac magnetic fields. 
While increasing particle density typically enhances sensitivity, interactions between particles often limit this improvement~\cite{HZPRX2020,TParXiv2025}.
Our protocol overcomes this challenge, maintaining HL precision while eliminating nonlinearity-induced shifts~\cite{JHarXiv2022,SCPR2025}.
For non-entangled states with linear readout, interparticle interactions degrade both signal contrast and accuracy. 
In contrast, spin cat states combined with interaction-based readout decouple the nonlinear term $\chi\hat{J}_z^2$ in the interrogating process.
This ensures that both the expectation values $\langle\hat{J}_z\rangle$ and $\langle\hat{J}_z^2\rangle$ remain independent of $\chi$, preserving both spectral resolution and measurement precision despite particle interactions.

In addition, within our SPDMBI-Ramsey framework, we propose a protocol specifically designed for ac magnetic field amplitude measurement that eliminates phase ambiguity.
Our protocol enables precise amplitude estimation through dc bias scanning while avoiding other spectroscopic shifts.
Unlike conventional fitting approaches, this method achieves superior dynamic range by exploiting the spectrum's anti-symmetry.
Through analytical analysis, we evaluate the measurement precision for both non-entangled and entangled many-body quantum states.  
For non-entangled states, the measurement precision follows SQL scaling as expected, though interparticle interactions degrade this performance. 
In contrast, entangled states like spin cat states can achieve Heisenberg-limited precision through proper interaction-based readout operations - maintaining this enhanced scaling even in the presence of interparticle interactions. 
Crucially, neither non-entangled nor spin cat states experience nonlinearity-induced frequency shifts, preserving measurement accuracy in both cases.

The desired entangled states have been successfully generated in various many-body quantum systems. 
For example, in an atomic Bose-Einstein condensate occupying two hyperfine levels, spin cat states can be prepared and measured through controlled manipulation of atom-atom interactions. 
These entangled states can be generated through dynamical evolution~\cite{HSScience2014,DLPRL2016,CLFOP2012,EDPRL2016,SSMPRA2018,CL2024} or adiabatic processes~\cite{CLPRL2006,CLPRL2009,JHPRA2018} under systems satisfying the symmetry~\eqref{H_sym1}, such as the one-axis twisting Hamiltonian $\hat{H}_\textrm{twist} = \chi\hat{J}_z^2 + \Omega_x\hat{J}_x$ and the XYZ model $\hat{H}_\textrm{XYZ} = \frac{2\chi_\textrm{XYZ}}{N}(\hat{J}_x\hat{J}_y\hat{J}_z+\hat{J}_z\hat{J}_y\hat{J}_x)$~\cite{XZPRL2024}.
The nonlinearity parameter $\chi$, whose magnitude and sign depend on the spatial overlap between spin components and s-wave scattering lengths, can be precisely tuned using spin-dependent forces~\cite{MFRNature2010,CFOPRL2013} or Feshbach resonance techniques~\cite{TParXiv2025,HSScience2014,MFRNature2010,WMPRL2014}. 
This work bridges fundamental research with practical applications, offering immediate benefits for developing high-sensitivity magnetometers~\cite{CLDRMP2017,JFBRMP2020,MWMRMP2020,YSGRMP2000}, atomic clocks~\cite{ADLRMP2015,LPRMP2018}, weak-force detectors~\cite{RSNC2017}, and noise spectroscopy detector~\cite{JBNP2011,IAPRA2016,SDCP2020}.

\acknowledgments{Sijie Chen and Jiahao Huang contribute equally. This work is supported by the National Natural Science Foundation of China (Grants No. 12025509, No. 12305022, and No. 92476201), the National Key Research and Development Program of China (Grant No. 2022YFA1404104), and the Guangdong Provincial Quantum Science Strategic Initiative (GDZX2305006 and GDZX2405002).}
\setcounter{equation}{0}
\setcounter{figure}{0}
\renewcommand{\theequation}{A\arabic{equation}}
\renewcommand{\thefigure}{A{\arabic{figure}}}
\section*{APPENDIX A: Symmetry-protected detection\label{SecSM1}}
In the main text, we have simply demonstrated that for a Hamiltonian of the form 
\begin{eqnarray}\label{HwI}
    \hat{H}_\textrm{I}(\delta,t)&=&f_1^\textrm{even}\hat{J}_x+f_2^\textrm{odd}\hat{J}_y+f_3^\textrm{odd}\hat{J}_z\nonumber\\
    & &+e_1^\textrm{even}\hat{J}_y\hat{J}_z+e_2^\textrm{odd}\hat{J}_z\hat{J}_x+e_3^\textrm{odd}\hat{J}_x\hat{J}_y\nonumber\\
    & &+g_1^\textrm{even}\hat{J}_x^2+g_2^\textrm{even}\hat{J}_y^2+g_3^\textrm{even}\hat{J}_z^2,
\end{eqnarray}
when the initial state is an eigenstate of the exchange operator $\hat{U}_\textrm{ex}^\dagger=e^{i\pi\hat{J}_x}$ (satisfying $C_m(0)=\pm C_{-m}(0)$), the population difference $J_z(\delta,t)=\bra{\Psi(\delta,t)}\hat{J}_z\ket{\Psi(\delta,t)}$ between between $\ket{\uparrow}$ and $\ket{\downarrow}$ states exhibits antisymmetric behavior about $\delta$ at all times $t$, such that $J_z(\delta,t)=-J_z(-\delta,t)$.
Here, $x_i^\textrm{even}(-\delta,t)=x_i^\textrm{even}(\delta,t)$ and $x_i^\textrm{odd}(-\delta,t)=-x_i^\textrm{odd}(\delta,t)$ are symmetric and antisymmetric functions versus $\delta$, in which $\delta$ is usually the difference between the target signal and the reference signal coming from destructive interference.
We will demonstrate that the eigenstate condition of the exchange operator $\hat{U}\textrm{ex}$ is equivalent to the initial state $\ket{\Psi_0} = \sum_m C_m(0)\ket{J,m}$ having either symmetric ($C_m(0) = +C_{-m}(0)$) or antisymmetric ($C_m(0) = -C_{-m}(0)$) coefficients. A rigorous proof of this equivalence will be provided.

Firstly, the adequacy, i.e. $\ket{\Psi_0}=\sum_mC_m\ket{J,m}$ is the eigenstate of $\hat{U}_\textrm{ex}^\dagger$ $\Rightarrow$ $C_m=\pm C_{-m}$.
Any initial state can be described by 
\begin{equation}
\ket{\Psi_0}=C_0\ket{J,0}+\sum_{m=1}^{J}(C_m\ket{J,m}+C_{-m}\ket{J,-m}).
\end{equation}
If $\ket{\Psi_0}$ is the eigenstate of $\hat{U}_\textrm{ex}^\dagger$, it would satisfy $\hat{U}_\textrm{ex}^{\dagger}\ket{\Psi_0}\bra{\Psi_0}\hat{U}_\textrm{ex}=\ket{\Psi_0}\bra{\Psi_0}$, which is equivalent to
\begin{widetext}
\begin{eqnarray}\label{eipro}
\hat{U}_\textrm{ex}^{\dagger}\ket{\Psi_0}\bra{\Psi_0}\hat{U}_\textrm{ex}\nonumber
&&=\ket{\Psi_0}\bra{\Psi_0}\\\nonumber&&=\left[C_0\ket{J,0}+\sum_{m=1}^{J}(C_m\ket{J,-m}+C_{-m}\ket{J,m})\right]\left[C_0^*\bra{J,0}+\sum_{m=1}^{J}(C_m^*\bra{J,-m}+C_{-m}^*\bra{J,m})\right]\\\nonumber
&&=\left[C_0\ket{J,0}+\sum_{m=1}^{J}(C_m\ket{J,m}+C_{-m}\ket{J,-m})\right]\left[C_0^*\bra{J,0}+\sum_{m=1}^{J}(C_m^*\bra{J,m}+C_{-m}^*\bra{J,-m})\right].\\
\end{eqnarray}
Here, we have used $\hat{U}_\textrm{ex}^\dagger\ket{J,m}=i^N\ket{J,-m}$ with the particle number $N$.
Since all spin coherent states with $\theta\in[0,\pi]$ and $\varphi\in[0,2\pi]$ form a set of over-complete bases,
therefore one can obtain $\ket{J,m}=\sum_{\theta,\varphi}C_{\theta,\varphi}\ket{\theta,\varphi}$ with $\ket{\theta,\varphi}=\left[\cos(\theta/2)e^{-i\varphi/2}\ket{\uparrow}\!+\!\sin(\theta/2)e^{i\varphi/2}\ket{\downarrow}\right]^{\otimes N}$, and
\begin{eqnarray}\label{UexJm}
\hat{U}_\textrm{ex}^\dagger\ket{J,m}&=&\sum_{\theta,\varphi}C_{\theta,\varphi}\{e^{i\pi\hat{\sigma}_x/2}\left[\cos(\theta/2)e^{-i\varphi/2}\ket{\uparrow}\!+\!\sin(\theta/2)e^{i\varphi/2}\ket{\downarrow}\right]\}^{\otimes N}\\\nonumber
&=&\sum_{\theta,\varphi}C_{\theta,\varphi}\{i\left[\cos(\theta/2)e^{-i\varphi/2}\ket{\downarrow}\!+\!\sin(\theta/2)e^{i\varphi/2}\ket{\uparrow}\right]\}^{\otimes N}=i^N\ket{J,-m}
\end{eqnarray}
\end{widetext}
with the Pauli matrix $\hat{\sigma}_x=\left(\begin{array}{ccc}0\quad1\\1\quad0\end{array}\right)$.
By comparing the coefficients of Eq.~\eqref{eipro}, we can get the last equal sign of Eq.~\eqref{eipro} holds if and only if $C_m=\pm C_{-m}$ as mentioned in the main text.

Secondly, the necessity, i.e.  $C_m=\pm C_{-m}$ $\Rightarrow$ $\ket{\Psi_0}=\sum_m C_m\ket{J,m}$ is the eigenstate of $\hat{U}_\textrm{ex}^\dagger$.
When $C_m=C_{-m}$, we have
\begin{small}
\begin{eqnarray}
    \small\hat{U}_\textrm{ex}^\dagger\ket{\Psi_0}&=&i^N\left[C_0\ket{J,0}+\sum_{m=1}^{J}(C_m\ket{J,-m}+C_{-m}\ket{J,m})\right] \nonumber\\
    &=& i^N\ket{\Psi_0}.
\end{eqnarray}
\end{small}
Similarly, for $C_m=-C_{-m}$, we have $C_0=0$ and
\begin{eqnarray}
    \hat{U}_\textrm{ex}^\dagger\ket{\Psi_0}&=&i^N\left[\sum_{m=1}^{J}(C_m\ket{J,-m}+C_{-m}\ket{J,m})\right]\nonumber\\
    &=&-i^N\ket{\Psi_0}.
\end{eqnarray}
Here we have also used $\hat{U}_\textrm{ex}^\dagger\ket{J,m}=i^N\ket{J,-m}$.
Thus we have proved $\ket{\Psi(0)}=\sum_{m=-J}^{J}C_m(0)\ket{J,m}$ with $C_m(0)=\pm C_{-m}(0)$ is the unique eigenstate solution of $\hat{U}_\textrm{ex}^\dagger$.

Further, as mentioned in the main text, the Hamiltonian satisfy
\begin{equation}\label{H_sym} 
\hat{U}_\textrm{ex}^{\dagger}\hat{H}_I(\delta,t)\hat{U}_\textrm{ex}=\hat{H}_I(-\delta,t).
\end{equation}
According to quantum Liouville equation
\begin{equation}\label{Liouv}
\frac{\partial\hat{\rho}(\delta,t)}{\partial t}=-i\left[\hat{H}_I(\delta,t),\hat{\rho}(\delta,t)\right],
\end{equation}
we have 
\begin{eqnarray}\label{Ld}
\frac{\partial\hat{U}_\textrm{ex}^\dagger\hat{\rho}(\delta,t)\hat{U}_\textrm{ex}}{\partial t}&=&\hat{U}_\textrm{ex}^\dagger\partial_t\hat{\rho}(\delta,t)\hat{U}_\textrm{ex}\\\nonumber
&=&-i\hat{U}_\textrm{ex}^\dagger\left[\hat{H}_I(\delta,t),\hat{\rho}(\delta,t)\right]\hat{U}_\textrm{ex}\\\nonumber
&=&-i\left[\hat{H}_I(-\delta,t),\hat{U}_\textrm{ex}^\dagger\hat{\rho}(\delta,t)\hat{U}_\textrm{ex}\right]
\end{eqnarray}
and
\begin{eqnarray}\label{L-d}
\frac{\hat{\rho}(-\delta,t)}{\partial t}=-i\left[\hat{H}_I(-\delta,t),\hat{\rho}(-\delta,t)\right].
\end{eqnarray}
Comparing Eqs.~\eqref{Ld} and \eqref{L-d}, and considering the input state $\ket{\Psi_0}$ and Hamiltonian $\hat{H}_I(\delta,t)$ uniquely determine $\hat{\rho}(\rho,t)$, we can obtain $\hat{U}_\textrm{ex}^\dagger\hat{\rho}(\delta,t)\hat{U}_\textrm{ex}=\hat{\rho}(-\delta,t)$ from $\hat{U}_\textrm{ex}^\dagger\hat{\rho}(\delta,0)\hat{U}_\textrm{ex}=\hat{\rho}(-\delta,0)$.
Therefore, we have
\begin{eqnarray}\label{Jz}
\langle\hat{J}_z(t)\rangle_{-\delta}&=&\textrm{Tr}[\hat{\rho}(-\delta,t)\hat{J}_z]=\textrm{Tr}[\hat{U}_\textrm{ex}^\dagger\hat{\rho}(\delta,t)\hat{U}_\textrm{ex}\hat{J}_z]\nonumber\\
&=&\textrm{Tr}[\hat{U}_\textrm{ex}\hat{J}_z\hat{U}_\textrm{ex}^\dagger\hat{\rho}(\delta,t)]=-\textrm{Tr}[\hat{J}_z\hat{\rho}(\delta,t)]\nonumber\\
&=&-\langle\hat{J}_z(t)\rangle_{\delta}
\end{eqnarray}
where $\textrm{Tr}[\hat{X}]$ represents tracing $\hat{X}$.
According to Eq.~\eqref{Jz} we have proved $\langle\hat{J}_z(t)\rangle_{\delta}$ is antisymmetric about $\delta$ as long as it satisfies Eq.~\eqref{HwI} for the initial state $\ket{\Psi(0)}=\sum_{m=-J}^{J}C_m(0)\ket{J,m}$ with $C_m(0)=\pm C_{-m}(0)$.

In particular for the Ramsey spectroscopy with protocol-I and protocol-II in the main text, the Hamiltonians in both protocols satisfy the following form
\begin{eqnarray}\label{HwI}
    \hat{H}_\textrm{I}(\delta,t)&=&f_1^\textrm{even}\hat{J}_x+f_3^\textrm{odd}\hat{J}_z+g_3^\textrm{even}\hat{J}_z^2.
\end{eqnarray}
%
%
The time-evolution of the system obeys the Schr\"{o}dinger equation (we set $\hbar=1$),
\begin{eqnarray}\label{SE}
i\frac{\partial{\ket{\Psi(\delta,t)}}}{\partial t}
=\hat{H}_\textrm{I}(\delta,t)\ket{\Psi(\delta,t)}
\end{eqnarray}
with $\ket{\Psi(\delta,t)}=\sum_{m=-J}^{J}C_m(\delta,t)\ket{J,m}$ denoting the system state at any time $t$.

Since $\hat{J}_x=(\hat{J}_++\hat{J}_-)/2$, $\hat{J}_y=(\hat{J}_+-\hat{J}_-)/(2i)$, $\hat{J}_z\ket{J,m}=m\ket{J,m}$, $\hat{J}_\pm\ket{J,m}=\lambda_m^{\pm}\ket{J,m\pm1}$ ($\lambda_m^{\pm}=\sqrt{(J\mp m)(J\pm m+1)}$), from Eq.~\eqref{SE}, we have 
\begin{widetext}
\begin{eqnarray}
i\sum_m\partial_tC_m(\delta,t)\ket{J,m}
&=&\frac{f_1^\textrm{even}(\delta,t)}{2}\left[\sum_{m=-J}^{J-1}C_{m}(\delta,t)\lambda_{m}^+\ket{J,m+1}\!+\!\sum_{m=-J+1}^{J}C_{m}(\delta,t)\lambda_{m}^-\ket{J,m-1}\right]\nonumber\\\nonumber
& &+f_3^\textrm{odd}(\delta,t)\sum_{m=-J}^JC_{m}(\delta,t)m\ket{J,m}+g_3^\textrm{even}(\delta,t)\sum_{m=-J}^JC_{m}(\delta,t)m^2\ket{J,m}\\\nonumber
&=&\frac{f_1^\textrm{even}(\delta,t)}{2}\left[\sum_{m'=-J+1}^{J}C_{m'-1}(\delta,t)\lambda_{m'-1}^+\ket{J,m'}\!+\!\sum_{m'=-J}^{J-1}C_{m'+1}(\delta,t)\lambda_{m'+1}^-\ket{J,m'}\right]\\
& &+f_3^\textrm{odd}(\delta,t)\sum_{m'=-J}^JC_{m'}(\delta,t)\ket{J,m'}+g_3^\textrm{even}(\delta,t)\sum_{m'=-J}^JC_{m'}(\delta,t)m'^2\ket{J,m'}.
\end{eqnarray}
\end{widetext}
So one can obtain $\partial_tC_m(\delta,t)=\pm\partial_tC_{-m}(-\delta,t)$ when $C_m(0)=\pm C_{-m}(0)$ for $m\in\{-J,\cdots,J\}$ and $t\geq0$.
Thus we have 
\begin{equation}
    C_{m}(\delta,t)=C_{-m}(-\delta,t)
\end{equation}
and 
\begin{eqnarray}
\langle\hat{J}_z(t)\rangle_{-\delta}&=&\sum_{m}m|C_m(-\delta,t)|^2\\\nonumber
&=&\sum_{-m}{-m}|C_{-m}(-\delta,t)|^2=\sum_{m}{-m}|C_{m}(\delta,t)|^2\\\nonumber
&=&-\langle\hat{J}_z(t)\rangle_{\delta}.
\end{eqnarray}
Therefore, we have proved $\langle\hat{J}_z(t)\rangle_{\delta}$ is antisymmetric about $\delta$ when the Hamiltonian satisfies Eq.~\eqref{HwI} for initial state $\ket{\Psi(0)}=\sum_{m=-J}^{J}C_m(0)\ket{J,m}$ with $C_m(0)=\pm C_{-m}(0)$.

Our SPDMBI theory is general. 
For example, for protocol-III in the main text, we find the GHZ state is robust against the imperfectness of pulses $(\Omega,\epsilon)$ [see Fig.~\ref{Fig4}~(e) and (g)] but the other spin cat states is not [see Fig.~\ref{Fig5}~(e) and (g)].
This is because although the readout stage of protocol-III does not satisfy the symmetry of $\hat{U}_\textrm{ex}$ with imperfect pulses $(\Omega,\epsilon)$, it still satisfies the symmetry under $\hat{U}_{\pi_y}=e^{-i\pi\hat{J}_y}$, i.e.,
\begin{equation}
    \hat{U}_{\pi_y}^\dagger\hat{H}_\textrm{Ram}^\text{III}(\delta,t)\hat{U}_{\pi_y}=\hat{H}_\textrm{Ram}^\text{III}(-\delta,t).
\end{equation}
Since the GHZ state also satisfies 
\begin{equation}
    \hat{U}_{\pi_y}^\dagger\ket{\textrm{GHZ}}\bra{\textrm{GHZ}}\hat{U}_{\pi_y}=\ket{\textrm{GHZ}}\bra{\textrm{GHZ}},
\end{equation}
but the spin cat state does not, i.e. $\hat{U}_{\pi_y}^\dagger\ket{\textrm{CAT}(\theta=\pi/8)}\bra{\textrm{CAT}(\theta=\pi/8)}\hat{U}_{\pi_y}\neq\ket{\textrm{CAT}(\theta=\pi/8)}\bra{\textrm{CAT}(\theta=\pi/8)}$, the GHZ state exhibits the robustness.
It should be mentioned that, the spin cat state $\ket{\textrm{CAT}(\theta=\pi/8,\phi=\pi/2)}$ satisfies the symmetry as GHZ state, but it cannot attain the HL.
Therefore, the GHZ state can not only suppress the spectral shift to promise measurement accuracy but also improve the measurement precision to HL.
\setcounter{equation}{0}
\setcounter{figure}{0}
\renewcommand{\theequation}{B\arabic{equation}}
\renewcommand{\thefigure}{B{\arabic{figure}}}

\section*{APPENDIX B: Analytical results for spin coherent state\label{SecSM2}}
\begin{figure*}[!htp]
	\includegraphics[width=1.6\columnwidth]{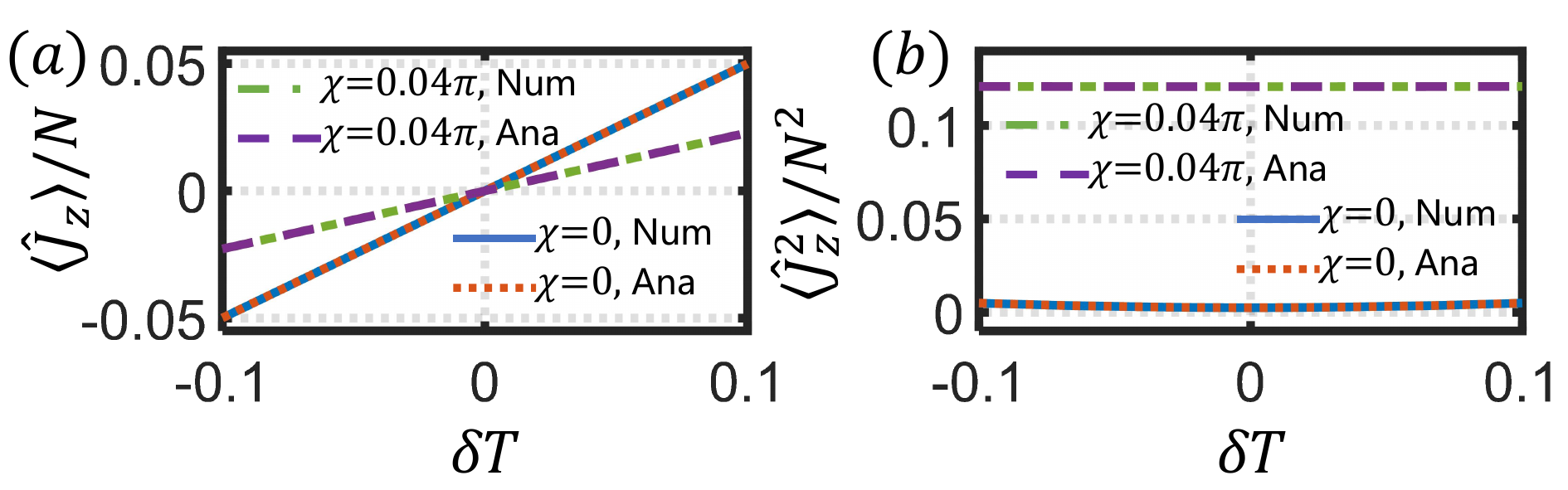}
	\caption{\label{B2}(color online).
Variation of (a) $\langle\hat{J}_z\rangle_f/N$ and (b)$\langle\hat{J}_z^2\rangle_f/N^2$ versus $\delta T$ for individual particles.
The analytical results (red dotted line, purple dashed line) are consistent with the numerical ones (blue solid line, green dotted-dashed line) for both $\chi=0$ (red dotted line, blue solid line) and $\chi=0.04\pi$ (purple dashed line, green dotted-dashed line).
Here, $T=1$ and $N=100$.}
\end{figure*}

In this section, we give the proof of Eqs.~\eqref{JzSCS} and ~\eqref{Jz2SCS} in the main text.
For the input SCS we consider $\ket{\Psi(0)}=\ket{\textrm{SCS}}_x=e^{-i\frac{\pi}{2}\hat J_y}\ket{J,J}=\left(\frac{\ket{\uparrow}+\ket{\downarrow}}{\sqrt{2}}\right)^{\otimes N}$,
the output state $\ket{\Psi}_f$ after the interrogation stage can be expressed as Eq.~\eqref{outstate}.
After a suitable unitary operation $\hat{U}_\textrm{re}=e^{-i\frac{\pi}{2}\hat{J}_x}$ performed on $\ket{\Psi}_f$ for recombination, the half-population difference is
\begin{eqnarray}\label{B_JzSCSn}
\langle\hat{J}_z\rangle_f&=&\bra{\Psi(T)}\hat{U}_\textrm{re}^{\dagger}\hat{J}_z\hat{U}_\textrm{re}\ket{\Psi(T)}\\\nonumber
&=&\bra{\Psi(T)}\hat{J}_y\ket{\Psi(T)}\\\nonumber
&=&\bra{\Psi(T)}\frac{\hat{J}_{+}-\hat{J}_{-}}{2i}\ket{\Psi(T)}
\end{eqnarray}
with 
$\ket{\Psi(T)}=\sum_{m=-J}^{J}C_m e^{-i\chi T m^2}e^{-i\delta T m}\ket{J,m}$, $C_m=\sqrt{\frac{(2J)!}{(J+m)!(J-m)!}}\left(\frac{1}{2}\right)^J$ and $\hat{J}_{\pm}=[\hat{J}_x\pm i\hat{J}_y]/2$.
Considering ${{\hat J}_ \pm }\left| m \right\rangle  = \lambda _m^ \pm \left| {m \pm 1} \right\rangle  = \sqrt {J(J + 1) - m(m \pm 1)} \left| {m \pm 1} \right\rangle $ into Eq.~\eqref{B_JzSCSn}, we have
\begin{widetext}
\begin{small}
\begin{eqnarray}\label{B_Jyoutn}
 {\hat J}_y\ket{\Psi(T)} &=& \frac{1}{2i}\sum_{m=-J}^{J} {C_m\left( {\lambda _m^ + \left| {J,m + 1} \right\rangle  - \lambda _m^ - \left| {J,m - 1} \right\rangle } \right)e^{{ - i\chi {T}{m^2}} }e^{-i\delta T}} \nonumber\\
 &=& \frac{1}{2i}\sum_{m' =  - J + 1}^J {C_{m' - 1} \lambda _{m' - 1}^ + e^{ - i\chi {T}{{(m' - 1)}^2}}e^{-i\delta T(m' - 1)}\left| {J,m'} \right\rangle } 
 - \frac{1}{2i}\sum_{m' =  - J}^{J - 1} {{C_{m' + 1}}\lambda _{m' + 1}^ - e^{ - i\chi {T}{{(m' + 1)}^2}}e^{-i\delta T(m' + 1) }\left| {J,m'} \right\rangle } \nonumber\\
 &=& \frac{1}{{2i}}\sum\limits_{m' =  - J + 1}^J {{C_{m' - 1}}\lambda _{m' - 1}^ + e^{ - i\chi {T}{{(m' - 1)}^2}} e^{-i\delta T(m' - 1)}\left| J,{m'} \right\rangle } \nonumber\\
 &&- \frac{1}{{2i}}\sum\limits_{m'' =  - J + 1}^J {{C_{m'' - 1}}\lambda _{m'' - 1}^ + e^{ - i\chi {T}{{(m'' - 1)}^2}} e^{i\delta T(m'' - 1)}\left| { J,- m''} \right\rangle } \nonumber\\
 &=& \frac{1}{{2i}}\sum\limits_{m' =  - J + 1}^J C_{m' - 1}\lambda _{m' - 1}^ + e^{ - i\chi {T}{{(m' - 1)}^2}} \left[e^{-i\delta T(m' - 1)}\left| {J,m'} \right\rangle  - e^{i\delta T(m' - 1)}\left| { J,- m'} \right\rangle \right],
\end{eqnarray}
\end{small}
which have used $\lambda_m^ +  = \lambda _{ - m}^ - $.
Inserting Eq.~\eqref{B_Jyoutn} and $_\textrm{out,n}\bra{\Psi}=\sum_{m=-J}^{J}C_m e^{2i\chi t_n m^2}e^{in\phi m}\bra{J,m}$ into Eq.~\eqref{B_JzSCSn}, we have 
\begin{equation}\label{B_JzSCSn1}
\langle\hat{J}_z\rangle_f(\chi) = \sin (\delta T)\sum_{m =  - J + 1}^J {C_{m - 1}}{C_m}\lambda _{m - 1}^{+}  e^{i \chi {T}(2m - 1)}
\end{equation}
Substituting ${C_{m - 1}}{C_m}\lambda _{m - 1}^ + = J{\left( {\frac{1}{2}} \right)^{2J - 1}}\frac{{(2J - 1)!}}{{(J + m - 1)!(J - m)!}}$ into Eq.~\eqref{B_JzSCSn1}, we have
\begin{eqnarray}
        \left\langle {{{\hat J}_z}} \right\rangle _f(\chi) &=&  \frac{N}{2}\sin (\delta T)\sum\limits_{m =  - J + 1}^J {{\left( {\frac{1}{2}} \right)}^{2J - 1}}\frac{{(2J - 1)!}}{{(J + m - 1)!(J - m)!}}e^{i\chi {T}(2m - 1)} \nonumber\\
        &=&  \frac{N}{2}\sin (\delta T){\left[ {\cos (\chi {T})} \right]^{2J - 1}},
    \end{eqnarray}
 which have used $\sum\limits_{m =  - J + 1}^J {{\left( {\frac{1}{2}} \right)}^{2J - 1}}\frac{{(2J - 1)!}}{{(J + m - 1)!(J - m)!}}e^{i\chi {T}(2m - 1)}  = \left( \frac{e^{- i\chi {T}} + e^{i\chi {T}}}{2} \right)^{2J - 1} = {\left[ {\cos (\chi {T})} \right]^{2J - 1}}$.

Similarly, we have
\begin{small}
\begin{eqnarray}
\langle\hat{J}_z^2\rangle_f(\chi)&=&\bra{\Psi(T)}\hat{U}_\textrm{re}^{\dagger}\hat{J}_z^2\hat{U}_\textrm{re}\ket{\Psi(T)}=\bra{\Psi(T)}\hat{J}_y^2\ket{\Psi(T)}=\left[{\hat J}_y\ket{\Psi(T)} \right]^\dagger{\hat J}_y\ket{\Psi(T)}\nonumber\\
&=&\frac{1}{2}\left[\sum\limits_{m' =  - J + 1}^J C_{m' - 1}^2(\lambda _{m' - 1}^ +)^2-\sum\limits_{m' =  - J + 1}^{J-1}C_{m' + 1}C_{m' - 1}\lambda _{m' + 1}^- \lambda _{m' - 1}^ + e^{4i\chi T m}\cos(2\delta T)\right] \nonumber\\
&=&\frac{J^2}{2}\sum_{m=-J+1}^{J}\left(\frac{1}{2}\right)^{2J-1}\frac{(2J-1)!}{(J+m-1)!(J-m)!}+\frac{J(m'+1/2)}{2}\sum_{m'=-J+1/2}^{J-1/2}\left(\frac{1}{2}\right)^{2J-1}\frac{(2J-1)!}{(J+m'-1/2)!(J-m'-1/2)!} \nonumber\\
& &-\frac{J(2J-1)}{4}\sum_{m=-J+1}^{J-1}\left(\frac{1}{2}\right)^{2J-2}\frac{(2J-2)!}{(J-m-1)!(J+m-1)!}e^{4i\chi t m}\cos(2\delta T) \nonumber\\
&=&\frac{N(N+1)}{8}-\frac{N(N-1)}{8}[\cos(2\chi T)]^{N-2}\cos(2\delta T).
\end{eqnarray}
\end{small}
\end{widetext}
Therefore, we have proved Eq.~\eqref{JzSCS} and Eq.~\eqref{Jz2SCS}.
In addition, we find our analytical results are consistent with the numerical results, as shown in {Fig.~\ref{B2}}.
\setcounter{equation}{0}
\setcounter{figure}{0}
\renewcommand{\theequation}{D\arabic{equation}}
\renewcommand{\thefigure}{D{\arabic{figure}}}
\section*{APPENDIX C: Analytical results for spin cat state\label{SecSM5}}
The spin cat state is a typical kind of macroscopic superposition of spin coherent states (MSSCS)~\cite{JHSR2016,JHPRA2018}, which is in the form 
\begin{equation}
    \ket{\textrm{MSSCS}(\theta,\varphi)}=N_c \left[\ket{\theta,\varphi}_\textrm{SCS}+\ket{\pi-\theta,\varphi}_\textrm{SCS}\right]
\end{equation}
with $N_c$ is the normalization and $\ket{\theta,\varphi}_\textrm{SCS}$ is the same as Eq.~\eqref{SCS_dk}.
Without loss of generality, we assume $\varphi=0$ and thus the MSSCS in the form 
\begin{eqnarray}
    \ket{\textrm{MSSCS}(\theta,0)}&\!=\!&\frac{1}{\sqrt{\mathcal{N}}} \sum_{m=-J}^{J}\!\left(C_m^{\theta,0}+C_m^{\pi-\theta,0}\right)\!\ket{J,m} \nonumber\\
    &\!=\!&\frac{1}{\sqrt{\mathcal{N}}} \sum_{m=-J}^{J}\!C_m^{\theta,0} \left(\ket{J,m}+\ket{J,-m}\right) \nonumber \\
\end{eqnarray}
with normalization coefficient $\mathcal{N}=\sum_{m=-J}^{J}\left|C_m^{\theta,0}+C_{-m}^{\theta,0}\right|^2$ which have used $C_{m}^{\pi-\theta,0}=C_{-m}^{\theta,0}$.
Specially, when the two superposition SCSs are orthogonal or quasi-orthogonal, the corresponding MSSCS can be regarded as a spin cat state which has been studied in detail~\cite{JHSR2016,JHPRA2018}.
Mathematically, the sufficient condition of spin cat states can be expressed as ~\cite{JHPRA2018}
\begin{equation}
    \theta\lesssim\theta_c\equiv\sin^{-1}\left(2\left\{\frac{[(J-1)!]^2}{2(2J)!}\right\}^\frac{1}{2J}\right).
\end{equation}
For total particle number $N \geq 6$, when $\theta \leq \pi/8$, the corresponding MSSCS can be regarded as a spin cat state which can be written as
\begin{eqnarray}
    \ket{\textrm{CAT}(\theta)}&\!=\!&\frac{1}{\sqrt{2}}\!\sum_{m=-J}^{J}\!\left(C_m^{\theta,0}+C_m^{\pi-\theta,0}\right)\!\ket{J,m} \nonumber\\
    &\!=\!&\frac{1}{\sqrt{2}}\!\sum_{m=-J}^{J}\!C_m^{\theta,0}\!\left(\ket{J,m}+\ket{J,-m}\right).
\end{eqnarray}
%
Here, we prove Eqs.~\eqref{CATo1b}, ~\eqref{Jz_CATb}, ~\eqref{Jz2_CATb}, ~\eqref{Jz_CAT} and ~\eqref{Jz2_CAT} as follows.

For protocol-II, after an interaction-based nonlinear operation $\hat{U}_\textrm{re}=e^{-i\delta t_r\hat{J}_y}e^{-i\frac{\pi}{2}{\hat{J}_{y}^2}}=e^{i\frac{\pi}{2}{\hat{J}_{x}}}e^{-i\frac{\pi}{2}{\hat{J}_{z}^2}}e^{-i\frac{\pi}{2}\hat{J}_x}$ imposed on $\ket{\Psi(T)}$ for recombination with $\chi=0$ and $t_r=\pi/(2\chi_r)$ here, the finial state is 
\begin{eqnarray}\label{C_CATf1}
	\ket{\Psi}_f &=& e^{-i\delta t_r\hat{J}_y}e^{-i\frac{\pi}{2}{\hat{J}_{x}}}e^{-i\frac{\pi}{2}{\hat{J}_{z}^2}}e^{i\frac{\pi}{2}{\hat{J}_{x}}}e^{-i\delta T\hat{J}_z}\ket{\textrm{CAT}(\theta)}\nonumber\\
 &=& e^{-i\delta t_r\hat{J}_y}\frac{{{{\left| {\psi \left( {\theta ,\delta T } \right)} \right\rangle }_1} + {{\left| {\psi \left( {\pi  - \theta ,\delta T } \right)} \right\rangle }_1}}}{{\sqrt {2} }}\\\nonumber
 &=&e^{-i\delta t_r\hat{J}_y}\ket{\Psi}_{f0}
\end{eqnarray}
with
\begin{equation}\label{C_psi1}
    \left| {\psi \left( {\theta ,\varphi } \right)} \right\rangle _1 
    = e^{-i\frac{\pi}{2}{\hat{J}_{x}}}e^{-i\frac{\pi}{2}{\hat{J}_{z}^2}}e^{i\frac{\pi}{2}{\hat{J}_{x}}} {\left| {\theta ,\varphi } \right\rangle _\textrm{SCS}}.
\end{equation}
according to 
\begin{widetext}
\begin{eqnarray}\label{C_psi1y}
     e^{i\frac{\pi}{2}{\hat{J}_{x}}} {\left| {\theta ,\varphi } \right\rangle _\textrm{SCS}}&=&{e^{iN{\phi _0}'}}\left| {\theta ',\varphi '} \right\rangle _\textrm{SCS} \nonumber\\
    &=&{\left( {\frac{1}{2}} \right)^J}{\left\{ {\left[ {\cos \left( {\frac{\theta }{2}} \right){e^{ - i\varphi /2}} + i\sin \left( {\frac{\theta }{2}} \right){e^{i\varphi /2}}} \right]\left|  \uparrow  \right\rangle  + \left[ { i \cos \left( {\frac{\theta }{2}} \right){e^{ - i\varphi /2}} + \sin \left( {\frac{\theta }{2}} \right){e^{i\varphi /2}}} \right]\left|  \downarrow  \right\rangle } \right\}^{ \otimes N}},
\end{eqnarray}
\begin{eqnarray}
    e^{ - i\frac{\pi }{2}\hat J_z^2} e^{iN{\phi _0}'}{\left| {\theta ',\varphi '} \right\rangle _\textrm{SCS}} &=& e^{iN{\phi _0}'}\sum\limits_{m =  - J}^J {C_m^{\theta ',\varphi '}e^{ - i\frac{\pi }{2}{m^2}}}\left| {J,m} \right\rangle \nonumber\\
    &=& e^{iN{\phi _0}'}\left[\sum_{m}^\textrm{even} {C_m^{\theta ',\varphi '}\left| {J,m} \right\rangle }  - i\sum_{m}^\textrm{odd} {C_m^{\theta ',\phi '}\left| {J,m} \right\rangle }\right] \nonumber\\
 &=& e^{iN{\phi _0}'}\frac{{{{\left| {\theta ',\varphi '} \right\rangle }_\textrm{SCS}} + {{\left| {\theta ',\varphi ' + \pi } \right\rangle }_\textrm{SCS}}}}{2} - i{e^{iN{\phi _0}'}}\frac{{{{\left| {\theta ',\varphi '} \right\rangle }_\textrm{SCS}} - {{\left| {\theta ',\varphi ' + \pi } \right\rangle }_\textrm{SCS}}}}{2} \nonumber\\
 &=& \frac{{{e^{ - i\pi /4}}{e^{iN{\phi _0}'}}{{\left| {\theta ',\varphi '} \right\rangle }_\textrm{SCS}} + {e^{i\pi /4}}{e^{iN{\phi _0}'}}{{\left| {\theta ',\varphi ' + \pi } \right\rangle }_\textrm{SCS}}}}{{\sqrt 2 }},
\end{eqnarray}
\begin{eqnarray}
    e^{ - i\frac{\pi }{2}{{\hat J}_x}} {e^{iN{\phi _0}'}}{\left| {\theta ',\varphi '} \right\rangle _\textrm{SCS}} &=&{\left[ {\cos \left( {\frac{\theta }{2}} \right){e^{-i\varphi /2}}\left|  \uparrow  \right\rangle  + \sin \left( {\frac{\theta }{2}} \right){e^{i\varphi /2}}\left|  \downarrow  \right\rangle } \right]^{ \otimes N}}={\left| {\theta , \varphi } \right\rangle _\textrm{SCS}},
\end{eqnarray}
and
\begin{eqnarray}
    e^{ - i\frac{\pi }{2}{{\hat J}_x}} {e^{iN{\phi _0}'}}{\left| {\theta ',\varphi ' + \pi } \right\rangle _\textrm{SCS}} &=&{\left[ {\sin \left( {\frac{\theta }{2}} \right){e^{i\varphi /2}}\left|  \uparrow  \right\rangle  - \cos \left( {\frac{\theta }{2}} \right){e^{ - i\varphi /2}}\left|  \downarrow  \right\rangle } \right]^{ \otimes N}} = {\left| {\theta-\pi , - \varphi } \right\rangle _\textrm{SCS}},
\end{eqnarray}
we have 
\begin{equation}\label{C_psi1_2}
    \left| {\psi \left( {\theta ,\varphi } \right)} \right\rangle _1= \frac{{{e^{ - i\pi /4}}{{\left| {\theta ,\varphi } \right\rangle }_\textrm{SCS}} + {e^{i\pi /4}}{{\left| {\theta-\pi , - \varphi } \right\rangle }_\textrm{SCS}}}}{{\sqrt 2 }}.
\end{equation}
Therefore, one can obtain $\ket{\Psi}_{f0}$ with $\chi=0$ as follows:
\begin{eqnarray}\label{C_CATf20}
\ket{\Psi}_{f0}&=&\frac{{{e^{ - i\pi \!/\!4}}{{\left| {\theta ,\delta T} \right\rangle }_\textrm{SCS}} \!+\! {e^{i\pi \!/\!4}}{{\left| {\theta-\pi , - \delta T } \right\rangle }_\textrm{SCS}}}}{2} + \frac{{{e^{ - i\pi\! /\!4}}{{\left| {\pi  \!-\! \theta ,\delta T} \right\rangle }_\textrm{SCS}} \!+\! {e^{i\pi \!/\!4}}{{\left( { - 1} \right)}^J}{{\left| {-\theta , -\delta T } \right\rangle }_\textrm{SCS}}}}{2} \nonumber\\
&=&\frac{{{e^{ - i\pi \!/\!4}}{{\left| {\theta , \delta T} \right\rangle }_\textrm{SCS}} \!+\! {e^{i\pi \!/\!4}}{{\left| {\theta-\pi , - \delta T } \right\rangle }_\textrm{SCS}}}}{2} + \frac{{{e^{ - i\pi\! /\!4}}{{\left| {\pi  \!-\! \theta ,\delta T} \right\rangle }_\textrm{SCS}} \!+\! {e^{i\pi \!/\!4}}{{\left( { - 1} \right)}^J}{{\left| {-\theta , -\delta T} \right\rangle }_\textrm{SCS}}}}{2} \nonumber\\
&=&\sum_{m=-J}^{J}A_m^{\theta,\delta T}\ket{J,m}
\end{eqnarray}
with $A_m^{\theta ,\varphi} = \frac{1}{2}\left[ {{e^{ - i\pi /4}}C_m^{\theta ,\varphi}+ {e^{i\pi /4}}{{\left( { - 1} \right)}^{J-m}}C_{ - m}^{\theta ,\varphi}} \right] $ $+\frac{1}{2} \left[ {{e^{ - i\pi /4}}C_{ - m}^{\theta , - \varphi} + {e^{i\pi /4}}{{\left( { - 1} \right)}^{J-m}}C_m^{\theta, - \varphi}} \right]$.
Moreover, according to $C_m^{\theta ,\phi } = C_m^{\theta ,0}{e^{ - im\phi }}$, we have
\begin{eqnarray}\label{C_Am}
A_m^{\theta,\varphi}&=&i^{J-m}\frac{{\left[ {{e^{ - i\pi /4}}{i^{ - (J-m)}}C_m^{\theta ,\varphi } + {e^{i\pi /4}}{i^{J-m}}C_m^{\theta , - \varphi }} \right]+ \left[ {{e^{ - i\pi /4}}{i^{ - (J-M)}}C_{ - m}^{\theta , - \varphi } + {e^{i\pi /4}}{i^{J-m}}C_{ - m}^{\theta ,\varphi }} \right]}}{2}\nonumber\\
&=& i^{J-m}\frac{{\left[ {{e^{ - i\pi /4}}{e^{ - i{J-m}\pi /2}}C_m^{\theta ,0}{e^{ - im\varphi }} + {e^{i\pi /4}}{e^{i(J-m)\pi /2}}C_m^{\theta ,0}{e^{im\varphi }}} \right]  }}{2}\nonumber\\
& &+i^{J-m}\frac{{ \left[ {{e^{ - i\pi /4}}{e^{ - i(J-m)\pi /2}}C_{ - m}^{\theta ,0}{e^{ - im\varphi }} + {e^{i\pi /4}}{e^{i(J-m)\pi /2}}C_{ - m}^{\theta ,0}{e^{im\varphi }}} \right]}}{2}\nonumber\\
&=& i^{J-m}\left( {C_m^{\theta ,0} + C_{ - m}^{\theta ,0}} \right)\cos \left( {\frac{{2m\varphi  + (J-m)\pi + \pi /2}}{2}} \right).
\end{eqnarray}
Hence, Eq.~\eqref{Jz_CATb} and Eq.~\eqref{Jz2_CATb} can be proved:
\begin{eqnarray}
\langle\hat{J}_z\rangle_f(\chi=0) &=& \cos(\delta t_r)\langle\hat{J}_z\rangle_{f0}(\chi=0)-\sin(\delta t_r)\langle\hat{J}_x\rangle_{f0}(\chi=0)\nonumber\\
&=&\cos(\delta t_r)\sum_{m =  - J}^J m \left|A_m\right|^{2}-\frac{\sin(\delta t_r)}{2}\sum_{m=-J}^{J-1}\lambda_m^{+}\left(A_mA_{m+1}^*+A_{-m}A_{-(m+1)}^*\right)\nonumber\\
&=& \cos(\delta t_r)\sum_{m = 1}^J m {\left( {C_m^{\theta ,0} + C_{ - m}^{\theta ,0}} \right)^2}\frac{{\cos \left( {2m\delta T + (J-m)\pi + \pi /2} \right) - \cos \left( { - 2m\delta T  + (J+m)\pi + \pi /2} \right)}}{2}\nonumber\\
& &+\sum_{m'=-J+1/2}^{J-1/2}i(-1)^{J-1/2-m'}\lambda_{m'-1/2}^+(C_{m'-1/2}+C_{-m'+1/2})(C_{m'+1/2}+C_{-m'-1/2})\cos(2m'\delta T)\nonumber\\
&=& \cos(\delta t_r)\left( { - 1} \right)^{J+1}\sum_{m = 1}^J (-1)^m m {\left( {C_m^{\theta ,0} + C_{ - m}^{\theta ,0}} \right)^2}\sin \left( { 2m\delta T } \right),
\end{eqnarray}
and
\begin{eqnarray}
\langle\hat{J}_{z}^2\rangle_f(\chi=0)&=& \cos^2(\delta t_r)\langle\hat{J}_z^2\rangle_{f0}(\chi=0)-\frac{\sin(\delta t_r)}{2}\langle\hat{J}_z\hat{J}_x+\hat{J}_x\hat{J}_z\rangle_{f0}(\chi=0)+\sin^2(\delta t_r)\langle\hat{J}_x\rangle_{f0}(\chi=0)\nonumber\\
&=& \cos^2(\delta t_r)\sum_{m =  - J}^J m^2 \left|A_m\right|^{2}-\frac{\sin(2\delta t_r)}{4}\sum_{m=-J}^{J-1}(2m+1)\lambda_m^+\left(A_{m+1}^*A_m-A_{-(m+1)}^*A_{-m}\right)\nonumber\\
& &+\frac{\sin^2(\delta t_r)}{4}\left[\sum_{m=-J+1}^J\lambda_{m-1}^+\lambda_m^-(|A_m|^2+|A_{-m}|^2)+\sum_{m=-J}^{J-2}\lambda_{m+1}^+\lambda_m^+(A_{m+2}^*A_m+A_{-(m+2)}^*A_{-m})\right]\nonumber\\
&=& \cos^2(\delta t_r)\sum_{m = 1}^J m^2 {\left( {C_m^{\theta ,0} + C_{ - m}^{\theta ,0}} \right)^2}+\frac{\sin^2(\delta t_r)}{4}\left[\sum_{m=-J+1}^J[J(J+1)-m(m-1)]|C_m+C_{-m}|^2\right.\nonumber\\
& &\left.+\sum_{m=-J}^{J-2}\lambda_{m+1}^+\lambda_m^+(C_m+C_{-m})(C_{m+2}+C_{-(m+2)})\cos(2\delta T)\right],
\end{eqnarray}
where $A_m$ is the abbreviation of $A_m^{\theta,\delta T}$ and $A_m^*$ is the conjugate of $A_m$.

In order to obtain Eq.~\eqref{CATo2} for general $\langle\hat{J}_{z}\rangle_f(\chi)$ and $\langle\hat{J}_{z}^2\rangle_f(\chi)$, we give the proof of 
\begin{equation}
    \hat{U}_\chi=e^{-i\frac{\pi}{2}\hat{J}_y^2}e^{-2i\chi t_n\hat{J}_z^2}e^{i\frac{\pi}{2}\hat{J}_y^2}=e^{-2i\chi t_n\hat{J}_z^2},
\end{equation}
which is equal to prove
\begin{equation}\label{C_Jz2}
   e^{ - i\frac{\pi }{2}\hat J_y^2} \hat J_z^2 e^{i\frac{\pi }{2}\hat J_y^2} = \hat J_z^2.
\end{equation}
For convenience, we rotate all coordinate axes around $y$ axis by $\pi/2$, and Eq.~\eqref{C_Jz2} is
%
\begin{equation}\label{C_Jx2}
e^{ - i\frac{\pi }{2}{{\hat J}_x}} e^ { - i\frac{\pi }{2}\hat J_y^2} \hat J_z^2 e^{i\frac{\pi }{2}\hat J_y^2} e^{i\frac{\pi }{2}{{\hat J}_x}} =
   {\left[ e^{ - i\frac{\pi }{2}\hat J_z^2}{{\hat J}_y}e^{i\frac{\pi }{2}\hat J_z^2}\right]^2} 
   =\left[ e^{ - i\frac{\pi }{2}\hat J_z^2} \frac{{{{\hat J}_ + } - {{\hat J}_ - }}}{2i}e^{i\frac{\pi }{2}\hat J_z^2} \right]^2
   =\hat J_y^2.
\end{equation}
%
Here, we define $\hat J_ \pm ^I = e^{ - it\hat J_z^2}{{\hat J}_ \pm }e^{it\hat J_z^2}$, and its deviation about $t$ is 
\begin{equation}
    \frac{{d\hat J_ \pm ^I}}{{dt}} =  - ie^{ - it\hat J_z^2}\left[ {\hat J_z^2,{{\hat J}_ \pm }} \right]e^{it\hat J_z^2} =  - i\hat J_ \pm ^I\left( {1 \pm 2{{\hat J}_z}} \right)
\end{equation}
which have used $\left[ {{{\hat J}_z},{{\hat J}_ \pm }} \right] =  \pm {{\hat J}_ \pm }$.
Therefore, we have $\hat J_ \pm ^I(t) = {{\hat J}_ \pm }{e^{ - it\left( {1 \pm 2{{\hat J}_z}} \right)}}$ and 
\begin{eqnarray}\label{C_Jx}
    e^{ - i\frac{\pi }{2}\hat J_z^2} {{\hat J}_y}e^{i\frac{\pi }{2}\hat J_z^2} &=& \frac{{\hat J_ + ^I\left( {\frac{\pi }{2}} \right) - \hat J_ - ^I\left( {\frac{\pi }{2}} \right)}}{2i}\nonumber\\
    &=&  \frac{{{{\hat J}_ + }{e^{ - i\pi {{\hat J}_z}}} - {{\hat J}_ - }{e^{i\pi {{\hat J}_z}}}}}{-2}.
\end{eqnarray}
According to Eq.~\eqref{C_Jx2} and Eq.~\eqref{C_Jx}, we can prove Eq.~\eqref{C_Jz2} as follows:
%
\begin{eqnarray}
e^{ - i\frac{\pi }{2}\hat J_y^2} \hat J_z^2 e^{i\frac{\pi }{2}\hat J_y^2}&=&e^{i\frac{\pi }{2}{{\hat J}_x}}e^{ - i\frac{\pi }{2}\hat J_z^2}\hat J_y^2 e^{i\frac{\pi }{2}\hat J_z^2}e^{-i\frac{\pi }{2}{{\hat J}_x}} \\\nonumber
&=&e^{i\frac{\pi }{2}{{\hat J}_x}}  \frac{{{{\hat J}_ + }{e^{ - i\pi {{\hat J}_z}}}{{\hat J}_ + }{e^{ - i\pi {{\hat J}_z}}} + {{\hat J}_ - }{e^{i\pi {{\hat J}_z}}}{{\hat J}_ - }{e^{i\pi {{\hat J}_z}}} - {{\hat J}_ + }{e^{ - i\pi {{\hat J}_z}}}{{\hat J}_ - }{e^{i\pi {{\hat J}_z}}} - {{\hat J}_ - }{e^{i\pi {{\hat J}_z}}}{{\hat J}_ + }{e^{ - i\pi {{\hat J}_z}}}}}{4}e^{-i\frac{\pi }{2}{{\hat J}_y}}\\\nonumber
 &=& e^{i\frac{\pi }{2}{{\hat J}_y}}\frac{{-\hat J_ + ^2 - \hat J_ - ^2 + {{\hat J}_ + }{{\hat J}_ - } + {{\hat J}_ - }{{\hat J}_ + }}}{4}e^{-i\frac{\pi }{2}{{\hat J}_y}} \\\nonumber
 &=& e^{i\frac{\pi }{2}{{\hat J}_x}}\hat J_y^2e^{-i\frac{\pi }{2}{{\hat J}_x}}\\\nonumber
 &=& \hat{J}_z^2.
\end{eqnarray}
Therefore, for $\chi\neq0$, we can just replace $A_m$ with $A_m'=A_me^{-i\chi T m^2}$.
Then we can get that
\begin{small}
\begin{eqnarray}
\langle\hat{J}_z\rangle_f(\chi) &=& \cos(\delta t_r)\langle\hat{J}_z\rangle_{f0}(\chi)-\sin(\delta t_r)\langle\hat{J}_x\rangle_{f0}(\chi)\nonumber\\
&=&\cos(\delta t_r)\sum_{m =  - J}^J m \left|A_m\right|^{2}-\frac{\sin(\delta t_r)}{2}\sum_{m=-J}^{J-1}\lambda_m^{+}e^{i\chi T(2m+1)}\left(A_mA_{m+1}^*+A_{-m}A_{-(m+1)}^*\right)\nonumber\\
&=& \cos(\delta t_r)\sum_{m = 1}^J m {\left( {C_m^{\theta ,0} + C_{ - m}^{\theta ,0}} \right)^2}\frac{{\cos \left( {2m\delta T + (J-m)\pi + \pi /2} \right) - \cos \left( { - 2m\delta T  + (J+m)\pi + \pi /2} \right)}}{2}\nonumber\\
& &-\frac{i\sin(\delta t_r)}{2}\sum_{m'=-J+1/2}^{J-1/2}(-1)^{J-1/2-m'}e^{2i\chi Tm'}\lambda_{m'-1/2}^+(C_{m'-1/2}+C_{-m'+1/2})(C_{m'+1/2}+C_{-m'-1/2})\cos(2m'\delta T)\nonumber\\
&=& \cos(\delta t_r)\left( { - 1} \right)^{J+1}\sum_{m = 1}^J (-1)^m m {\left( {C_m^{\theta ,0} + C_{ - m}^{\theta ,0}} \right)^2}\sin \left( { 2m\delta T } \right)\nonumber\\
& &+\sin(\delta t_r)\sum_{m'=1/2}^{J-1/2}(-1)^{J-1/2-m'}\sin(2\chi Tm')\lambda_{m'-1/2}^+(C_{m'-1/2}+C_{-m'+1/2})(C_{m'+1/2}+C_{-m'-1/2})\cos(2m'\delta T),\nonumber\\
\end{eqnarray}
\end{small}
and
\begin{eqnarray}
\langle\hat{J}_{z}^2\rangle_f(\chi)&=& \cos^2(\delta t_r)\langle\hat{J}_z^2\rangle_{f0}(\chi)-\frac{\sin(\delta t_r)}{2}\langle\hat{J}_z\hat{J}_x+\hat{J}_x\hat{J}_z\rangle_{f0}(\chi=0)+\sin^2(\delta t_r)\langle\hat{J}_x\rangle_{f0}(\chi)\nonumber\\
&=& \cos^2(\delta t_r)\sum_{m =  - J}^J m^2 \left|A_m\right|^{2}-\frac{\sin(2\delta t_r)}{4}\sum_{m=-J}^{J-1}(2m+1)\lambda_m^+\left(A_{m+1}^*A_m-A_{-(m+1)}^*A_{-m}\right)e^{i\chi T(2m+1)}\nonumber\\
& &+\frac{\sin^2(\delta t_r)}{4}\left[\sum_{m=-J+1}^J\lambda_{m-1}^+\lambda_m^-(|A_m|^2+|A_{-m}|^2)+\sum_{m=-J}^{J-2}\lambda_{m+1}^+\lambda_m^+(A_{m+2}^*A_m+A_{-(m+2)}^*A_{-m})e^{4i\chi T(m+1)}\right]\nonumber\\
&=& \cos^2(\delta t_r)\sum_{m = 1}^J m^2 {\left( {C_m^{\theta ,0} + C_{ - m}^{\theta ,0}} \right)^2}\nonumber\\
& &-\frac{\sin(2\delta t_r)}{4}\sum_{m'=1/2}^{J-1/2}\sin(2\chi Tm')(2m')\lambda_{m'-1/2}^+(C_{m'-1/2}+C_{-m'+1/2})(C_{m'+1/2}+C_{-m'-1/2})\sin(\delta T)\\\nonumber
& &+\frac{\sin^2(\delta t_r)}{4}\left[\sum_{m=-J+1}^J[J(J+1)-m(m-1)]|C_m+C_{-m}|^2\right.\nonumber\\
& &\left.+\sum_{m=-J}^{J-2}\cos[4\chi T(m+1)]\lambda_{m+1}^+\lambda_m^+(C_m+C_{-m})(C_{m+2}+C_{-(m+2)})\cos(2\delta T)\right].
\end{eqnarray}
\end{widetext}
Our analytical results are consistent with numerical result, as shown in {Fig.\ref{C1}}.
\begin{figure}[!htp]
\includegraphics[width=\columnwidth]{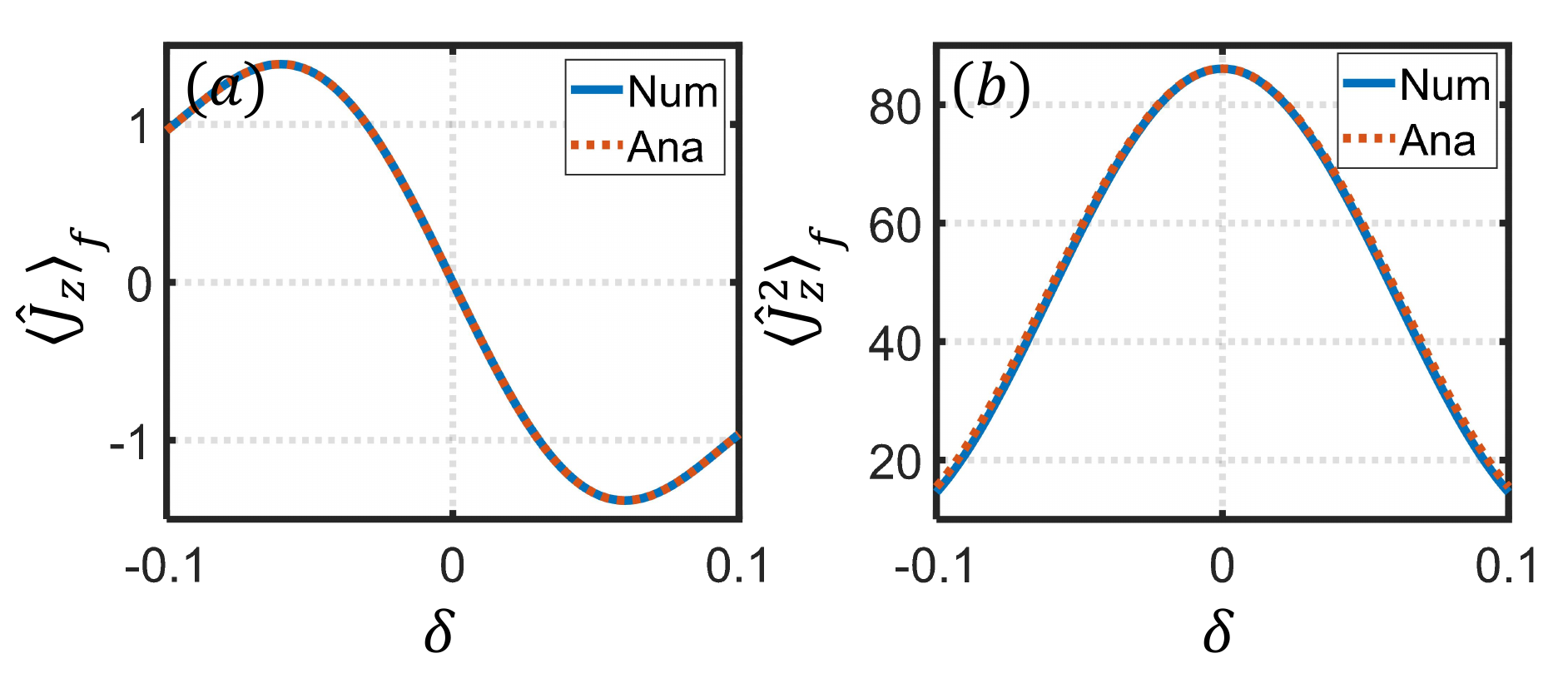}
\caption{\label{C1}(color online).
The $\langle\hat{J}_z\rangle_f$ (a) and $\langle\hat{J}_z^2\rangle_f$ (b) versus the detuning $\delta$ with numerical (blue solid line) and analytical (red dotted line) method for $\ket{\textrm{CAT}(\theta=\pi/8)}$.
Here, $T=1$, $\chi=0.02\pi$, $\chi_r=0.04\pi$, $t_r=\pi/(2\chi_r)$ and particle number $N=20$.
}
\end{figure}
In particular, for GHZ state with $C_m=\left\{\begin{array}{ccc}1/\sqrt{2},&m=\pm J\\0,&\textrm{others}\end{array}\right.$, we have
$\langle\hat{J}_z\rangle_f(\chi)=\langle\hat{J}_z\rangle_f(\chi=0)$ and $\langle\hat{J}_z^2\rangle_f(\chi)=\langle\hat{J}_z^2\rangle_f(\chi=0)$.
%
%
\setcounter{equation}{0}
\setcounter{figure}{0}
\renewcommand{\theequation}{D\arabic{equation}}
\renewcommand{\thefigure}{D{\arabic{figure}}}
\section*{APPENDIX D: Frequency locking with many-body quantum lock-in amplifier\label{SecSM1}}
In our protocol, we consider that the frequency of ac signal is known and the amplitude is to be measured.
Here, we can use the many-body quantum lock-in amplifier to obtain the frequency of ac field~\cite{MZPRX2021}.
In this section, we introduce the frequency locking progress with many-body quantum lock-in amplifier.

\begin{figure}[!htp]
\includegraphics[width=\columnwidth]{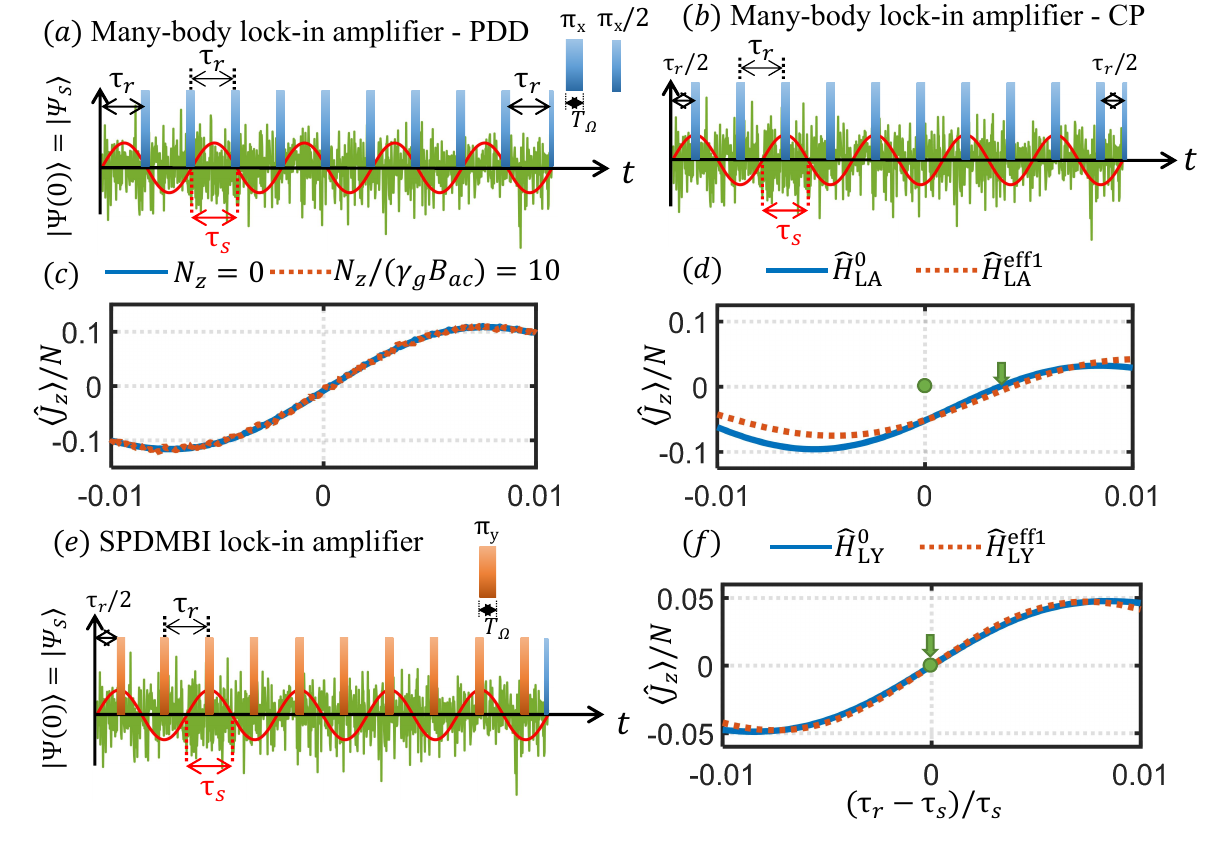}
\caption{\label{D1}(color online).
(a)-(b) The many-body quantum lock-in amplifier scheme for PDD (a) and CP (b) sequences.
(c) The robustness against white noise $N_o(t)=N_zG(t)$.
(d) The many-body lock-in spectroscopy about $(\tau_r-\tau_s)/\tau_s$ calculated by $\hat{H}_L$ (blue solid line) and $\hat{H}_{L'}^\textrm{eff1}$ (red dotted line).
The zero point shifts when considering the pulse length $T_{\Omega}$ and particle-particle interaction $\chi$ (green arrow and green circle).
(e) The SPDMBI lock-in amplifier scheme.
(f) The precision spectroscopy about $(\tau_r-\tau_s)/\tau_s$ calculated by $\hat{H}_\textrm{LY}^{0}$ (blue solid line) and $\hat{H}_\textrm{LY}^\textrm{eff1}$ (red dotted line).
The zero point still hold when considering the pulse length $T_{\Omega}$ and particle-particle interaction $\chi$ (green arrow and green circle).
Here, $\gamma_g=1$, $B_\textrm{ac}=1$, $N=20$, $\omega=200\pi$, $T_{\Omega}=0.2\pi/\omega_s$ and pulse number $L=100$.
}
\end{figure}

The many-body lock-in amplifier can measure ac signal via Carr-Purcell-Meiboom-Gill (CPMG) sequence or period dynamical decoupling (PDD) sequence with high signal noise ratio (SNR). 
The coupling between the probe system and the external field can be described by the 
Hamiltonian ${\hat{H}}_\textrm{int}(t)=M(t) \hat{J}_{z}$ with $M(t)=\emph{S}(t)+\emph{N}_{o}(t)$ consists of the target signal $\emph{S}(t)=A\sin(\omega_s t)$ and the stochastic noise $\emph{N}_{o}(t)$.
To implement the quantum lock-in measurement, one can mix the system with an induced modulation signal $\hat{H}_\textrm{ref}=\Omega_\pi(t)\hat{J}_x$designed as a sequence of periodic $\pi$ pulses along $x$ axis that does not commute with ${\hat{H}}_\textrm{int}(t)$~\cite{SKNature2011,MZPRX2021,SCCP2024}.
In experiments, the $\pi$ pulse sequences can usually be approximated as square waves with equidistant spacing $\tau_r$ and pulse length $T_{\Omega}$, i.e.,
$\Omega_\pi(t)=\pi/T_\Omega$ when $|t-(l-\lambda)\tau_r|\leq T_{\Omega}/2$, otherwise $\Omega_\pi(t)=0$,
%
where $l=1,2,\cdots,L$ with $L$ the pulse number, and $(1-\lambda)\tau_r$ is the time when the first $\pi$ pulse is applied.
After $L\tau_r$ evolution time, a $\pi/2$ pulse along $x$ direction is imposed for readout, which also satisfies our SPDMBI.
We select the PDD [see {Fig.\ref{D1}(a)}] and CPMG [see {Fig.\ref{D1}(b)}] sequences with $\lambda=0$ and $\lambda=1/2$ respectively~\cite{SKNature2011,CLDRMP2017,MZPRX2021,SCCP2024,MZQF2024}.
The whole system Hamiltonian reads 
\begin{equation}\label{HL}
\hat{H}_\textrm{LA}^{0}(t)=\chi\hat{J}_z^2+M(t)\hat{J}_z+\Omega_\pi(t)\hat{J}_x.
\end{equation}
For convenience, one can move into the interaction picture respect to $\Omega_\pi(t)\hat{J}_{x}$, and the Hamiltonian becomes
\begin{equation}\label{HIlock}
\hat{H}_{LA}(t)=\chi\hat{J}_\alpha^2(t)+M(t)\hat{J}_\alpha(t),
\end{equation}
where $\hat{J}_\alpha(t)=\cos[\alpha(t)]\hat{J}_z+\sin[\alpha(t)]\hat{J}_y$ and $\alpha(t)=\int_0^t\Omega_{\pi}(t')dt'$.

In the limit of $T_\Omega \to 0$, the $\pi$ pulses are ideal and are described by $\Omega_\pi(t)=\pi\sum_{l=1}^{L}\delta_D(t-(l-\lambda)\tau_r)$, where $\delta_D(t)$ is the Dirac function.
One can obtain $\sin[\alpha(t)]=0$ and 
\begin{eqnarray}\label{ht}
\cos[\alpha(t,\lambda)]
=\left\{
\begin{array}{rcl}
1  &&{2k\tau_\textrm{r}\leq t+\lambda\tau_\textrm{r} \leq(2k+1)\tau_\textrm{r}},\nonumber\\
-1  &&{(2k+1)\tau_\textrm{r}\leq t+\lambda\tau_\textrm{r} \leq 2(k+1)\tau_\textrm{r}
	}
\end{array} \right.
\end{eqnarray}
which is a square wave function.
According to Fourier series expansion $\cos[\alpha(t,\lambda)]=\sum_{k=1,\textrm{odd}}\frac{4}{k\pi}\sin(k\omega_rt+k\lambda\pi)$, one can simplify the Hamiltonian of Eq.~\eqref{HIlock} as
\begin{eqnarray}\label{HLeff0}
\hat{H}_{L'}\approx\chi\hat{J}_z^2+\frac{2\gamma_g B_{ac}}{\pi}\sin[(\omega_s-\omega_r)t+\lambda\pi]
\end{eqnarray}
ignoring the high frequency terms with $\omega_r=\pi/\tau_r$~\cite{SCCP2024,MZQF2024}.
For CPMG sequence with $\lambda=1/2$, we have
\begin{eqnarray}\label{HLCP}
\hat{H}_{L'}^\textrm{eff,CP}&=&\frac{1}{L\tau_r}\int_{0}^{L\tau_r}\{\chi\hat{J}_z^2+\frac{2\gamma_g B_{ac}}{\pi}\sin[(\omega_s-\omega_r)t]\}dt\nonumber\\
&=&\chi\hat{J}_z^2+\frac{2\gamma_g B_\textrm{ac}}{L\pi}\frac{\sin^2(L\omega_s\delta_\tau/2)}{\sin(\omega_s\delta_\tau/2)}\hat{J}_z
\end{eqnarray}
with $\delta_\tau=(\tau_r-\tau_s)$ and $\tau=\pi/\omega$.
Similarly, for PDD sequence with $\lambda=0$, we have
\begin{eqnarray}\label{HLPDD}
\hat{H}_{L'}^\textrm{eff,PDD}&=&\frac{1}{L\tau_r}\int_{0}^{L\tau_r}\{\chi\hat{J}_z^2+\frac{2\gamma_g B_{ac}}{\pi}\cos[(\omega_s-\omega_r)t]\}dt\nonumber\\
&=&\chi\hat{J}_z^2+\frac{2\gamma_g B_\textrm{ac}}{L\pi}\frac{\sin(L\omega_s\delta_\tau)}{\omega_s\delta_\tau}\hat{J}_z.
\end{eqnarray}
There is no doubt that the input states satisfy $C_m(0)=C_{-m}(0)$, and in the interrogating process, the effective Hamiltonian $\hat{H}_{L,\tau}^\textrm{eff}$ also satisfies Eq.~\eqref{HwI}.
One can choose $\pi/2$ pulse along $x$ axis $\hat{U}=e^{-i\pi/2\hat{J}_x}$ reading for $\ket{\Psi_0}=\ket{\Psi_S}$ so that the readout Hamiltonian $\Omega\hat{J}_x$ also satisfies our SPDMBI.
Moreover, the many-body lock-in amplifier within SPDMBI is also robust against white noise $N_zG(t)\hat{J}_z$ [see {Fig.\ref{D1}(c)}] where $G(t)$ is a set of Gaussian random distributions with a variance of $1$ and a mean of $0$. 
In experiments, however, there will inevitably be some errors such as the pulse length $T_{\Omega}\neq0$ limited by the Rabi frequency, the particle-particle interaction $\chi$, inaccurate reading time control, pulse length $T_{\Omega}$, detuning $B_\textrm{dc}\hat{J}_z$ (equivalent to dc noise or dc bias)~\cite{WZPRL2019}, white noise, dephasing and particle losses.
For $T_{\Omega}\neq0$, we have $\cos[\alpha(t)]=\sum_{k=1}^{+\infty}a_k\cos(k\omega_r t)$ and $\sin[\alpha(t)]=\sum_{k=1}^{+\infty}b_k\sin(k\omega_r t)$ with $a_k=\frac{2[1-(-1)^k]}{k\pi}\left[\sin\left(\frac{k\pi}{2}-\frac{k\pi T_{\Omega}}{2\tau_r}\right)+\frac{\cos\left(\frac{k+1}{2}\pi+\frac{k\pi T_{\Omega}}{2\tau_r}\right)}{1-\left(\frac{\tau_r}{kT_{\Omega}}\right)^2}\right]$ and $b_k=kT_{\Omega}a_k/\tau_r$.
Therefore, one can obtain the effective Hamiltonian of Eq.~\eqref{HIlock} under the conditions of $|\omega-\omega_{r}|\ll\omega$ and evolution time $T\gg2\pi/\omega$ can be written as
\begin{eqnarray}\label{HITO}
\hat{H}_I(t) &\approx& \chi(a_s\hat{J}_z^2 + b_s\hat{J}_y^2)+ \gamma_g B_\textrm{ac} a_1/2\sin[(\omega_s-\omega_r) t]\hat{J}_z \nonumber\\
& &+\gamma_g B_\textrm{ac} b_1/2\cos[(\omega_s-\omega_r) t]\hat{J}_y
\end{eqnarray}
with $a_s=\sum_ka_k^2$ and $b_s=\sum_kb_k^2$.
Similarly, for CPMG sequence with $\lambda=1/2$, we have
\begin{widetext}
\begin{eqnarray}\label{HLeff1}
\hat{H}_{LA}^\textrm{eff1}(\delta_{\tau},T_{\Omega})&=&\frac{1}{L\tau_r}\int_{0}^{L\tau_r}\{\chi(a_s\hat{J}_z^2 + b_s\hat{J}_y^2) + B_\textrm{ac} a_1/2\sin[(\omega_s-\omega_r) t]\hat{J}_z \nonumber+B_\textrm{ac} b_1/2\cos[(\omega_s-\omega_r) t]\hat{J}_y\}dt\nonumber\\
&=& \frac{\chi}{2}(a_s\hat{J}_z^2 + b_s\hat{J}_y^2) +\frac{a_1 \gamma_g B_\textrm{ac}}{2} \frac{\sin^2(L\omega_s\delta_\tau/2)}{L\omega_s\delta_\tau/2}\hat{J}_z+ \frac{b_1 \gamma_g B_\textrm{ac}}{2} \frac{\sin(L\omega_s\delta_\tau)}{L\omega_s\delta_\tau}\hat{J}_y.
\end{eqnarray}
\end{widetext}
In order to further confirm Eq.~\eqref{HLeff1}, we give a comparison between the spectra calculated by Eq.~\eqref{HL} and Eq.~\eqref{HLeff1} when $\chi=0.001\omega_s$ and $T_{\Omega}=0.2\pi/\omega_s$ for CPMG sequence, as shown in {Fig.~\ref{D1}~(d)}.
The last term of Eq.~\eqref{HLeff1} obviously does not satisfy Eq.~\eqref{HwI}, so the spectral symmetry will be shifted, which can be even amplified by $\chi$.

In order to suppress the shift, we choose a series of equally spaced $\pi$ pulses in the $y$ direction as a reference signal, that is, the reference Hamiltonian is $\hat{H}_\textrm{ref}=\Omega_\pi(t)\hat{J}_y$, as shown in {Fig.~\ref{D1}~(e)}.
Similarly, in the interaction picture with respect to $\hat{H}_\textrm{ref}=\Omega_\pi(t)\hat{J}_{y}$, the original Hamiltonian 
\begin{equation}\label{HL1}
    \hat{H}_\textrm{LY}^0(t)=\chi\hat{J}_z^2+M(t)\hat{J}_z+\Omega_\pi(t)\hat J_y
\end{equation}
transforms into
\begin{equation}\label{HIlock}
 \hat{H}_\textrm{LY}(t)=\chi\hat{J}_\alpha'^2(t)+M(t)\hat{J}_\alpha'(t),
\end{equation}
where $\hat{J}_\alpha(t)=\cos[\alpha(t)]\hat{J}_z-\sin[\alpha(t)]\hat{J}_x$ and $\alpha(t)=\int_0^t\Omega_{\pi}(t')dt'$.
Thus the effective Hamiltonian for Eq.~\eqref{HIlock} becomes
\begin{eqnarray}\label{Heff11}
\small \hat{H}_\textrm{LY}^{\textrm{eff1}} &=& \frac{\chi}{2}(a_s\hat{J}_z^2 + b_s\hat{J}_x^2)  +\frac{a_1 \gamma_g B_\textrm{ac}}{2} \frac{\sin^2(L\omega_s\delta_\tau/2)}{L\omega_s\delta_\tau/2} \hat{J}_z\nonumber \\
&& - \frac{b_1 \gamma_g B_\textrm{ac}}{2} \frac{\sin(L\omega_s\delta_\tau)}{L\omega_s\delta_\tau}\hat{J}_x,
\end{eqnarray}
which satisfies Eq.~\eqref{H_sym}.
We also numerically compare the spectra calculated by Eq.~\eqref{HL1} and Eq.~\eqref{HLeff1} when $\chi=0.001\omega_s$ and $T_{\Omega}=0.2\pi/\omega_s$ for CPMG sequence, as shown in {Fig.~\ref{D1}~(f)}.

\begin{figure*}[!htp]
\includegraphics[width=1.6\columnwidth]{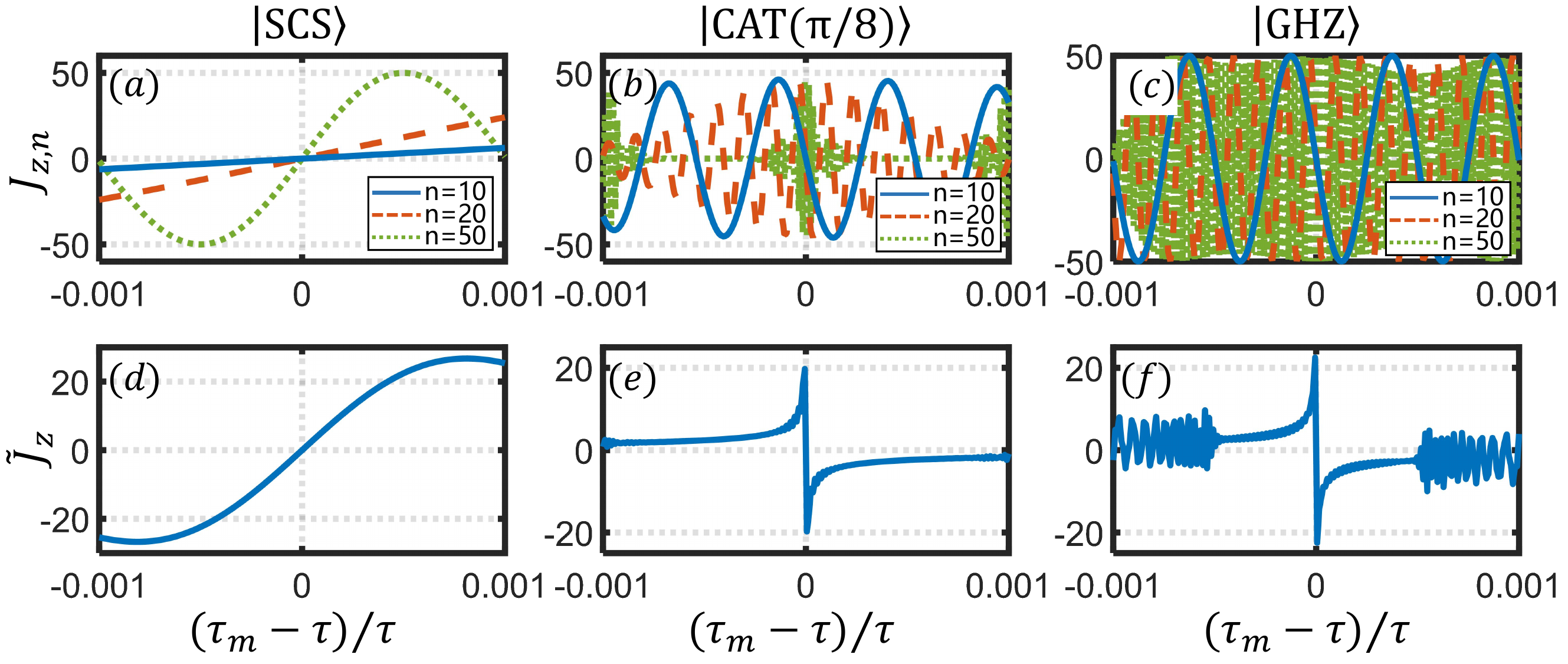}
\caption{\label{D2}(color online).
The many-body lock-in amplifier for different input states: (a)(d) SCS , (b)(e) spin cat state, and (c)(f) GHZ state.
(a)-(c): The measurement signal $J_{z,n}$ versus the modulation $(\tau_m-\tau)/\tau$ for (a) SCS, (b) spin cat state, and (c) GHZ state with $n=10$ (blue solid line), $n=20$ (red dashed line) and $n=50$ (green dotted line).
(d)-(f): The time-average measurement signal $\tilde{J}=\sum_{n=1}^{50}J_{z,n}$ versus the modulation $(\tau_m-\tau)/\tau$ for (d) SCS, (e) spin cat state, and (f) GHZ state.
Here, particle number $N=100$.
}
\end{figure*}

Moreover, the symmetric entangled state can also improve the measurement precision.
Here, we choose SPDMBI lock-in protocol to demonstrate the results for different symmetric input states.
In the interaction picture with respecting to $\hat{H}_\textrm{mix}'=\Omega_\pi(t)\hat{J}_y$, the time-evolution obeys
\begin{eqnarray}\label{IE}
i\frac{\partial{\ket{\Psi(t)}_\textrm{I}}}{\partial{t}}
=B(t)\left[\cos(\alpha_\textrm{L}(t))\hat{J}_\textrm{z}-\sin(\alpha_\textrm{L}(t))\hat{J}_\textrm{x}\right]{\ket{\Psi(t)}_\textrm{I}}\nonumber\\
\end{eqnarray}
with $\alpha_\textrm{L}(t)=\int^{t}_0 \Omega_\textrm{L}(t')dt'=\lfloor\frac{t}{\tau_m}+\frac{1}{2}\rfloor\pi$ with $\lfloor x \rfloor$ denoting the maximum integer not greater than $x$ when the pulse length $T_{\Omega}\to 0$.
Therefore, we have $\sin\left(\alpha_\textrm{L}(t)\right)=0$ and $\hat{H}_I=\gamma_g B(t)\cos\left(\int_{0}^{t}\Omega_\textrm{L}(t)dt\right)\hat{J}_z$.
Initializing each quantum system into the state $\ket{\Psi}_\textrm{in}$, in the interaction picture, the instantaneous state at time $t_n=2n\tau_m$ becomes
\begin{eqnarray}\label{FI}
\ket{\Psi(t_n)}_\textrm{I}=e^{-i\phi_{\textrm{L},n}\hat{J}_\textrm{z}}\ket{\Psi}_\textrm{in}
\end{eqnarray}
with
\begin{eqnarray}\label{phi_L}
\phi_{\textrm{L},n}&=&\int_0^{t_n}\frac{1}{2}B(t')\cos[\alpha(t')]dt'\nonumber\\
&=&\frac{2\gamma_g B_\textrm{ac}}{\omega}\sin\left[\frac{n\omega\cdot(\tau_m-\tau)}{2}\right]\nonumber\\
&\times&\left[1+\sin\left(\frac{\omega\cdot(\tau_m-\tau)}{2}\right)\right]\frac{\sin[n\omega\cdot(\tau_m-\tau)/2]}{\sin[\omega\cdot(\tau_m-\tau)/2]}\nonumber\\
\end{eqnarray}
and $n=1,2,3,\cdots$ to suppress DC noise~\cite{CLDRMP2017}.
In particular, $\phi_{\textrm{L},n}\approx\frac{2\gamma_g B_\textrm{ac}}{\omega}\sin\left[\frac{n\omega\cdot(\tau_m-\tau)}{2}\right] \frac{\sin[n\omega\cdot(\tau_m-\tau)/2]}{\sin[\omega\cdot(\tau_m-\tau)/2]}$ is antisymmetric with respect to $(\tau_m-\tau)=0$ when $\left|\frac{\omega\cdot(\tau_m-\tau)}{2}\right| \ll 1$.

For the input state $\ket{\Psi}_\textrm{in}$, we observe $\hat{J}_z$ with an expected value of $_{\textrm{f},n}\bra{\Psi}\hat{J}_z\ket{\Psi}_{\textrm{f},n}$ and $\ket{\Psi}_{\textrm{f},n}=\hat{U}e^{-i\phi\hat{J}_z}\ket{\Psi}_\textrm{in}$.
For example, initializing the quantum system into the $\ket{\textrm{SCS}}_x$, we have $\langle\hat{J}_z\rangle_{\textrm{f},n}=\frac{N}{2}\sin(\phi_{\textrm{L},n})$ with $\hat{U}=e^{-i\frac{\pi}{2}\hat{J}_x}$, as shown in {Fig.~\ref{D2}~(a)}.
For $\ket{\textrm{CAT}(\theta)}$, we have $\langle\hat{J}_z\rangle_{\textrm{f},n}=-\sum_{m=-J}^{J}m\left|C_m^J(\theta)\right|^2\sin(2m\phi_{\textrm{L},n})$ with $\hat{U}=e^{-i\frac{\pi}{2}\hat{J}_x^2}$, see {Fig.~\ref{D2}~(b)}.
For $\ket{\textrm{GHZ}}$, we have $J_{z,n}=\langle\hat{J}_z\rangle_{\textrm{f},n}=-\frac{N}{2}\sin(N\phi_{\textrm{L},n})$ with $\hat{U}=e^{-i\frac{\pi}{2}\hat{J}_x^2}$, see {Fig.~\ref{D2}~(c)}.
We find that the time-averaged signal $\tilde{J}_z$ is also antisymmetric about $(\tau_m-\tau)=0$ near the lock-in point, as shown in {Fig.~\ref{D2}~(d)~(e) and (f)}.
Therefore, we can obtain $\tau$ according to the symmetry, and design a periodic dynamical decoupling (PDD) with the pulse interval $\tau_m=\tau$ for the maxmimum accumulation of ac field.

\setcounter{equation}{0}
\setcounter{figure}{0}
\renewcommand{\theequation}{E\arabic{equation}}
\renewcommand{\thefigure}{E{\arabic{figure}}}
\section*{APPENDIX E: Analytical proof for $\ket{\Psi(t_n)}$ in the ac field sensing\label{SecSM5}}
In this section, we give the proof of Eq.~\eqref{psi_out} in the main text.
The Schr\"{o}dinger equation for an ensemble of two-level bosonic particles evolving under the Hamiltonian $\hat{H}=\gamma_g B(t)\hat{J}_z+\Omega(t)\hat{J}_x$ can be written as ($\hbar=1$)
\begin{eqnarray}\label{A_SE}
i\frac{\partial{\ket{\Psi(t)}_S}}{\partial t}
=\left[\gamma_g B(t)\hat{J}_{z}+\Omega({t})\hat{J}_{x}\right] \ket{\Psi(t)}_S
\end{eqnarray}
In the first accumulation process, we have $\Omega(t)=\pi\sum_{j=1}^n\delta(t-j\tau_m)$ for $t<t_n$ with $\omega_m$ is the repetition frequency of reference pulse sequence.
For convenience, we move into the interaction picture with respect to $\Omega(t)\hat{J}_x$, the state evolution for $\ket{\Psi(t)}_I=e^{i\int_{0}^{t}\Omega(t)dt\hat{J}_x}\ket{\Psi(t)}_S$ can be expressed as 
\begin{equation}\label{A_IE}
i\frac{\partial{\ket{\Psi(t)}_I}}{\partial t}=\hat{H}_I\ket{\Psi(t)}_I
\end{equation}
with 
\begin{eqnarray}\label{A_HI}
\hat{H}_I&\!=\!&e^{i\int_{0}^{t}\Omega(t)dt \hat{J}_x} \left(\gamma_g B(t) \hat{J}_z\right) e^{-i\int_{0}^{t}\Omega(t)dt\hat{J}_x}\nonumber\\
&\!=\!&\gamma_g B(t)\!\left[\cos\!\left(\!\int_{0}^{t}\Omega(t)dt\!\right)\hat{J}_z+\sin\!\left(\!\int_{0}^{t}\Omega(t)dt\!\right)\hat{J}_y\right].\nonumber\\
\end{eqnarray}
We have $\int_{0}^{t}\Omega(t)dt=\lfloor\frac{\omega_m t}{\pi}\rfloor\pi$ with $\lfloor\frac{\omega_m t}{\pi}\rfloor$ is the pulse number before time $t$.
Therefore, we have $\sin\left(\int_{0}^{t}\Omega(t)dt\right)=0$ and $\hat{H}_I=\gamma_g B(t)\cos\left(\int_{0}^{t}\Omega(t)dt\right)\hat{J}_z$.
In the case of $\omega_m=\omega$, neglecting the fast oscillating term, we have $\hat{H}_\textrm{eff}^\textrm{ac}=\frac{2\gamma_g B_\textrm{ac}}{\pi}\hat{J}_z\approx\hat{H}_I$~\cite{SKNature2011,MZPRApplied2020}.
In the second accumulation process, we have $\Omega(t)=0$ and $\int_{0}^{t}\Omega(t)dt=(2n-1)\pi$ for $t>t_n$.
Hence according to Eq.~\eqref{A_HI}, we have $\hat{H}_\textrm{eff}^\textrm{dc}=-\gamma_g B_\textrm{dc}\hat{J}_z=\hat{H}_I$.
Then, we can obtain the output state after the interrogation stage in the interaction picture is 
\begin{eqnarray}\label{psi_out_I}
\ket{\Psi}_\textrm{out,n}^I&=&e^{-i\hat{H}_\textrm{eff}^\textrm{dc}t_{n}}e^{-i\hat{H}_\textrm{eff}^\textrm{ac}t_{n}}\ket{\Psi}_\textrm{in}\nonumber\\
& &\times e^{i\pi\hat{J}_x}e^{-in\phi_\textrm{ac}\hat{J}_z}e^{-i\pi\hat{J}_x}\nonumber\\
& &\times e^{i\pi\hat{J}_x}\ket{\Psi}_\textrm{in}\nonumber\\
&=&e^{in\phi\hat{J}_z}\ket{\Psi}_\textrm{in},
\end{eqnarray}
Further, the output state after the interrogation stage in the Schr\"{o}dinger picture in Eq.~\eqref{psi_out} can be given as
\begin{eqnarray}\label{A_psi_out_S}
    \ket{\Psi(t_n)}|_{\chi=0}&=&e^{-i\int_{0}^{t_n}\Omega(t)dt\hat{J}_x}\ket{\Psi}_\textrm{out,n}^I\nonumber\\
    &=&e^{-i(2n-1)\pi\hat{J}_x}e^{in\phi\hat{J}_z}\ket{\Psi}_\textrm{in}\nonumber\\
    &=&e^{i\pi\hat{J}_x}e^{in\phi\hat{J}_z}e^{-i\pi\hat{J}_x}e^{i\pi\hat{J}_x}\ket{\Psi}_\textrm{in}\nonumber\\
    &=&e^{-in\phi\hat{J}_z}e^{i\pi\hat{J}_x}\ket{\Psi}_\textrm{in}
\end{eqnarray}
which have used $e^{-i2n\pi\hat{J}_x}$ is an identity matrix.
When taking into account the interparticle interaction, we have 
\begin{eqnarray}\label{A_psi_out_S_chi}
    \ket{\Psi(t_n)}=e^{-2i\chi t_n\hat{J}_z^2}e^{-in\phi\hat{J}_z}e^{i\pi\hat{J}_x}\ket{\Psi}_{\textrm{in}}.
\end{eqnarray}
\setcounter{equation}{0}
\setcounter{figure}{0}
\renewcommand{\theequation}{F\arabic{equation}}
\renewcommand{\thefigure}{F{\arabic{figure}}}
\section*{APPENDIX F: Influences of imperfect $\pi/2$ pulse for preparing the initial state\label{SecSM6}}
\begin{figure*}[!htp]
\includegraphics[width=\textwidth]{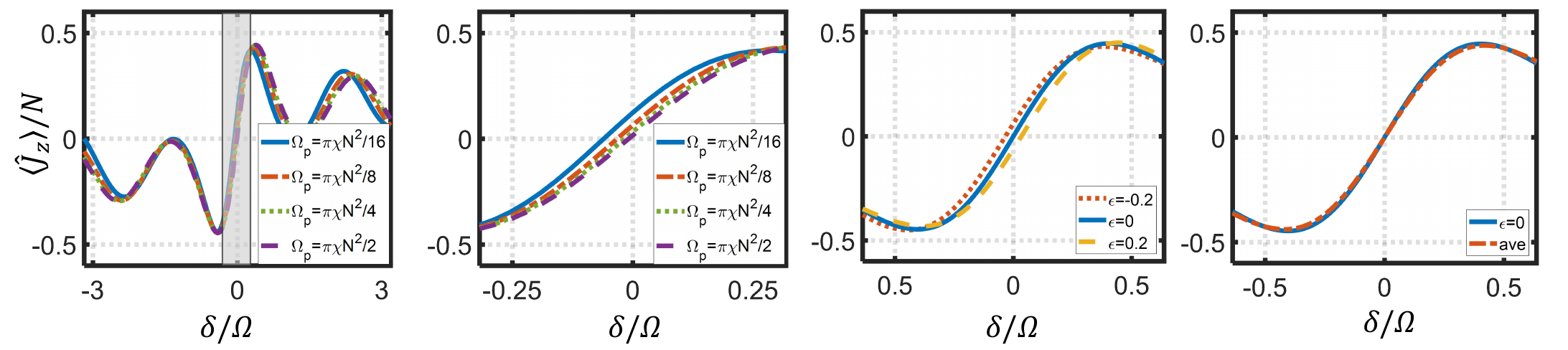}
\caption{\label{F2}(color online).
The final scaled half population difference $\langle\hat{J}_z\rangle$ versus detuning $\delta$ with different imperfect $\pi/2$ pulses.
(a)(b) Imperfect $\pi/2$ pulses are chosen with Rabi frequency $\Omega_p=\pi N^2\chi/16$ (blue solid line), $\Omega_p=\pi N^2\chi/8$ (red dashed-dotted line), $\Omega_p=\pi N^2\chi/4$ (green dotted line) and $\Omega_p=\pi N^2\chi/2$ (purple dashed line) with pulse duration $t_p=\pi/(2\Omega_p)$.
Here, $N=20$, $\Omega=2\pi$, $\chi=0.05\Omega$ and $t_R=0.5$.
(c) The deviation to a perfect $\pi/2$ for $\epsilon=-0.2$ (red dotted line), $\epsilon=0$ (blue solid line) and $\epsilon=0.2$ (yellow dashed line).
(d) The comparison of $\langle\hat{J}_z(\epsilon=0)\rangle_\delta$ (blue solid line) and $(\langle\hat{J}_z(\epsilon=-0.2)\rangle_\delta+\langle\hat{J}_z(\epsilon=0.2)\rangle_\delta)/2$ (red dotted line).
}
\end{figure*}
In our protocol, we need a symmetric initial state, for example, the spin coherent state $\ket{\Psi(0)}=\ket{\textrm{SCS}}_x=e^{-i\pi/2\hat{J}_y}\ket{J,J}$.
Of course, in the $\Lambda$ configuration systems, the $\ket{\Psi_S}$ can also be easily prepared by coherent population trapping~\cite{RFNPJ2021,ZAWD2017,EEMOS2010,SCCP2024}.
In this section, we consider the first preparation method by a $\pi/2$ pulse along $y$ axis applied on the spin-down state $\ket{J,J}$.
Then we will discuss the influence of imperfect $\pi/2$ pulse for preparing the initial state to our SPDMBI protocol.

There are two dominant challenges for preparing the initial state that should be taken into account.
The first challenge is the interparticle interaction $\chi\hat{J}_z^2$ and the detuning $\delta\hat{J}_z$ can affect the $\pi/2$ pulse.
The Hamiltonian for generating the $\pi/2$ pulse can be written as
\begin{equation}\label{Hp}
\hat{H}_p=\chi\hat{J}_z^2+\delta\hat{J}_z+\Omega_p\hat{J}_y
\end{equation}
with the Rabi frequency $\Omega_p$ during the pulse.
Hence the prepared state after the pulse can be expressed as
\begin{equation}\label{psi_p}
\ket{\psi}_p=e^{-i\chi t_p\hat{J}_z^2-i\delta t_p\hat{J}_z-i\Omega_p t_p\hat{J}_y}\ket{J,J},
\end{equation}
where $t_p$ is the duration of the pulse.
Here, when $t_p=\pi/(2\Omega_p)$ is sufficiently small, we have
\begin{eqnarray}\label{psi_p1}
\ket{\psi}_p &\approx& e^{-i\pi\chi/(2\Omega_p)\hat{J}_z^2}e^{-i\pi\delta/(2\Omega_p) t_p\hat{J}_z}e^{-i\pi/2\hat{J}_y}\ket{J,J}\nonumber\\
&\approx& e^{-i\pi\chi/(2\Omega_p)\hat{J}_z^2}e^{-i\pi\delta/(2\Omega_p)\hat{J}_z}\ket{\Psi_S}.
\end{eqnarray}
Therefore, the condition for realizing the $\pi/2$ pulse is $\pi\chi/(2\Omega_p)(N/2)^2+\pi\delta/(2\Omega_p)(N/2)\ll 1$, i.e., the condition for realizing
the $\pi/2$ pulse is~\cite{JHarXiv2022}
\begin{equation}\label{Omega_p}
\Omega_p\gg\pi\chi N^2/8+\pi\delta N/4.
\end{equation}
In Fig.~\ref{F2}~(a) and (b), we show the final signal $\langle\hat{J}_z\rangle_{\delta}$ versus $\delta$ detuning using the input state prepared by $\pi/2$ pulses of different $\Omega_p$ for Rabi spectroscopy.

The second challenge is the precision time control of $t_p=\pi/(2\Omega_p)$.
Even with large Rabi frequency $\Omega_p$, if the time duration $t_p$ is not precisely controlled to $t_p=\pi/(2\Omega_p)$, there still be some imperfections.
Imperfect $\pi/2$ pulse due to the imprecise control of pulse duration can be characterized by
\begin{equation}\label{t_p}
    e^{-i\frac{\pi}{2}(1+\epsilon)\hat{J}_y}
\end{equation}
where $\epsilon$ is the relative deviation from $\pi/2$ pulse.
The prepared state after the pulse can be expressed as
\begin{eqnarray}\label{psi_tp}
    \ket{\Psi(\epsilon,0)}&=&e^{-i(1+\epsilon)\frac{\pi}{2}\hat{J}_y}\ket{J,J}\nonumber\\
    &=&\left[\left(\frac{\cos(\pi\epsilon/4)}{\sqrt{2}}-\frac{\sin(\pi\epsilon/4)}{\sqrt{2}}\right)\ket{\uparrow}\right.\nonumber\\
    & &\left.+\left(\frac{\cos(\pi\epsilon/4)}{\sqrt{2}}+\frac{\sin(\pi\epsilon/4)}{\sqrt{2}}\right)\ket{\downarrow}\right]^{\otimes N}
\end{eqnarray}
which is not satisfy $C_m=\pm C_{-m}$ when $\epsilon\neq 0$.
However, the equal probability mixed states of $\ket{\Psi(\epsilon,0)}$ and $\ket{\Psi(-\epsilon,0)}$ satisfy $\hat{U}_\textrm{ex}^\dagger\hat{\rho}\hat{U}_\textrm{ex}=\hat{\rho}$, which means that mean signal $(\langle\hat{J}_z(\epsilon)\rangle_\delta+\langle\hat{J}_z(-\epsilon)\rangle_\delta)/2$ is antisymmetric with respect to $\delta=0$, see Fig.~\ref{F2}~(c) and (d).
Moreover, in practice, one can scan the pulse duration to obtain the condition for perfect $\pi/2$ pulse ($\epsilon=0$) by analyzing the antisymmetry of the signal.
So the influences of the imperfect pulse can be easily eliminated in experiments.


\begin{thebibliography}{99}
%
\bibitem{ADLRMP2015}
A. D. Ludlow, M. M. Boyd, J. Ye, E. Peik, and P. O. Schmidt, Optical atomic clocks, Rev. Mod. Phys. \textbf{87}, 637 (2015).
%
\bibitem{ADCRMP2009}
A. D. Cronin, J. Schmiedmayer, and D. E. Pritchard, Optics and interferometry with atoms and molecules, Rev. Mod. Phys. \textbf{81}, 1051 (2009).
%
\bibitem{KHRMP2010}
K. Hammerer, A. S. S$\emptyset$rensen, and E. S. Polzik, Quantum interface between light and atomic ensembles, Rev. Mod. Phys. \textbf{82}, 1041 (2010).
%
\bibitem{MGKRMP2018}
M. G. Kozlov, M. S. Safronova, J. R. Crespo L\'{o}pez-Urrutia, and P. O. Schmidt, Highly charged ions: Optical clocks and applications in fundamental physics, Rev. Mod. Phys. \textbf{90}, 045005 (2018).
%
\bibitem{LPRMP2018}
L. Pezz\`{e}, A. Smerzi, M.~K. Oberthaler, R. Schmied, and P. Treutlein, Quantum metrology with nonclassical states of atomic ensembles, Rev. Mod. Phys. \textbf{90}, 035005 (2018).
%
%
\bibitem{YSGRMP2000}
Ya. S. Greenberg, Application of superconducting quantum interference devices to nuclear magnetic resonance, Rev. Mod. Phys. \textbf{72}, 329 (2000).
%
\bibitem{CLDRMP2017}
C. L. Degen, F. Reinhard and P. Cappellaro, Quantum sensing, Rev. Mod. Phys. \textbf{89}, 035002 (2017)
%
\bibitem{JFBRMP2020}
J. F. Barry, J. M. Schloss, E. Bauch, M. J. Turner, C. A. Hart, L. M. Pham, and R. L. Walsworth, Sensitivity optimization for NV-diamond magnetometry, Rev. Mod. Phys. \textbf{92}, 015004 (2020).
%
\bibitem{MWMRMP2020}
M. W. Mitchell and S. P. Alvarez, Colloquium: Quantum limits to the energy resolution of magnetic field sensors, Rev. Mod. Phys. \textbf{92}, 021001 (2020).
%
\bibitem{SBRMP2025}
S. Bose, I. Fuentes, Andrew. A. Geraci, S. M. Khan, S. Qvarfort, M. Rademacher, M. Rashid, M. Toro\v{s}, Hendrik Ulbricht et al., Massive quantum systems as interfaces of quantum mechanics and gravity, Rev. Mod. Phys. \textbf{97}, 015003 (2025).
%
\bibitem{CCPRX2025}
C. Cassens, B. Meyer-Hoppe, E. Rasel, and C. Klempt, Entanglement-Enhanced Atomic Gravimeter, Phys. Rev. X \textbf{15}, 011029 (2025).
%
\bibitem{JHAPR2024}
J. Huang, M. Zhuang, C. Lee, Entanglement-enhanced quantum metrology: From standard quantum limit to Heisenberg limit, Appl. Phys. Rev. \textbf{11}, 031302 (2024).
%
%
\bibitem{FCRMP2014}
F. Caruso, V. Giovannetti, C. Lupo, and S. Mancini, Quantum channels and memory effects, Rev. Mod. Phys. 86, 1203 (2014).
%
\bibitem{SCPR2025}
S. Chen, Jiahao Huang, M. Zhuang, and C. Lee, Ramsey spectroscopy via pymmetry-protected destructive many-body interferometry (2025).
%
\bibitem{TParXiv2025}
T. Petrucciani, A. Santoni, C. Mazzinghi, D. Trypogeorgos, F. S. Cataliotti, M. Inguscio, G. Modugno, A. Smerzi, L. Pezz\'{e}, and M. Fattori, Mach-Zehnder atom interferometry with non-interacting trapped Bose Einstein condensates, arXiv:2504.17391 (2025).
%
\bibitem{JHarXiv2022}
J. Huang, S. Chen, M. Zhuang, and C. Lee, Robust Interaction-Enhanced Sensing via Antisymmetric Rabi Spectroscopy, arXiv:2208.03179 (2022).
%
\bibitem{HPB2007}
Heinz-Peter Breuer, and F. Petruccione, The Theory of Open Quantum Systems (Oxford, 2007; online edn, Oxford Academic, 1 Feb. 2010).
%
\bibitem{DMAIPA2020}
D. Manzano, A short introduction to the Lindblad master equation, AIP Advances \textbf{10}, 025106 (2020).
%
\bibitem{UDPRL2009}
U. Dorner, R. Demkowicz-Dobrzanski, B. J. Smith, J. S. Lundeen, W. Wasilewski, K. Banaszek, and I. A. Walmsley, Optimal Quantum Phase Estimation, Phys. Rev. Lett 102, 040403 (2009).
%
\bibitem{JHSR2016}
J. Huang, X. Qin, H. Zhong, Y. Ke, and C. Lee, Quantum metrology with spin cat states under dissipation, Sci. Rep. \textbf{5}, 17894 (2016).
%
\bibitem{JLNC2019}
J. Li, A.~K. Harter, J. Liu et al., Observation of parity-time symmetry breaking transitions in a dissipative Floquet system of ultracold atoms, Nat. Commun. \textbf{10}, 855 (2019).
%
\bibitem{CMarXiv2025}
C. Marconi, G. M\"{u}ller-Rigat, J. Romero-Pallej\`{a}, J. Tura and A. Sanpera, Symmetric quantum states: a review of recent progress, arXiv:2506.10185 (2025).
%
\bibitem{MKPRA1993}
M. Kitagawa and M. Ueda, Squeezed spin states, Phys. Rev. A \textbf{47}, 5138 (1993).
%
\bibitem{PYNPJ2025}
P. Yang, G. Bao, J. Chen, W. Du, J. Guo, and W. Zhang, Quantum locking of intrinsic spin squeezed state in Earth-field-range magnetometry, npj Quantum Inf \textbf{11}, 36 (2025). 
%
\bibitem{MRPRA2007}
M. Rodr\'{i}guez, S. R. Clark, and D. Jaksch, Generation of twin Fock states via transition from a two-component Mott insulator to a superfluid, Phys. Rev. A \textbf{75}, 011601(R) (2007).
%
\bibitem{XLScience2017}
X. Luo, Y. Zou, L. Wu, Q. Liu, M. Han, M. K. Tey, and L. You, Deterministic entanglement generation from driving through quantum phase transitions, Science \textbf{355}, 620-623 (2017).
%
\bibitem{SSPRB2021}
S. Sharma, V. A. S. V. Bittencourt, A. D. Karenowska, and S. V. Kusminskiy, Spin cat states in ferromagnetic insulators, Phys. Rev. B \textbf{103}, L100403 (2021).
%
\bibitem{JHPRA2022}
J. Huang, H. Huo, M. Zhuang, and C. Lee, Efficient generation of spin cat states with twist-and-turn dynamics via machine optimization, Phys. Rev. A \textbf{105}, 062456 (2022).
%
\bibitem{FDPRA2025}
F. Du, M. Ma, Z. Bai, and Q. Tan, Generation of arbitrary high-dimensional qudit-based entangled states, Phys. Rev. A \textbf{111}, 032604 (2025).
%
\bibitem{CWH1976}
C. W. Helstrom, Quantum Detection and Estimation Theory (academic Press, New York, 1976).
%
\bibitem{MSAPX2016}
M. Szczykulska, T. Baumgratz, and A. Datta, Multi-parameter quantum metrology, Adv. Phys.: X \textbf{1}, 621 (2016).
%
\bibitem{NB2021}
N. Bouleau, An Intuitive Introduction to Error Structures. In: The Mathematics of Errors. (Springer, Cham, 2021).
%
\bibitem{JHPRA2018}
J. Huang, M. Zhuang, B. Lu, Y. Ke, and C. Lee, Achieving Heisenberg-limited metrology with spin cat states via interaction-based readout, Phys. Rev. A \textbf{98} 012129 (2018).
%
\bibitem{SSMPRA2018}
S. S. Mirkhalaf, S. P. Nolan, and S. A. Haine, Robustifying twist-and-turn entanglement with interaction-based readout, Phys. Rev. A \textbf {97}, {053618} (2018).
%
\bibitem{MZPRX2021}
M. Zhuang, J. Huang, and C. Lee, Many-body quantum lock-in amplifier, PRX Quantum \textbf{2}, 040317 (2021).
%
\bibitem{SKNature2011}
S. Kotler, N. Akerman, Y. Glickman, A. Keselman, and R. Ozeri, Single-ion quantum lock-in amplifier, Nature \textbf{473}, 61–65 (2011).
%
\bibitem{SSScience2017}
S. Schmitt, T. Gefen, T. Unden, G. Wolff, C. M\"{u}ller, J. Scheuer, B. Naydenov, M. Markham, S. Pezzagna, J. Meijer, I. Schwarz, M. Plenio, A. Retzker, L. P. McGuinness, and F. Jelezko, Submillihertz magnetic spectroscopy performed with a nanoscale quantum sensor, Science \textbf{356} 834-837 (2017).
%
\bibitem{JMBScience2017}
J. M. Boss, K. S. Cujia, J. Zopes, and C. L. Degen, Quantum sensing with arbitrary frequency resolution, Science \textbf{356} 837-840 (2017).
%
\bibitem{EDHPRApplied2022}
E. D. Herbschleb, I. Ohki, K. Morita, Y. Yoshii, H. Kato, T. Makino, S. Yamasaki, and N. Mizuochi, Low-Frequency Quantum Sensing, Phys. Rev. Appl. \textbf{18}, 034058 (2022).
%
\bibitem{EDNature2001}
E. A. Donley, N. R. Claussen, S. L. Cornish, J. L. Roberts, E. A. Cornell, and C. E. Wieman, Dynamics of collapsing and exploding Bose-Einstein condensates, Nature \textbf{412}, 295 (2001).
%
\bibitem{GPNJP2018}
G. Pelegr\'{\i}, J. Mompart and V. Ahufinger, Quantum sensing using imbalanced counter-rotating Bose-Einstein condensate modes, New J. Phys. \textbf{20}, 103001 (2018).
%
\bibitem{TVNSR2021}
T. V. Ngo, D. V. Tsarev, R. K. Lee and A. P. Alodjants, Bose-Einstein condensate soliton qubit states for metrological applications, Sci. Rep. \textbf{11},
19363 (2021).
%
\bibitem{BNPRA2011}
B. Y. Ning, J. Zhuang, J. Q. You and W. Zhang, Enhancement of spin coherence in a spin-1 Bose-Einstein condensate by dynamical decoupling approaches, Phys. Rev. A \textbf{84}, 013606 (2011).
%
\bibitem{PAISR2016}
P. A. Ivanov, N. V. Vitanov and K. Singer, High-precision force sensing using a single trapped ion, Sci. Rep. \textbf{6}, 28078 (2016).
%
\bibitem{LDPRAppl2021}
L. Dong, I. Arrazola, X. Chen and J. Casanova, Phase-Adaptive Dynamical Decoupling Methods for Robust Spin-Spin Dynamics in Trapped Ions, Phys. Rev. Appl. \textbf{15}, 034055 (2021).
%
\bibitem{MJBPRA2009}
M. J. Biercuk, H. Uys, A. P. VanDevender, N. Shiga, W. M. Itano, and J. J. Bollinger, Experimental Uhrig dynamical decoupling using trapped ions, Phys. Rev. A \textbf{79}, 062324 (2009).
%
\bibitem{KAGScience2021}
K. A. Gilmore, M. Affolter, R. J. Lewis-Swan, D. Barberena, E. Jordan, A. Maria Rey, and J. J. Bollinger, Quantum-enhanced sensing of displacements and electric fields with two-dimensional trapped-ion crystals, Science, \textbf{373}, 673-678 (2021).
%
\bibitem{FWMS2021}
F. Wolf and P. O. Schmidt, Quantum sensing of oscillating electric fields with trapped ions, Measurement: Sensors \textbf{18}, 100271 (2021).
%
\bibitem{DFPRB2015}
D. Farfurnik, A. Jarmola, L. M. Pham, Z. H. Wang, V. V. Dobrovitski, R. L. Walsworth, D. Budker, and N. Bar-Gill, Optimizing a dynamical decoupling protocol for solid-state electronic spin ensembles in diamond, Phys. Rev. B \textbf{92}, 060301 (2015).
%
\bibitem{JHSEL2012}
J. H. Shim, I. Niemeyer, J. Zhang and D. Suter, Robust dynamical decoupling for arbitrary quantum states of a single NV center in diamond, Europhys. Lett. \textbf{99}, 40004 (2012).
%
%
\bibitem{ZQNPJ2022}
Z. Qiu, A. Hamo, U. Vool, T. X. Zhou and A. Yacoby, Nanoscale electric field imaging with an ambient scanning quantum sensor microscope, npj Quantum Information \textbf{8}, 107 (2022).
%
\bibitem{CLSR2017}
C. Lei, S. Peng, C. Ju, M. H. Yung and J. Du, Decoherence Control of Nitrogen-Vacancy Centers, Sci. Rep. \textbf{7}, 11937 (2017).
%
\bibitem{HZPRX2020}
H. Zhou, J. Choi, S. Choi, R. Landig, A. M. Douglas, J. Isoya, F. Jelezko, S. Onoda, H. Sumiya et al., Phys. Rev. X \textbf{10}, 031003 (2020).
%
\bibitem{HSScience2014}
H. Strobel, W. Muessel, D. Linnemann, T. Zibold, D. B. Hume, L. Pezz\'{e}, A. Smerzi, and M. K. Oberthaler, Science \textbf{345}, 424 (2014).
%
\bibitem{DLPRL2016}
D. Linnemann, H. Strobel, W. Muessel, J. Schulz1, R. J. Lewis-Swan, K. V. Kheruntsyan, and M. K. Oberthaler, Quantum-Enhanced Sensing Based on Time Reversal of Nonlinear Dynamics, Phys. Rev. Lett. \textbf{117}, 013001 (2016).
%
\bibitem{CLFOP2012}
C. Lee, J. Huang, H. Deng, H. Dai, and J. Xu, Nonlinear quantum interferometry with Bose condensed atoms, Frontiers of Physics \textbf{7}(1): 109-130 (2012).
%
\bibitem{EDPRL2016}
E. Davis, G. Bentsen, and M. Schleier-Smith, Approaching the Heisenberg Limit without Single-Particle Detection, Phys. Rev. Lett. \textbf{116}, 053601 (2016).
%
\bibitem{CL2024}
C. Luo, H. Zhang, V. P. W. Koh, J. D. Wilson, A. Chu, M. J. Holland, A. M. Rey, and J. K. Thompson, Momentum-exchange interactions in a Bragg atom interferometer suppress Doppler dephasing, Science \textbf{384} 551-556 (2024).
%
\bibitem{CLPRL2006}
C. Lee, Adiabatic Mach-Zehnder Interferometry on a Quantized Bose-Josephson Junction, Phys. Rev. Lett. \textbf{97}, 150402 (2006).
%
\bibitem{CLPRL2009}
C. Lee, Universality and Anomalous Mean-Field Breakdown of Symmetry-Breaking Transitions in a Coupled Two-Component Bose-Einstein Condensate, Phys. Rev. Lett. \textbf{102}, 070401 (2009).
%
\bibitem{XZPRL2024}
X. Zhang, Z. Hu, and Y. Liu, Fast generation of GHZ-like states using collective-spin XYZ model, Phys. Rev. Lett. \textbf{132}, 113402 (2024).
%
\bibitem{MFRNature2010}
M. F. Riedel, P. B\"{o}hi, Y. Li, T. W. H\"{a}nsch, A. Sinatra, and P. Treutlein, Atom-chip-based generation of entanglement for quantum metrology, Nature \textbf{464}, 1170 (2010).
%
\bibitem{CFOPRL2013}
C. F. Ockeloen, R. Schmied, M. F. Riedel, and P. Treutlein, Quantum Metrology with a Scanning Probe Atom Interferometer, Phys. Rev. Lett. \textbf{111}, 143001 (2013).
%
\bibitem{WMPRL2014}
W. Muessel, H. Strobel, D. Linnemann, D. B. Hume, and M. K. Oberthaler, Scalable Spin Squeezing for Quantum-Enhanced Magnetometry with Bose-Einstein Condensates, Phys. Rev. Lett. \textbf{113}, 103004 (2014).
%
\bibitem{RSNC2017}
R. Shaniv and R. Ozeri, Quantum lock-in force sensing using optical clock Doppler velocimetry, Nat. Commun. \textbf{8}, 14157 (2017).
%
\bibitem{SDCP2020}
S. Dorscher, A. Al-Masoudi, M. Bober, R. Schwarz, R. Hobson, U. Sterr, and C. Lisdat, Dynamical decoupling of laser phase noise in compound atomic clocks, Commun. Phys. \textbf{3}, 185 (2020).
%
%
\bibitem{IAPRA2016}
I. Almog, G. Loewenthal, J. Coslovsky, Y. Sagi and N. Davidson, Dynamic decoupling in the presence of colored control noise, Phys. Rev. A \textbf{94}, 042317 (2016).
%
\bibitem{JBNP2011}
J. Bylander, S. Gustavsson, F. Yan, F. Yoshihara, K. Harrabi, G. Fitch, D. G. Cory, Y. Nakamura, J.-S. Tsai, and W. D. Oliver, Noise spectroscopy through dynamical decoupling with a superconducting flux qubit, Nat. Phys. \textbf{7}, 565–570 (2011).
%
\bibitem{SCCP2024}
S. Chen, M. Zhuang, J. Huang, and C. Lee, Quantum double lock-in amplifier, Commun. Phys. \textbf{7}, 189 (2024).
%
\bibitem{MZQF2024}
M. Zhuang, S. Chen, J. Huang and C. Lee. Quantum lock-in measurement of weak alternating signals, Quantum Front \textrm{3}, 4 (2024). 
%
\bibitem{WZPRL2019}
Zhen-Yu Wang, Jacob E. Lang, Simon Schmitt, Johannes Lang, Jorge Casanova, Liam McGuinness, Tania S. Monteiro, Fedor Jelezko, and Martin B. Plenio, Randomization of Pulse Phases for Unambiguous and Robust Quantum Sensing, Phys. Rev. Lett. \textbf{122}, 200403 (2019).
%
\bibitem{MZPRApplied2020}
M. Zhuang, J. Huang and C. Lee, Phys. Rev. Applied. \textbf{13}, 044049 (2020).
%
\bibitem{RFNPJ2021}
R. Fang, C. Han, X. Jiang, Y. Qiu, Y. Guo, M. Zhao, J. Huang, B. Lu, and C. Lee , Temporal analog of Fabry-P\'{e}rot resonator via coherent population trapping, npj Quantum Inf \textbf{7}, 143 (2021). 
%
\bibitem{ZAWD2017}
Z.~A. Warren, Coherent Population Trapping and Optical Ramsey Interference for Compact Rubidium Clock Development, (2017).
%
\bibitem{EEMOS2010}
E. E. Mikhailov, T. Horrom, N. Belcher, and I. Novikova, Performance of a prototype atomic clock based on $lin||lin$ coherent population trapping resonances in Rb atomic vapor, Soc. Am. B \textbf{27}, 417 (2010).
%
%
\end{thebibliography}
\end{document}